\documentclass[aps,prb,amsmath,amssymb,reprint,author-year,author-numerical,floatfix]{revtex4-2}
\usepackage{graphicx}
\graphicspath{{./images/}}
\usepackage{dcolumn}
\usepackage{bm}
\usepackage{mathtools}
\usepackage{amssymb}
\expandafter\let\csname equation*\endcsname\relax
\expandafter\let\csname endequation*\endcsname\relax
\usepackage{amsmath}
\usepackage{amsbsy}
\usepackage{booktabs}
\usepackage{verbatim}
\usepackage{array}
\usepackage{braket}
\newcolumntype{P}[1]{>{\centering\arraybackslash}p{#1}}
\usepackage[colorlinks=true,citecolor=blue]{hyperref}

\AtBeginDocument{\renewcommand{\natexlab}[1]{#1}}%

\begin{document}

\preprint{AIP/123-QED}

\title[CAI, VAN DE PUT, and FISCHETTI]{Effects of phonon confinement on electron transport in Si nanowire and armchair-edge graphene nanoribbon transistors: 
            A dissipative quantum-transport study}

\author{Bimin Cai}
\affiliation{Department of Materials Science and Engineering, The University of Texas at Dallas\\
             800 W. Campbell Rd., Richardson, TX 75080}
\author{Maarten L. Van de Put}
\altaffiliation[Present address: ]{imec, Kapeldreef 75, B-3001 Leuven, Belgium}
\affiliation{Department of Materials Science and Engineering, The University of Texas at Dallas\\
             800 W. Campbell Rd., Richardson, TX 75080}
\author{Massimo V. Fischetti}
\email[email: ]{max.fischetti@utdallas.edu.}
\affiliation{Department of Materials Science and Engineering, The University of Texas at Dallas\\
             800 W. Campbell Rd., Richardson, TX 75080}            

\date{\today}

\begin{abstract}
Electronic transport in low-dimensional structures, such as thin bodies, nanosheets, nanoribbons and nanowires, is strongly affected by electron and phonon confinement, in addition to interface roughness.  Here we use a quantum-transport formulation based on empirical pseudopotentials and the Master equation to study the effect of the phonon boundary conditions on the electron transport in field effect transistors (FETs) based on a small cross-section (3$\times$3 cells) Si nanowire (NW) and a 10 armchair graphene nanoribbon (10-aGNR). For the dispersion of the confined phonons we employ a simple empirical model based on the folding of the bulk phonon dispersion that approximates the results of the elastic-continuum model at long wavelengths. We consider two extreme cases for their boundary conditions: clamped boundary conditions (CBCs) or free-standing (FSBCs). We find that phonon confinement affects more severely the Si nanowires than graphene
nanoribbons. In particular, for 3$\times$3 SiNW-FETs, CBCs result in a higher room-temperature electron mobility than FSBCs, a result consistent with
what previously reported. On the contrary, in the off-equilibrium conditions seen in gate-all-around (GAA) 3$\times$3 SiNW-FETs with 7~nm gate-length, FSBCs yield 
a higher on-current than what is obtained assuming CBCs. However, for 10-aGNR-FETs, both the electron mobility and the on-current are higher when assuming FSBCs.
\end{abstract}

\maketitle

\section{\label{sec:level1}{Introduction}}
\noindent The purpose of this work is to consider the effect of phonon confinement on the charge-transport properties in metal-oxide-semiconductor field-effect 
transistors (MOSFETs) with channels based on one-dimensional (1D) semiconductor structures, such as nanowires (NWs) or nanoribbons (NRs). 
Interest in such structures is motivated by the short-channel effects that become increasingly severe when scaling MOSFETs to the nanometer
size, since devices based on 1D) channels provide better gate electrostatic control. 

Given the mature state of silicon technology, Si-based devices have been considered in this context.
The well-controlled growth/synthesis and doping of silicon nanowires (SiNWs)~\cite{ycui2000,Ycui2001,Ycui2003,JJie2008,nfukata2009}, has led to the fabrication 
of gate all-around (GAA) SiNW-FETs with sub-10~nm-diameter NWs~\cite{nsingh2006,sbang2009,mlee2009}. Similarly, 
motivated by the high carrier mobility observed in graphene~\cite{Bolotin_2008}, attention has been paid to graphene-based FETs, in particular to 
devices based on graphene nanoribbons with armchair edges (aGNRs), since they exhibit a bandgap~\cite{Son_2006}, as it is required in 
nanoelectronics applications. Recently, sub-1-nm-wide aGNRs with seven (7-aGNRs) or nine (9-aGNRs) dimer lines along the width direction have been successfully 
grown via on-surface synthesis~\cite{ruffieux2012electronic,huang2012spatially,Dinh_2024}, opening the way to the fabrication of FETs based on precisely controlled
aGNRs. The state of the art has been reviewed recently in Refs.~\cite{Tian_2023} and \cite{Lou_2024}. 

In parallel with experimental work, a large amount of theoretical work on electronic transport in these devices has been performed to identify the best 
low-dimensional structures and materials for transistor applications~\cite{Gunst_2016,Ponce_2020,Afzalian_2021,Campi_2023,Cresti_2023}. In particular, 
SiNW-FETs have been studied theoretically by Wang {\it et al.}~\cite{Wang_2004}, Ng {\it et al.}~\cite{Ng_2007}, Jin and coworkers~\cite{sJin2007}, Schenk 
and Luisier~\cite{Schenk_2008}, Rurali {\it et al.}~\cite{Rurali_2008}, Luisier and Klimeck~\cite{Luisier_2009}, and many 
others~\cite{Dong_2011,fang2016pseudopotential,Georgiev_2017}. Similarly, electronic transport in aGNRs and the performance of aGNR-FETs have been studied
theoretically very extensively~\cite{liang2007ballistic,ouyang2007scaling,Fiori_2007,ouyang2008carrier,yoon2008performance,lu2010local,yoon2011role,yoon2012dissipative,
zhao2009computational,betti2011strong,grassi2013mode,fischetti2013pseudopotential,yousefvand2017analytical}.

Some of these studies are based on the semiclassical Boltzmann transport equation, as recently reviewed in chapters 37, 39, and 40 of Ref.~\cite{Springer_2022}. 
This choice is based on the consideration that the interaction of electrons with lattice vibrations and, most important, Coulomb interactions among channel electrons and the high-density electron gases in the contacts (source, drain, gate(s)) usually leads to a loss of coherence. Nevertheless, the quantum confinement in 1D and 2D structures and the short channel lengths of interest render quantum simulations, if not necessary, at least interesting since, for example, they may suggest the ideal maximum performance of the devices under study, as in ballistic simulations. Such approach is most commonly implemented using the non-equilibrium
Green's function (NEGF) method~\cite{Kadanoff_1962,Keldysh_1965}, as implemented in the NEMO computer program~\cite{Lake_1997}, 
with a most recent example provided by the fully {\it ab initio} NEGF study of electronic transport in 2D materials reported in Ref.~\cite{Luisier_2023}. 

Unfortunately, even in such state-of-art implementation, the computational burden required to treat inelastic electron-phonon scattering is so large that
significant approximations are necessary; for example, by considering only zone-center phonons~\cite{Luisier_2023}, or performing the simulations
ignoring dissipation altogether. Such is the case for many of the studies of the theoretical performance of aGNR-FETs that employing the NEGF method 
self-consistently with the three-dimensional (3D) Poisson equation~\cite{ouyang2007scaling,liang2007ballistic,yoon2008performance,zhao2009computational,grassi2009investigation}, 
However, work that extends this method to account for inelastic scattering have highlighted the importance of carrier-phonon scattering even in short-channel 
aGNR-FETs~\cite{grassi2013mode,ouyang2008carrier,yoon2011role,yoon2012dissipative}. Even in studies that account for phonon-induced 
dissipation, often computational simplicity forces the assumption of treating scattering with acoustic phonons as an elastic process or assuming a 
a bulk phonon dispersion, ignoring their confinement. However, the existence of the acoustic phonon confinement has been observed experimentally in GaAs NWs~\cite{Mante_2018} and Si nanomembranes,~\cite{torres2004observations}, for example.

The dispersion of phonons confined in low-dimensional structures has been studied extensively in the past. Nishiguchi~\cite{Nishiguchi_1997} has provided 
a detailed analysis of the symmetry of the acoustic vibrational modes in rectangular-cross-section SiNWs (summarized in Appendix~\ref{Appen.A}, 
mapping Nishiguchi's analysis to the model we have employed). Confined phonons in thin Si films and their effects on the electron mobility have been studied by
Donetti and coworkers~\cite{Donetti_2006}. Similar studies have been performed for SiNWs by Ramayya and coworkers~\cite{Ramayya_2008}, 
by Tienda-Luna {\it et al.}~\cite{tienda2013effect}, all studies being based on the elastic continuum model (also employed by Mickevi\v{c}ius and Mitin~\cite{mickevivcius1993acoustic} for GaAs NWs and in Refs.~\cite{Bannov_1994,Ridley_2000,Perez_1997} in the context of heterostructures and quantum wells), and by Karamitaheri and coworkers~\cite{Karamitaheri_2013} using, instead, the valence force model~\cite{Kane_1985,Paul_2010}.
Normal mode decomposition~\cite{Ye_2015a} and molecular dynamics~\cite{Islam_2017} have also been used to study the vibrational properties of GNRs. 

The advances of density functional theory (DFT) have also rendered almost `routine' {\it ab initio} the calculation 
of the vibrational properties of GNRs, examples being provide by Refs.~\cite{Gillen_2009,Zhang_2014}. Studies of confined optical
modes have been reported in Refs.~\cite{Chamberlain_1993} (superlattices), \cite{Xao_2008} (quantum cascade lasers), \cite{Ridley_2000}
(polar heterostructures), \cite{Perez_2000} (nonpolar optical modes in heterostructures), and \cite{Comas_2007} (polar NWs). However, not much
work has been performed on the study of the effect of confined phonons in low-dimensional FETs using a quantum-transport formalism. 

In this paper, to overcome some of the limitations mentioned above, we study the effect of phonon confinement on charge transport in GAA SiNW- and aGNR-FETs,
paying particular attention to the boundary conditions that determine their confinement. Being interested in details about the
electron-phonon interaction, we treat quantum transport following a different approach based on the Pauli Master equation (PME), as previously described 
by one of us~\cite{Fischetti98jap,Fischetti99prb}, within an atomistic formulation of the electronic structure of the system based on empirical 
pseudopotentials~\cite{Vandeput19computer}. The use of the PME, while correct only when dealing with devices with an active region (the channel) shorter
than the electron phase-coherence length~\cite{vanHove_1954,Zwanzig_1964}, does nevertheless permit a physically correct treatment of inelastic carrier-phonon
scattering, accounting, for example, for the full dependence of the scattering matrix elements on the phonon wave vector.
 
To keep the numerical effort at a tractable level, we describe the confined phonons using an empirical model based on the folding of the dispersion of 
the bulk phonons to the 1D Brillouin Zone (BZ) and on the elastic continuum model at long wavelengths. Hopefully, this will allow us to obtain a qualitatively,
if not quantitatively, correct idea of the effect of phonon confinement on charge transport in these small nanostructures.  

The paper is organized as follows: Section~\ref{sec:theory} presents the theoretical model we employ, starting with a brief overview of the Pauli Master equation 
and the procedure we use to solve the Schr\"{o}dinger equation in an open system described atomistically by empirical pseudopotentials. We then  present the
empirical model used to treat confined phonons. Section~\ref{sec:results} focuses on the results of our research: the phonon spectra, the electron/confined-phonon
scattering rates, and the simulation of GAA SiNW-FETs and aGNR-FETs, comparing the effect of different phonon boundary conditions (BCs) at the surface of the
structure. We summarized our main findings in Sec.~\ref{sec:conclusion}.
\section{\label{sec:theory}{Methods}}
\subsection{\label{subsec:scattering rate}{The Pauli Master equation}}
\noindent As discussed in Refs.~\cite{Fischetti99prb,Fischetti98jap}, the PME gives the correct approach to equilibrium of a system whose dynamics is given 
by the Liouville-von~Neumann equation in the interaction picture~\cite{vonNeumann_1927,Landau_1965},
\begin{equation}
    \frac{\partial {\boldsymbol{\rho}}}{\partial t} = \frac{i}{\hbar} \left [ {\boldsymbol{\rho}},\bf{H}_{\rm int} \right ] + 
                 \left ( \frac{\partial {\boldsymbol{\rho}}}{\partial t} \right )_{\rm res}  \ .
\label{eq:Liouville}
\end{equation}
In this equation, $\hbar$ is the reduced Planck's constant, 
${\boldsymbol{\rho}}$ is the single-electron reduced density matrix, $\bf{H}_{\rm int}$ is the perturbation Hamiltonian describing the 
electron-phonon interaction and the second term on the right-hand side represents the injection/extraction of particles from/into the contacts. The validity 
of Eq.~\ref{eq:Liouville}) is limited to cases in which: {\it i}) $\bf{H}_{\rm int}$ describes the interaction of the system with a `bath' with a large number 
of degrees of freedom (the bath of thermal phonons, in our case); {\it ii}) the initial off-equilibrium state of the system belongs to one of the two classes 
of states considered by van~Hove type~\cite{vanHove_1954}; and {\it iii}) the perturbation Hamiltonian is weak (in the van~Hove sense~\cite{vanHove_1954}). 

In our case of open systems ({\it i.e.}, transistors in which electrons are exchanged between the active region and 
the contacts), this is the class of electrons that are injected into the device with a coherence length longer that the channel. In this case, 
in the basis of eigenstates $\ket{\mu}$ of the eigenstates of the unperturbed Hamiltonian (that accounts for the electron kinetic energy and the electrostatic potential), the off-diagonal elements of the density matrix can be ignored and for the diagonal elements ($\rho_{\mu} = \rho_{\mu \mu}$), at steady state 
Eq.~(\ref{eq:Liouville}) takes the simpler form~\cite{Fischetti99prb,Fischetti98jap}: 
\begin{equation}
    \sum_{E_\nu} \left ( W_{\mu \nu}\rho_\nu - W_{\nu\mu}\rho_\mu \right ) = 
    |A_\mu|^2\upsilon_{\mu,z} \left [ \rho_{\mu} - f_{\rm FD}(E_{\mu}) \right ] \ ,
\label{eq:steady_equation}
\end{equation}
where $W_{\mu\nu}$ is the transition rate from state $\ket{\mu}$ to state $\ket{\nu}$ and $E_{\mu}$ is the energy of the state $\ket{\mu}$.  
The last term on the right hand side describes the injection of carriers 
from the contacts, considered as infinite reservoirs of particles with an equilibrium Fermi-Dirac distribution $f_{\rm FD}(E)$ adjusted self-consistently to
maintain charge neutrality, as described in Refs.~\cite{Fischetti99prb,Fischetti98jap}. The quantity $|A_\mu|$ is the normalization constant of the plane waves 
in the contact, $\upsilon_{\mu,z}$ is the component of the group velocity of the state $\ket{\mu}$ along the transport direction ($z$ axis). For weak interaction
and complete collisions (the van~Hove limit), the transition rates $W_{\mu\nu}$ can be obtained using Fermi's Golden rule:
\begin{equation}
W_{\mu \nu} = \frac{2\pi}{\hbar} \left \vert \braket{\mu|\bf{H}_{\rm int}|\nu} \right \vert^2 
                                        \delta( E_{\mu} - E_{\nu} \pm \hbar \omega ) \ ,
\label{eq:fermis}
\end{equation}
where, in the case of electron-phonon interactions, $\hbar \omega$ is the phonon energy and the upper/lower sign refers to emission/absorption.
Note that we lump into the label $\mu$ that identifies the traveling state not only the eigenvalue of the unperturbed Hamiltonian 
(in general a continuous quantum number that becomes discrete when solving the problem numerically), but also other quantum numbers, such as spin or the
injecting contact (left-going or right-going states). 

Equation~(\ref{eq:steady_equation}) is a linear system consisting of $M$ such equations, where $M$ is the total number of distinct electronic states 
$\ket{\mu}$ injected from all contacts. In Sec.~\ref{subsubsec:el-pho scattering} we discuss the calculation of the transition probabilities 
$W_{\mu \nu}$ in the case of electron/confined-phonon scattering.

The states $\ket{\mu}$ used as basis-functions for the reduced density matrix are the eigenstates of the unperturbed Hamiltonian, $\bf{H}_{0}$, that
accounts for the electrostatic potential of the devices. That is,
\begin{equation}
{\bf{H}}_{0} \ket{\mu} = E_{\mu} \ket{\mu} ,
\label{eq:H0}
\end{equation} 
where ${\bf{H}}_{0} = {\bf{T}} + V_{\rm lat}({\bf {r}}) + V_{\rm H}({\bf r})$, $\bf{T}$ is the kinetic-energy operator, $V_{\rm lat}(\bf r)$ the lattice
periodic (pseudo)potential, and $V_{\rm H}({\bf r})$ the Hartree potential; that is, the electrostatic potential in the device obtained from a 
self-consistent solution of the 3D Poisson equation and the electron charge $\rho({\bf r})=-e\sum_{\mu} \rho_{\mu} \vert \braket{{\bf r}|\mu} \vert^2$
(where $e$ is the magnitude of the electron charge).

The physical foundations and the recent numerically efficient implementation of method we use to solve Eq.~(\ref{eq:H0}) are described at length in
Ref.~\cite{Fischetti_2011} and Refs.~\cite{Vandeput19computer,Maarten18APSmeeting}, respectively. Therefore, here we simply refer the reader to these
publications for the rather intricate detailed description of the method.
 
Since the charge density is determined by the occupation $\rho_{\mu}$, the solution of Eq.~(\ref{eq:steady_equation}), in principle an iterative method is required to find the self-consistent solution. In the following, given the assumption of weak scattering (short channels) implicit when using the PME,
as starting point we employ the solution of the ballistic problem (that is, of Eq.~(\ref{eq:H0})) using the quantum transmitting boundary method 
(QTBM)~\cite{Lent90JAP}, as described in Ref.~\cite{Vandeput19computer}, and consider the PME as a `post-processing' step, stopping at the first iteration. 
We shall show below in Sec.~\ref{subsec:10aGNR_self_consistency} that, in the case of a 10-aGNR-FET, when performing a few self-consistent iterations, the
Hartree potential does show significant changes (almost exclusively at the drain side of the device) and the current does not change appreciably when 
performing several iterative steps. 
\subsection{\label{subsec:phonon dispersion}{Phonons in one-dimensional structures}}
\subsubsection{\label{subsubsec:1Dphonons}{An empirical model}}
\noindent To deal with electron/confined-phonon scattering, we must first approximate the dispersion of phonons in one-dimensional systems using a reasonable but
numerically convenient model. In principle, the phonon spectrum could be obtained from {\it ab initio} calculations (such as from density functional theory, DFT) 
or from some semi-empirical model, such as the modified force-field model~\cite{Paul_2010}, the adiabatic bond charge model~\cite{Weber_1977}, or the valence-shell
model~\cite{Kunc_1979}. First-principle methods would be numerically very expensive, given the large size of the supercells and the presence of the applied bias would severely complicate calculations based on computer programs such as EPW~\cite{EPW_2010}. 

On the other hand, the use of semi-empirical model would require the identification of each phonon branch (longitudinal or transverse, acoustic or optical), another demanding task. Therefore, we prefer to focus on implementing a numerically  'light' and simple model that captures, even if only qualitatively, the basic physical picture (phonon dispersion and 
electron-phonon matrix elements) without creating an excessive numerical burden. The model we employ is based on an empirical form for the ion displacement field 
-- derived from the elastic/dielectric continuum model~\cite{Landau_1970,Ridley_2000} in the limit of wide structures (wide ribbons and nanowires with a 
large cross-sectional area), as shown in Appendix~\ref{Appen.A} -- and, at short wavelength, on the folding of the bulk phonon dispersion into the 1D BZ. 
This is required by the fact that the validity of the  elastic-continuum model is limited to long-wavelength ionic vibrations. This is seen in the unphysical
large frequency of modes whose wavelength is shorter than the lattice constant~\cite{Donetti_2006}. This may not be a problem
when focusing on low-field transport, but it is a serious issue when dealing with high-field transport in devices. 
The details are shown in Sec.~\ref{subsec:phonon dispersion} and in Sec.~\ref{sec:results} we compare our results with those obtained using DFT, showing  
that indeed our model captures correctly the basic physics.

Since atomic vibrations at the surfaces of the NWs or edges of the aGNRs are affected by the surrounding medium (vacuum or, typically,
gate insulators), we consider the two extreme cases for the boundary conditions at surfaces/edges: clamped boundary conditions (CBCs, as when assuming that the 1D structure is surrounded by a `hard' gate insulator, such as Al$_2$O$_3$ or HfO$_2$) or free-standing (FSBCs, for a `soft' gate insulator, such as
SiO$_2$). Of course, the real conditions is a complicated intermediate situation. For example, as done by Ridley {\it et al.}~\cite{Ridley_1993,Chamberlain_1993} 
in the case of III-V quantum wells, optical phonons are better described by CBCs, while for acoustic phonons the situation is less clear and depends strongly on the nature of the surrounding medium.\\

\noindent {\it 1A. Clamped-surface boundary conditions}\\

\noindent Using Cartesian axes with the $z$ axis along the transport (axial) direction of the NW or NR, $y$ axis along the width of the NR, or 
the cross-section of the NW on the $(x,y)$ plane, the divergence of the ionic displacement, $\nabla \cdot {\bf u}(\bf {r})$ for confined phonon labeled 
by quantum numbers $n$ and $m$ can be approximated by:
\begin{equation}
\nabla \cdot {\bf u}_{n,m}({\bf {r}}) = i \ q_{n,m} \ \left ( \frac{\hbar}{2 \rho \omega} \right )^{1/2} \ e^{iQz} \ \mathcal{G}_{n,m}(x,y)  \ ,
\label{eq:displ_1} 
\end{equation} 
where $\rho$ is the bulk mass density of the  crystal (per unit volume in the case of NWs, per unit area in the case of NRs), $\omega$ is the phonon frequency, $q_{n,m}$ is the magnitude of the wave vector ${\bf q}_{n,m} = ({\bf k}_{n,m},Q)$, having indicated with ${\bf k}_{n,m}$ 
the quantized wave vector on the transverse $(x,y)$ plane. The function $\mathcal{G}_{n,m}$, which we will refer to as to the `phonon shape function', 
represents the divergence of the standing waves normalized to the width $W$ of the nanoribbon or the area $A$ of the nanowire with cross section 
$L_{x} \times L_{y}$:
\vspace{0.15cm}
\begin{equation}
\mathcal{G}_{n,m}(x,y) \ = \left \{
\begin{array}{ll}
\left ( \dfrac {2}{W}\right )^{1/2} \cos \left ( \dfrac{n \pi y}{W} \right ) & \mbox{(aGNR)} \\
\left ( \dfrac {4}{A}\right )^{1/2}  \cos \left ( \dfrac{n \pi x}{L_{x}} \right ) 
              \cos \left ( \dfrac{m \pi y}{L_{y}} \right ) & \mbox{(SiNW)} \\
\end{array}
\right. 
\label{eq:shape_function_clamped_surface}
\end{equation}\\

\vspace{-0.15cm}
\noindent The derivation of these expressions is given in Appendix~\ref{Appen.A} for the simpler case of aGNRs, although a similar procedure can be followed also in the case
of NWs. Obviously, in the first expression the index $m$ plays no role, since we are assuming the GNRs formed by cutting a 2D sheet with 2D phonons. 
Appendix~\ref{Appen.A} shows that these boundary conditions do not allow any mode with $n$=0 or $m=0$, so Eq.~(\ref{eq:shape_function_clamped_surface}) 
is valid only for non-vanishing $n$ and $m$.\\

\noindent {\it 1B. Freestanding-surface boundary conditions}\\

\noindent In the opposite case of FSBCs, the elastic continuum model requires the strain, ${\boldsymbol{\sigma}} = {\bf C} \cdot \nabla {\bf u}$ (${\bf C}$ 
is the fourth-rank stiffness tensor and the notation $\nabla {\bf u}$ stands for the second-rank tensor 
$\partial_{i} u_{j}$, with $i,j = 1,3$) to vanish the edge/surface of the structure; that is ${\boldsymbol{\sigma}} \cdot {\bf n}_{\rm S}=0$, 
where ${\bf n}_{\rm S}$ is the normal to the surface. For transversally isotropic materials (with a single transverse sound velocity), 
$C_{ijkl} \approx \sigma_{ij} \delta_{kl}$ (where $\boldsymbol{\sigma}$ is the stress tensor) 
and this condition becomes $\partial_{l} u_{l}=0$ at the surface. Therefore, Eq.~(\ref{eq:shape_function_clamped_surface}) is replaced by:
\begin{equation}
\mathcal{G}_{n,m}(x,y) \ = \left \{
\begin{array}{ll}
\left ( \dfrac {2}{W}\right )^{1/2} \sin \left ( \dfrac{n \pi y}{W} \right ) & \mbox{aGNR} \\
\left ( \dfrac {4}{A}\right )^{1/2} \sin \left ( \dfrac{n \pi x}{L_{x}} \right ) 
                 \sin \left ( \dfrac{m \pi y}{L_{y}} \right ) & \mbox{SiNW} \\
                
\end{array}
\right. 
\label{eq:shape_function_free_standing}
\end{equation}
Note that assuming this simplified model for FSBCs, the mode with $n=0$ (NRs) (or $n=0$, $m=0$ for rectangular NWs), a purely longitudinal dilatational
acoustic-like mode, has a vanishing divergence of the displacement ${\bf u}$ and plays no role in any scattering process. On the contrary, from the 
discussion of Appendix~\ref{Appen.A}, we see that such a mode exists and we must assume that $\mathcal{G}_{0,0}(x,y) = 1/W^{1/2}$ (for GNRs) 
or $=1/A^{1/2}$ (for NWs).

The number of phonon branches we need to consider is determined by the fact that  
the unit cell of rectangular cross-section (100) SiNWs contains 4 atoms. Since there are $N_{\rm x}$$\times$$N_{\rm y}$ cells in each
supercell (where $N_{\rm x}$($N_{\rm y}$) represents $N$ number of cell along $x (y)$ direction), the supercell contains $N_{\rm a}=4N_{\rm x}N_{\rm y}$
atoms. This results in a total of 12$N_xN_y$ ($3 N_{\rm a}$) phonon branches. Of these, 4$N_x N_y$ are flexural waves and are ignored: Since the structures
we consider here are mirror-symmetric, these flexural modes do not couple to electrons~\cite{Fischetti_2016}, as mentioned above. Both the longitudinal and the 
shear waves have 4$N_x N_y$ modes, respectively, half of which are  acoustic phonons, and the remaining modes are optical~\cite{cnote3}. When assuming FSBCs, 
$0\leq n(m)\leq 2N_x(2N_y)-1$) and even-even combinations of indices $n$ and $m$ correspond to transverse symmetric shear waves, while odd-odd combinations correspond to longitudinal antisymmetric dilatational waves. These even-even/odd-odd assignments are the opposite under CBCs ($1\leq n(m) \leq 2N_x(2N_y)$). 
Flexural modes arise from different parity combinations of $n$ and $m$ under either boundary condition.

For an $N$-aGNR, the supercell contains $2N$ atoms and $6N$ modes, with $4N$ modes coupling to electrons: $2N$ flexural ('ZA') phonons, evenly split between optical and acoustic types. For FSBCs ($0 \leq n \leq N-1$), even $n$ corresponds to transverse shear waves, and odd $n$ to longitudinal dilatational waves. Once more, for CBCs ($1 \leq n \leq N$), these even/odd are the opposite.
\subsubsection{\label{Equation of phonons}{Phonon dispersion}}
\noindent As we discussed above, to extend our model beyond the long-wavelength region of validity of the elastic-continuum model, we fold the dispersion 
of the bulk phonons into the first 1D BZ by mapping the axial ($z$) component $Q$ of the phonon phonon wave vector inside the 
1D BZ when it is larger than $\pi/a$, where $a$ is the size of the unit cell of the nanowire or nanoribbon along the axial direction. Thus, for the 
dispersion of acoustic phonons, we split a single mode with sub-band indices $(n,m)$ into two modes:
\begin{equation}
\omega_{n,m}^{\rm (ac)}(Q) = \left \{ 
    \begin{array}{ll}
        \omega_{\rm BZ} \sin \left ( \dfrac{aq}{4} \right )               & Q \in\left[0,\frac{\pi}{a}\right]       \\
        \\
        \omega_{\rm BZ} \sin \left ( \dfrac{a \widetilde{q}}{4} \right )  & Q \in\left[\frac{\pi}{a},\frac{2\pi}{a} \right ]
    \end{array}
    \right.  \ ,            
\label{eq:dispersion_ac_sinw}
\end{equation}
with ${\bf q}=({\bf k}_{n,m},Q)$, $\widetilde{\bf q}=({\bf k}_{n,m},2\pi/a-Q)$,
$n$ and $m$ both odd or both even and $n,m \le 2 N_{\rm x,y}$ (NWs) or $n \le N$ (NRs), and $\omega^{ac}_{BZ}$ is the dilatational 
or shear waves of acoustic phonon frequency at the edge of the BZ. Similarly, for optical phonons, the `folded' version:
\begin{equation}
\omega_{n,m}^{\rm (op)}(Q) = \left \{ 
    \begin{array}{ll}
        \dfrac{\omega_{\Gamma} + \omega_{\rm BZ}}{2} + \dfrac{\omega_{\Gamma} - \omega_{\rm BZ}}{2} \cos \left ( \dfrac{aq}{2} \right )\\ \\
                     \hspace{4.0cm} \left ( \mbox{for } Q \in\left[0,\frac{\pi}{a}\right] \right ) \\ \\
        \dfrac{\omega_{\Gamma} + \omega_{\rm BZ}}{2} + \dfrac{\omega_{\Gamma} - 
                          \omega_{\rm BZ}}{2} \cos \left ( \dfrac{a \widetilde{q}}{2} \right ) \\ \\
                     \hspace{4.0cm} \left (\mbox{for } Q \in\left[\frac{\pi}{a},\frac{2\pi}{a} \right ] \right )
    \end{array}
    \right. \ ,             
\label{eq:op_dispersion_sinw}
\end{equation}
for $n,m \le 2 N_{\rm x,y}$ (only even-even and odd-odd combinations), where $\omega_\Gamma$ and $\omega^{op}_{BZ}$ are the frequencies of the dilatational or shear waves of optical phonons at the center and the edge of the BZ, respectively.\\

Recalling that ${\bf q}_{n,m} = ({\bf k}_{n,m},Q)$, the variable $q_{n,m}$ denotes the magnitude of the `total' wave vector of the confined phonons, that is:
\begin{equation}
q_{n,m}^{2} = k^{2}_{n,m} + Q^{2} = \left \{
\begin{array}{ll}
\dfrac{n^{2}\pi^{2}}{W^{2}} + Q^{2} & \mbox{(aGNR)}\\
\\ 
\dfrac{n^{2}\pi^{2}}{L_{\rm x}^{2}} + \dfrac{m^{2}\pi^{}}{L_{\rm y}^{2}} + Q^{2}  & \mbox{(SiNW)} \\
\end{array}
\right. \ ,
\label{eq:total_wave_vector}
\end{equation}
thus splitting $q_{n,m}^{2}$ into its axial and transverse components, $Q^{2}$ and $k_{n,m}^{2}$, respectively.
\subsubsection{\label{subsubsec:el-pho scattering}{Electron/confined-phonon scattering}}
\noindent As mentioned above, we use first-order perturbation theory (Fermi's golden rule) to calculate the electron-phonon scattering rates, $W_{\mu \nu}$. 
We also employ the adaptive discretization of the energy spectrum and wavefunction normalization $\braket{\mu|\mu}_{\Omega}=L_{\mu}$, where $L_{\mu}$ is a 
state-dependent length obtained from the procedure given by Eqs.~(20) and (21) of Ref.~\cite{Vandeput19computer}. Using this normalization, the rate at which 
an electron in state $\ket{\nu}$ with wavefunction $\braket{{\bf r}|\nu}=\Psi_{\nu}({\bf r})$ emits or absorbs a phonon is given by:
\begin{multline}
\frac{1}{\tau^{\rm (ac)}_{\nu}} =
 \frac{\Delta_{\rm dil/sh}^{2}}{2\hbar\rho}  
    \sum_{\mu,n,m}
      \frac{ w_{f} \Delta E'_{\mu} \mathcal{D}_{\rm el}(E_{\mu}) \Delta E'_{\nu} \mathcal{D}_{\rm el}(E_{\nu}) q^{2}_{n,m} }
           { \omega(q_{n,m}) \left \vert \frac{ {\rm d}\omega(q_{n,m}) }{ {\rm d}Q } \right \vert  } \\
    \times \left \vert \int_{\Omega} \ {\rm d}{\bf r} \ \Psi^{\ast}_{\mu}({\bf r}) \ e^{iQ_{n,m}z} \ \mathcal{G}_{n,m}(x,y) \ \Psi_{\nu}({\bf r})
             \vert \right \vert^{2} \\ 
               \times \left ( N_{q,n,m} + \frac{1}{2} \pm \frac{1}{2} \right ) \ .                                      
\label{eq:acoustic_contact_normal}
\end{multline}
In this expression, the upper/lower sign denotes absorption/emission, $\Delta_{\rm dil/sh}$ is the deformation potential for scattering with confined acoustic phonons (dilatational/shear modes), and $\rho$ is the bulk mass density (2D for aGNRs, 3D for SiNWs). The `weight factor' $w_{\rm f}$ represents the fraction of the (discretized) energy range spanned by the final state that overlaps with the range spanned by the initial state, and $\Delta E'$ is the energy interval of states used to discretize the problem. The quantity $\mathcal{D}_{\rm el}(E)$ is the 1D electron density of states at the energy $E$ in the 
contacts. The indices $n$ and $m$ denote the phonon branches of the nanowire and it is implied that only the index $n$ is used for nanoribbons, a notation
we follow throughout. Since we are considering only states at discrete energies with intervals $\Delta E'$, energy conservation is enforced by looking for
possible values of 
$q_{n,m} = \left ( k^2 _{n,m} + Q^2 \right )^{1/2}$ that satisfies the energy-conserving condition:
\begin{equation}
    \hbar\omega(q_{n,m}) = \pm (E_\mu - E_\nu) \ ,
\label{eq:abs_emit_equation}
\end{equation}
(where the upper/lower sign denotes absorption/emission) up to $\pm \Delta E'/2$. Finally, $N_{q,n,m}$ denotes the Bose-Einstein occupation of the phonons,
assumed to be at equilibrium at the lattice temperature. The shape function $\mathcal{G}_{n,m}(x,y)$ is given by Eqs.~(\ref{eq:shape_function_clamped_surface})
or (\ref{eq:shape_function_free_standing}), depending on the phonon boundary conditions, CBCs or FSBCs, respectively.

In a similar fashion, the nonpolar scattering with optical phonons can be expressed as:
\begin{multline}
\frac{1}{\tau^{\rm (nop)}_{\nu}} = \frac{(DK)_{\rm op}^{2}}{2 \hbar \rho} 
    \sum_{\mu,n,m}
      \frac{ w_{f} \Delta E'_{\mu} \mathcal{D}_{\rm el}(E_{\mu}) \Delta E'_{\nu} \mathcal{D}_{\rm el}(E_{\nu})}
           { \omega(q_{n,m}) \left \vert \frac{ {\rm d}\omega(q_{n,m}) }{ {\rm d}Q } \right \vert  } \\
    \times \left \vert \int_{\Omega} \ {\rm d}{\bf r} \ \Psi^{\ast}_{\mu}({\bf r}) \ e^{iQ_{n,m}z} \ \mathcal{G}_{n,m}(x,y) \ \Psi_{\nu}({\bf r})
             \vert \right \vert^{2} \\ 
               \times \left ( N_{q,n,m} + \frac{1}{2} \pm \frac{1}{2} \right )                                      
\label{eq:optic_contact_normal}
\end{multline}
where $(DK)_{\rm op}$ is a constant deformation potential for scattering with optical phonons.
\section{\label{sec:results}{Results}}
\subsection{\label{subsec:device structure}{Device structure}}
\begin{figure}[tb]
{\vbox{
\centerline{\includegraphics[width=8.0cm]{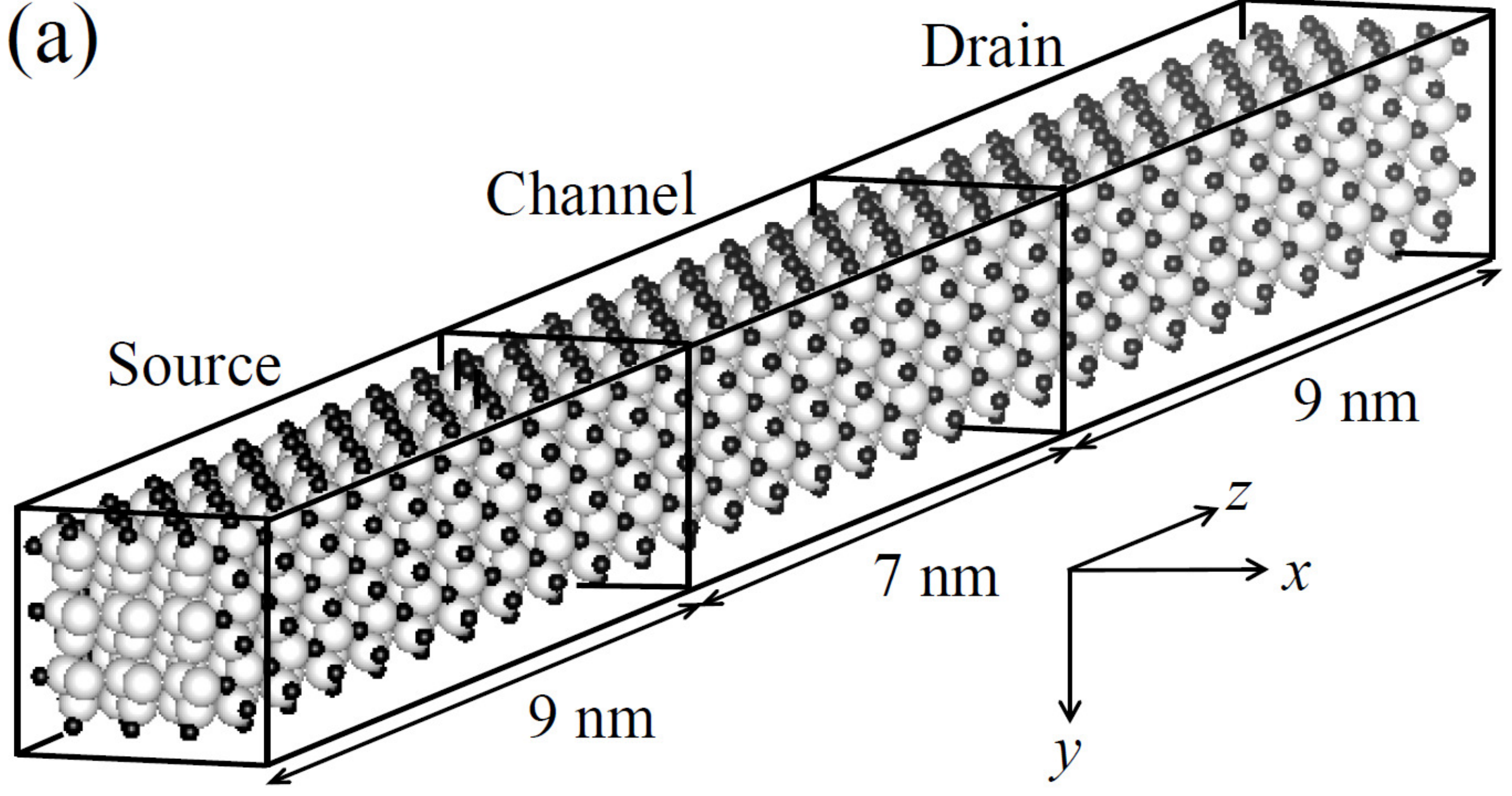}}
\vspace{0.25cm}
\centerline{\includegraphics[width=8.0cm]{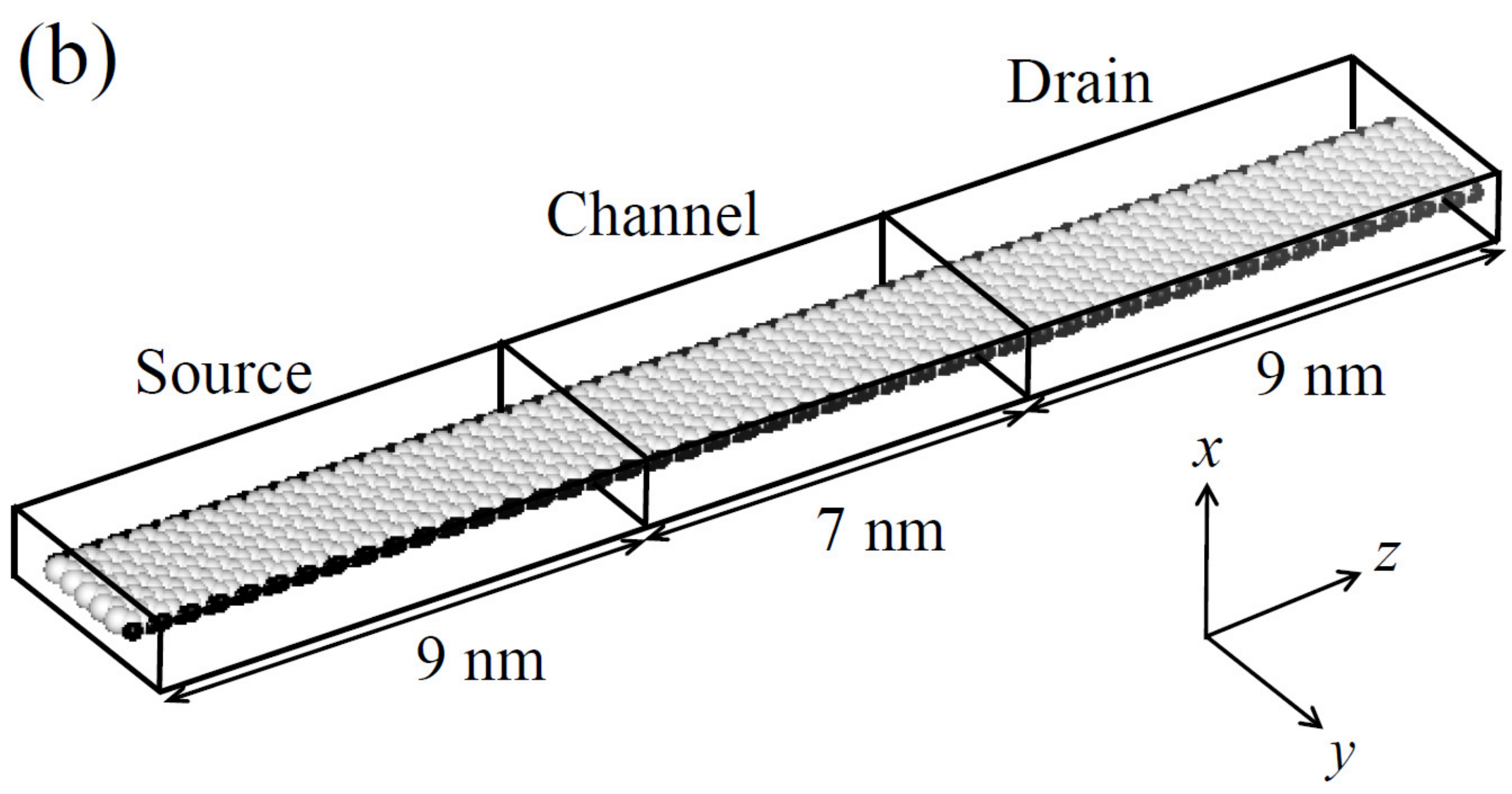}}
}}
\caption{\label{fig:device_structure} Schematic representation (qualitative only, not to scale) of the structures considered here: 
                 (a) the GAA 3$\times$3 SiNW-FET and (b) the GAA 10-aGNR-FET. The gray spheres represent Si and C atoms in (a) and (b), respectively 
                 whereas the dark dots represent the terminating H atoms. The region enclosed in the 'box' includes the semiconductor and the vacuum used 
                 to mimic a gate insulator with an effective SiO$_2$-equivalent thickness of 0.7~nm. Ohmic contacts are assumed at the edges of the 
                 source and drain extensions and an ideal 7.0~nm-long all-around metal gate is assumed to surround the channel region.}
\end{figure}

\noindent We have considered the GAA SiNW-FETs and aGNR-FETs illustrated schematically in Fig.~\ref{fig:device_structure}. For the SiNW-FET, the channel is a SiNW
with a cross section of 3$\times$3 cell cubic cells (1.15~nm$\times$1.15~nm on the $(x,y)$ plane) with the $z$ axis along the (001) direction and surfaces
terminated by hydrogen atoms. The calculated band gap is 2.93~eV. For aGNR-FETs, the channel is a 10-aGNR (1.11~nm wide along the $y$ direction) with edges
terminated by hydrogen atoms and a calculated band gap of 1.3~eV. To mimic the gate insulator, these structures are embedded in vacuum with the gate contact separated from the channel by a distance that yields a SiO$_2$-equivalent thickness of 0.7~nm. Although such a large physical thickness affects the characteristics
of the devices (mainly, transconductance and subthreshold slope), our goal is to compare the effects of different phonon boundary conditions, not the intrinsic
performance of these FETs. In both devices the channel is left undoped, while the source and drain regions are assumed to be n-type-doped with a `conservative'
donor concentration of $6.7 \times 10^{5}$~{cm}$^{-1}$ for the SiNW-FET and $1.8 \times 10^{5}$~{cm}$^{-1}$ for the aGNR-FET, although a much larger carrier
density can be obtained by employing various methods of modulation doping~\cite{Fanciulli_2016,Verbeeck_2017,Choi_2021}. The simulated region is 25~nm long, 
with a channel length of 7~nm, and it contains 2820 atoms for the 3$\times$3 SiNW-FET (1692 silicon and 1128 hydrogen atoms) rand 1416 atoms for the 10-aGNR-FET
(1180 carbon and 236 hydrogen atoms). 
\subsection{\label{subsec:phonon_dispersion_calculation}{Phonon dispersion}}
\noindent Table~\ref{phonon_parameters} lists the acoustic and optical phonon energies and deformation potentials we have employed. 
These values result in a realistic sound velocities at $\Gamma$, $\upsilon_s = a \omega^{\rm (dil;ac)}_{\rm BZ}/4 \approx 9.3 \times 10^5$~cm/s for dilatational acoustic phonons and $\approx 4.1 \times 10^5$~cm/s for shear acoustic phonons in the 3$\times$3 SiNW. As we discuss below, the sound velocity 
(or, better, the phonon density of states that it implies) is a key element that affects electronic transport. In the case of the SiNWs, there are 108 
phonon branches in total, one-third of them being flexural modes. Of the remaining 72 modes, 36 are longitudinal waves and 36 shear waves, divided equally 
into acoustic and optical modes. 

Moving to the 10-aGNRs, the acoustic and optical phonons energy listed in Table~\ref{phonon_parameters} also result in a realistic sound velocity at $\Gamma$,
$\upsilon_s = a \omega^{\rm (dil;ac)}_{\rm BZ}/4 \approx 20 \times 10^5$~cm/s for dilatational acoustic phonons and $\approx 10 \times 10^5$~cm/s for
shear acoustic phonons. There are 60 phonon branches in total. Of these, only 40 modes couple to the electrons, while 20 modes are acoustic phonons and 
the other 20 modes are optical phonons.
 
\begin{figure}[tb]
\centerline{\includegraphics[scale=1]{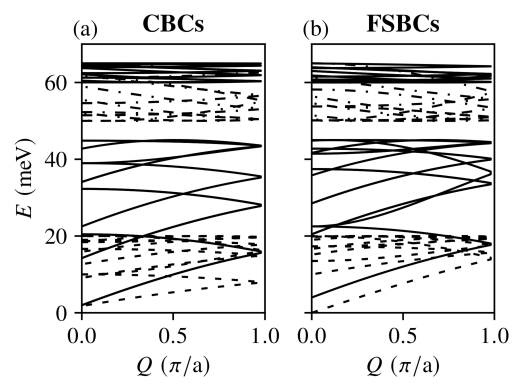}}
\caption{\label{fig:SiNW_phonon_BC_empi} Phonon dispersion for a 3$\times$3 SiNW obtained using our empirical model in the case of CBCs (a) and of FSBCs (b). 
           The lines below about 50~meV represent acoustic dilatational (solid lines) and shear modes (dashed lines), while the high-energy branches 
           (above about 50~meV) represent optical dilatational (solid lines) and shear modes (dashed lines). The additional 36 flexural branches are not 
           shown, since they are assumed to be decoupled from electrons.}
\end{figure}
\begin{figure}[tb]
\centerline{\includegraphics[scale=1]{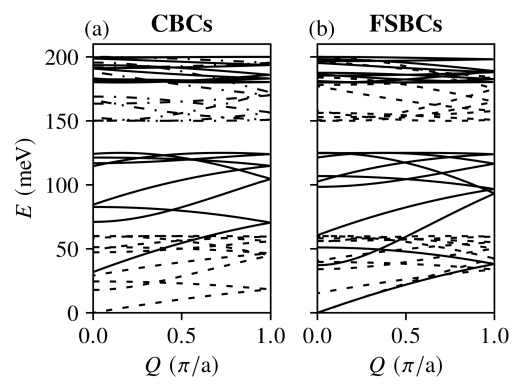}}
\caption{\label{fig:10aGNR_phonon_BC_empi}
           As in Fig.~\ref{fig:SiNW_phonon_BC_empi}, but for the 10-aGNRs. In this case, 20 additional flexural branches are not shown since, once more, 
           they are assumed to be decoupled from electrons.}   
\end{figure}
\begin{figure}[tb]
\centerline{\includegraphics[scale=1]{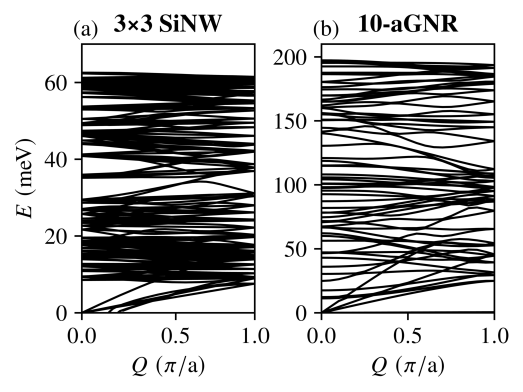}}
\caption{\label{fig:DFT_phonon_all} Phonon dispersion obtained using DFT with surfaces terminated by H to mimic free-standing 
          boundary conditions for the ionic displacement (FSBCs) in (a) 3$\times$3 SiNWs and (b) 10-aGNRs. Note the many low-energy
          flexural branches that do not couple to electrons in these atomically mirror-symmetric structures. The negative squared frequencies 
          (plotted as negative frequencies for illustration purposes only) seen in the last frame are the result of a computational artifact.}
\end{figure}

Figures~\ref{fig:SiNW_phonon_BC_empi} and \ref{fig:10aGNR_phonon_BC_empi} show the phonon dispersions of the 3$\times$3 SiNW and the 10-aGNR, respectively.
\begin{table*}[tb]
\centering
\caption{Phonon energies and deformation potentials for SiNWs and aGNRs}
\label{phonon_parameters}
\begin{tabular}{lcccc}
\toprule
\bf{Quantity}                       &              \bf{Symbol}                 & \bf{Value (SiNWs)}  & \bf{Value (aGNRs)}    &  \bf{Units} \\
\midrule
Dilatational acoustic phonon energy & $\hbar \omega^{\rm (dil;ac)}_{\rm BZ}$   &       45            &        12             &      meV     \\
Shear acoustic phonon energy        & $\hbar \omega^{\rm (sh;ac)}_{\rm BZ}$    &       20            &        60             &      meV     \\
Optical phonon energy at $\Gamma$   & $\hbar \omega_{\Gamma}$                  &       65            &        200            &      meV     \\
Dilatational optical phonon energy  & $\hbar \omega^{\rm (dil;op)}_{\rm BZ}$   &       60            &        180            &      meV     \\
Shear optical phonon energy         & $\hbar \omega^{\rm (sh;op)}_{\rm BZ}$    &       50            &        150            &      meV     \\
Dilatation deformation potential    & $\Delta_{\rm dil}$                       &        9            &        12.3           &      eV      \\
Shear deformation potential         & $\Delta_{\rm sh}$                        &        1            &        0.8            &      eV      \\   
Optical deformation potential       & $(DK){\rm op}$                           &  1.75$\times 10^8$  &    3.1$\times 10^8$   &     eV/cm    \\ 
\bottomrule
\end{tabular}
\end{table*}
It is interesting to compare these results with the dispersion obtained using DFT. This is shown in Fig.~\ref{fig:DFT_phonon_all}. 
These results have been obtained using Quantum ESPRESSO (QE)~\cite{giannozzi2009quantum} with the Perdew–Burke–Ernzerhof (PBE)~\cite{perdew1996generalized}
generalized gradient approximation (GGA) exchange-correlation potential. The 3$\times$3 SiNW and the 10-aGNR have been assumed to be terminated by 
hydrogen atoms to mimic the case of FSBCs. Attempts to mimic CBCs by terminating the surfaces/edges with heavy elements have caused several severe artifacts.
Indeed, one obvious issue with the spectra shown in Fig.~\ref{fig:DFT_phonon_all}, especially for the 3$\times$3 SiNW, is the presence of negative
(squared) frequencies for the flexural phonons at long wavelengths. This is a common artifact resulting from the necessity of employing very large
supercells. Termination by heavy elements rendered this issue even more severe. Being interested only in the qualitative nature of the results, 
we have not attempted to resolve this issue. In addition to the presence of the flexural modes, not shown in Figs.~\ref{fig:SiNW_phonon_BC_empi} 
and \ref{fig:10aGNR_phonon_BC_empi}, the most obvious shortcoming of our empirical model is the appearance of an optical/acoustic gap at energies 
between $\sim$45 to 50~meV for 3$\times$3 SiNW (Fig.~\ref{fig:SiNW_phonon_BC_empi}) and $\sim$120 to 150~meV for 10-aGNR (Fig.~\ref{fig:10aGNR_phonon_BC_empi}), 
a gap that is not seen in the results shown in Fig.~\ref{fig:DFT_phonon_all}. This results from the fact that our `folded' model for the phonon confinement 
ignores the elastic interaction among the confined branches. However, the low-frequency range of the vibrational spectrum, important in low-field transport, 
is well captured by our model.  
\subsection{\label{subsec:scattering calculation}{Electron/confined-phonon interaction: Effects of the boundary conditions}}
\noindent In principle, as mentioned in Sec.~\ref{sec:theory}, the full Schr\"{o}dinger/PME/Poisson problem should be solved self-consistently. 
This is quite a numerically intensive task: On the one hand, the solution of the open-BC Schr\"{o}dinger/pseudopotential/QTBM problem is very efficient, 
thanks to the high efficiency of the method discussed in Ref.~\cite{Vandeput19computer}. Similarly, the solution of the Poisson equation can be obtained 
without any effort. The numerically expensive task is the calculation of the matrix elements (the overlap integrals) that enter the expression for the 
scattering rates, Eq.~(\ref{eq:acoustic_contact_normal}) and (\ref{eq:optic_contact_normal}). This amounts to evaluating numerically $M \times M$ integrals
(about $10^6$ for a typical number of states $M \approx 10^3$ obtained discretizing the energy range (-0.05,0.4)~eV) 
over $\approx 10^4-10^5$ spatial mesh points. Since the use of the PME amounts implicitly in assuming 
weak scattering (in the van~Hove sense), most of the results presented here are obtained performing only one iterative `step', treating the PME as 
a `post-processing' step, using the states $\ket{\mu}$ obtained from the ballistic solution. In Sec.~\ref{subsec:10aGNR_self_consistency} we shall
discuss the effect of the self-consistency showing that, as expected, the drain current of a 10-aGNR-FET does not change appreciably when performing 
several iterative steps.\\ 
 
\begin{figure}[tb]
\centerline{\includegraphics[width=8.5cm]{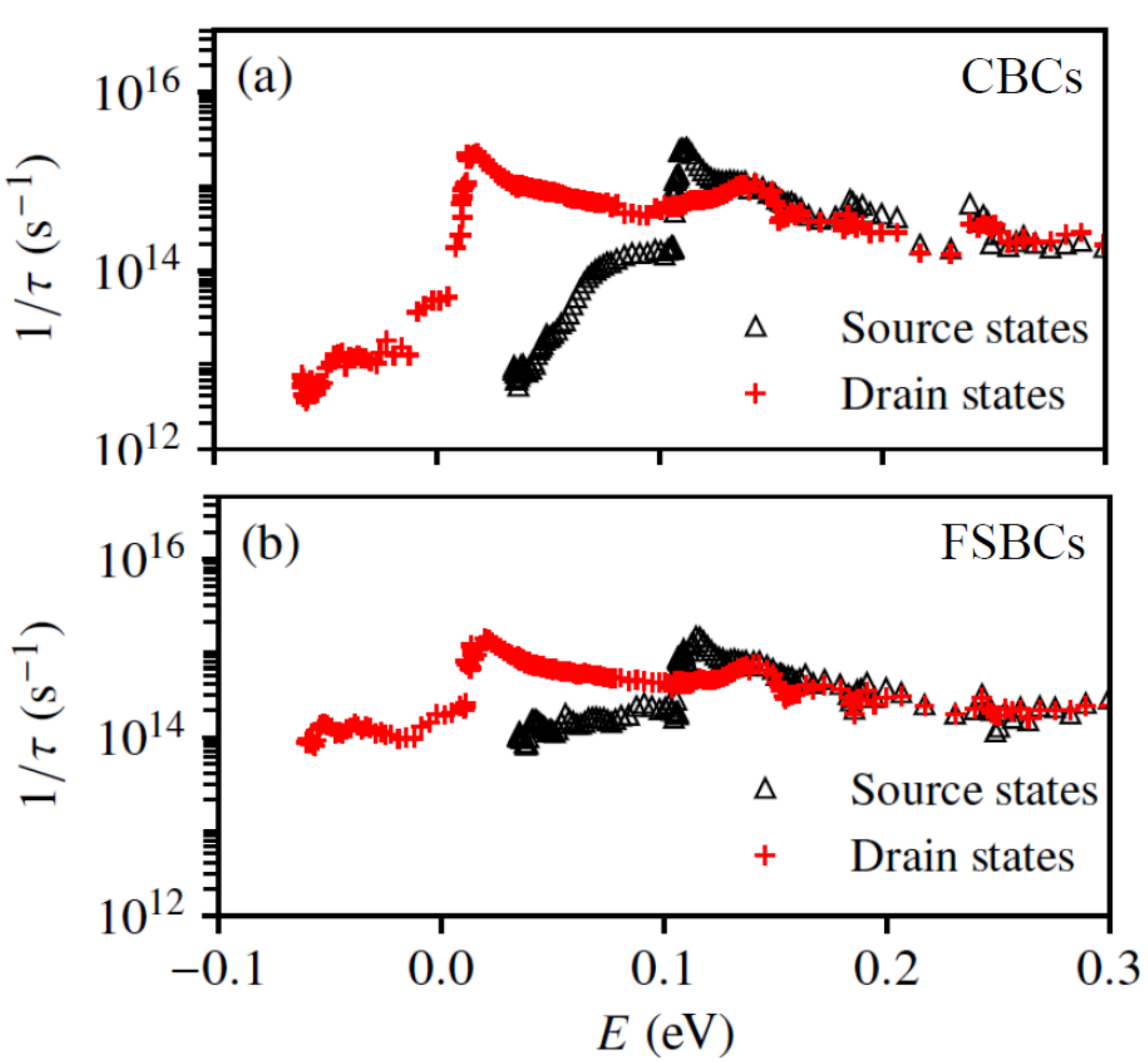}}
\caption{\label{fig:SiNW_scattering_BCs} The total electron/confined-phonon scattering rates for the 3$\times$3 SiNW-FET and different phonon 
          boundary conditions. The rates for states injected from the source are represented by black open triangles; the rates for states injected 
          from the drain are represented by red crosses. The data have been obtained for the device with applied bias of $V_{\text{DS}} = 0.1$ V and 
          $V_\text{GS}=-0.05$ V.}
\end{figure}
\begin{figure}[tb]
\centerline{\includegraphics[width=8.5cm]{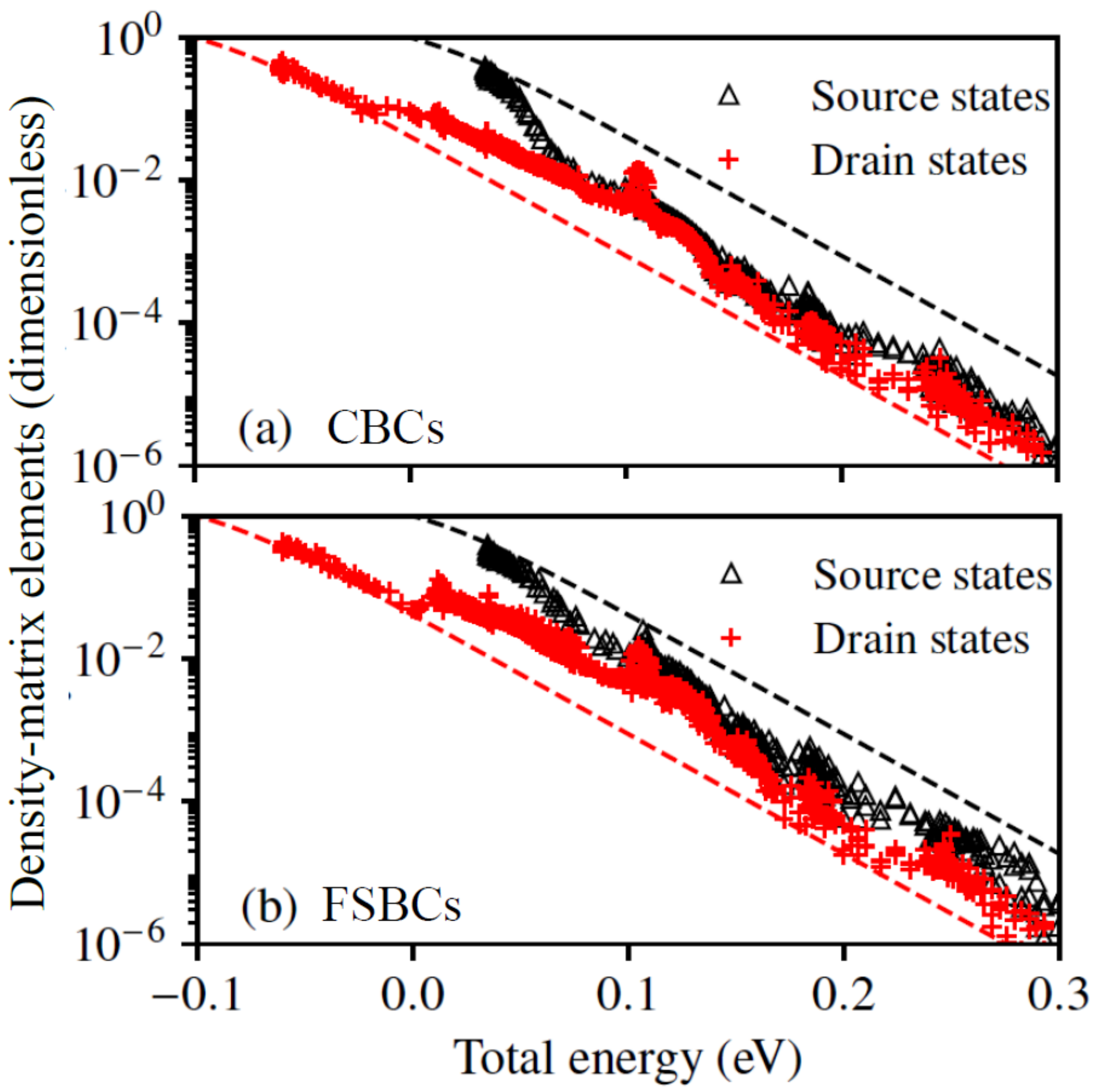}}
\caption{\label{fig:density_matrix_occupation} 
          Diagonal elements of the density matrix of the 3$\times$3 SiNW-FET plotted as a function of the total electron energy, assuming CBCs (a) and FSBCs (b).
          The black and red symbols represent the occupation of the states injected from the source and the drain while the black and red lines show the 
          occupation of the states assuming ballistic transport. The data have been obtained for assuming an applied bias of $V_{\rm DS} = 0.1$~V and 
          $V_{\rm GS}=-0.05$~V.} 
\end{figure}
\begin{figure}[tb]
\centerline{\includegraphics[scale=1]{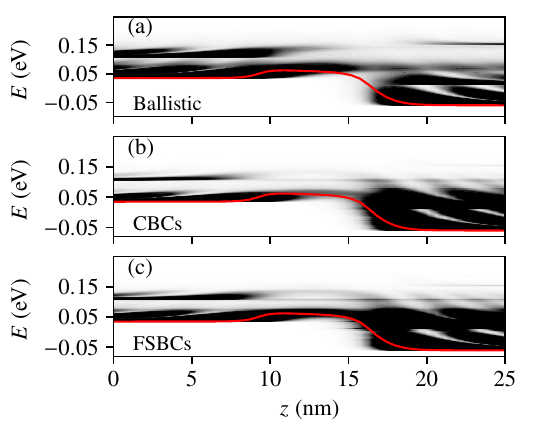}}
\caption{\label{fig:energy-resolved-density} 
         Energy-resolved density for the 3$\times$3 SiNW-FET shown in the case of the ballistic limit (a) and in the presence of 
         electron/confined-phonon scattering for the two cases of CBCs (b) and FSBCs (c) for $V_{\rm DS} = 0.1$~V and $V_{\rm GS}=-0.05$~V. 
         The red solid line represents the conduction band minimum.} 
\end{figure} 
\begin{figure}[tb]
\centerline{\includegraphics[width=8.5cm]{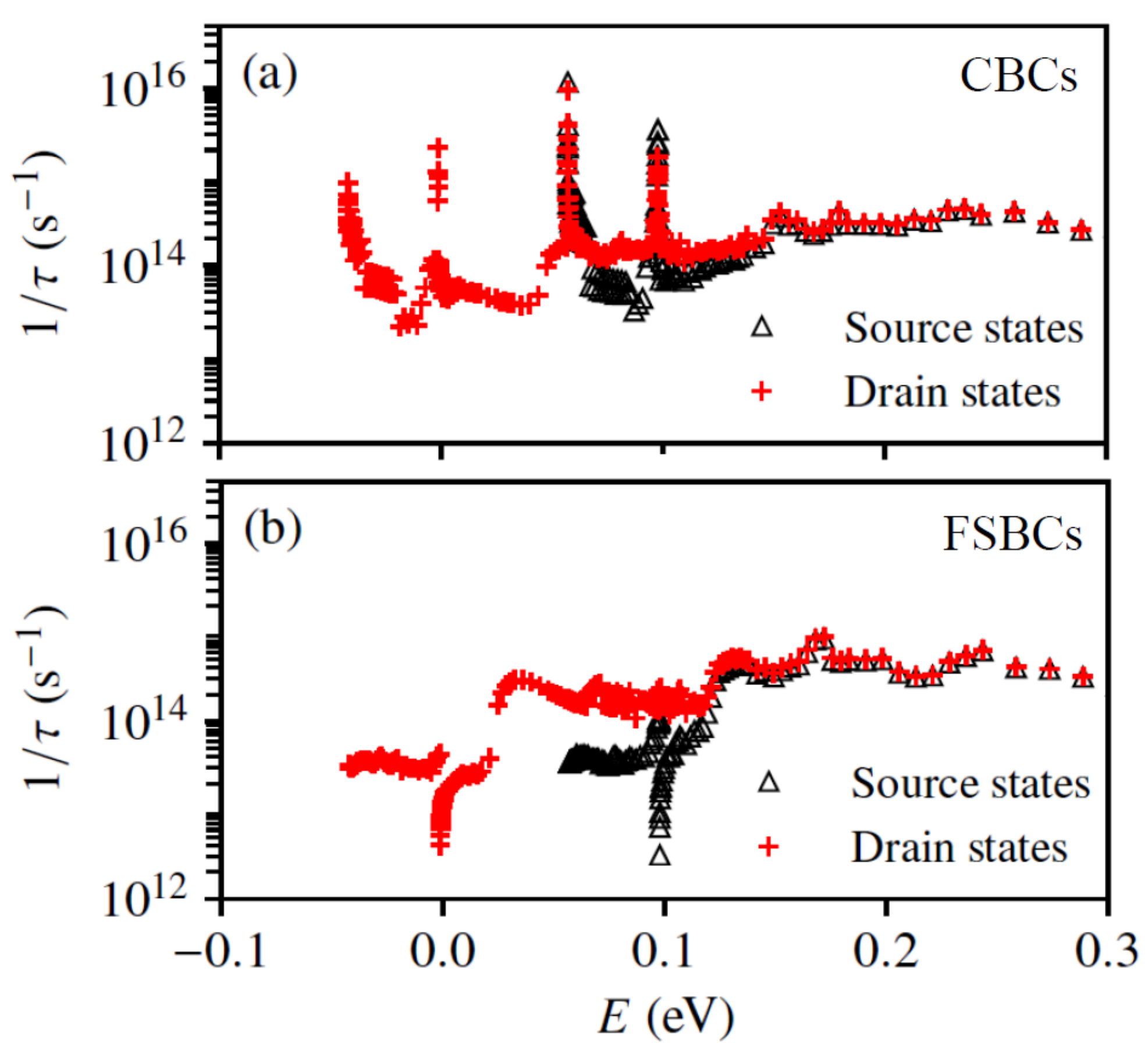}}
\caption{\label{fig:10aGNR_scattering_BCs} 
         As in Fig.~\ref{fig:SiNW_scattering_BCs}, but for the 10-aGNR-FET at $V_{\rm DS} = 0.1$ V and $V_{\rm GS}=-0.1$ V. The sharp features about 50~meV
         above the conduction-band minimum are associated with the second conduction band.}
\end{figure}
\begin{figure}[tb]
\centerline{\includegraphics[width=8.5cm]{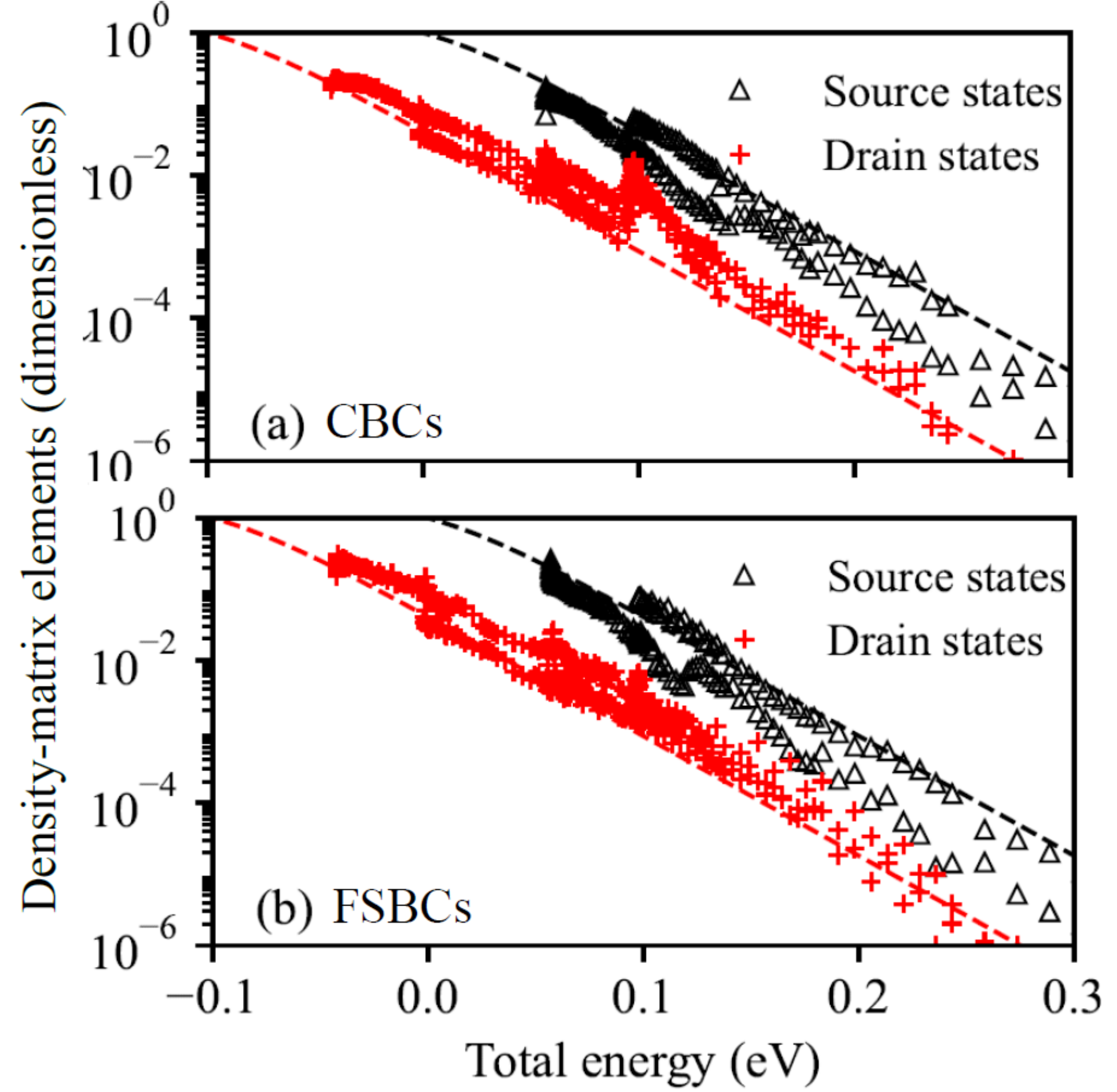}}
\caption{\label{fig:10aGNR_occupation}
         As in Fig.~\ref{fig:density_matrix_occupation}, but for the 10-aGNR-FET at an applied bias of $V_{\rm DS}=$ 0.1~V and $V_{\rm GS}=$-0.1~V.
         The presence of two conduction band, also seen in Fig.~\ref{fig:10aGNR_scattering_BCs}, results in the different occupation of the two bands 
         for states injected from the source and from the drain.} 
\end{figure}
\begin{figure*}[tb]
\centerline{\includegraphics[scale=1]{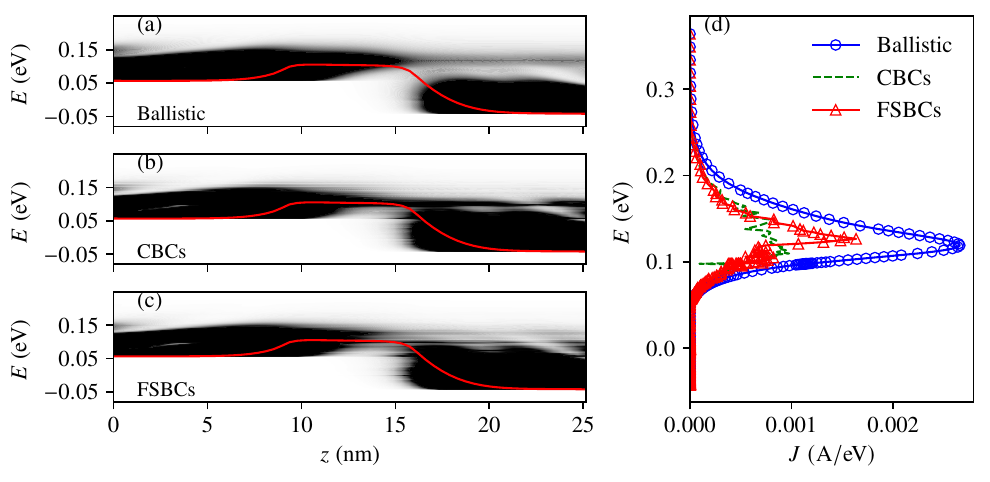}}
\caption{\label{fig:10aGNR_spectral_density_current} 
         Energy-resolved density for the 10-aGNR-FET in the case of the ballistic limit (a) and in the presence of electron/confined-phonon interactions 
         with CBCs (b) and FSBCs (c), with $V_{\rm DS} = 0.1$~V and $V_{\rm GS}=-0.1$~V. The red solid lines represent the energy of the conduction band 
         minimum. The energy-resolved current for these three cases is shown in (d). The blue solid line with circles represents the ballistic case;
         the green dashed line and the solid red line with triangles show the current obtained accounting for electron/confined-phonon scattering assuming 
         CBCs and FSBCs, respectively.} 
\end{figure*}

\noindent {\it C1. 3$\times$3 SiNW-FETs}\\

\noindent The electron/confined-phonon scattering rates are calculated using Eqs.~(\ref{eq:acoustic_contact_normal}) and (\ref{eq:optic_contact_normal})
with a bulk mass density $\rho_{\rm 3D} = 2.33 \times 10^3$~kg/m$^3$ and the deformation potentials from Refs.~\cite{fischetti_1996,fischetti1988monte} and 
listed in Table~\ref{phonon_parameters}. Note that processes that in bulk Si would correspond to intervalley scattering are implicitly accounted for, 
since the empirical pseudopotential band structure we employ folds the valleys into the 1D Brillouin zone. (We do not show here the band structure of 
square-cross-section SiNWs obtained from the empirical pseudopotentials of Ref.~\cite{Zunger_1993}, since it can be seen in Fig.~14 of Ref.~\cite{Fischetti_2011}.) The phonon are assumed to remain at equilibrium at 300~K. 

In Fig.~\ref{fig:SiNW_scattering_BCs} we show the scattering rates for different BCs plotted as a function of the electron kinetic energy. 
This is defined as an average over the density of initial electronic states: 
\begin{equation}
    \frac{1}{\tau^{ave}_\mu} = \frac{\sum_{s=1}^{N(E_\mu)} \mathcal{D}^s_{el} \left(E_\mu\right)/\tau_s^{\rm (ac/op)}(E_\mu) }
                                    {\sum_{E_\mu}\mathcal{D}^s_{el}\left(E_\mu\right)},
\label{eq:average_rate}
\end{equation}
where $s$ denotes subbands and $1/\tau_s^{\rm (ac/op)}(E_\mu)$ is obtained from Eqs.~({\ref{eq:acoustic_contact_normal}) and (\ref{eq:optic_contact_normal}}). 
The smaller scattering rate at low electron energies ($\lesssim$ 0.1~eV) seen in the case of CBCs compared to FSBCs can be understood from
Fig.~\ref{fig:SiNW_phonon_BC_empi} and from symmetry considerations regarding the overlap integral between the electronic states and the phonon shape function, 
$\mathcal{G}_{n,m}(x,y)$. When assuming FSBCs, the two acoustic branches present at low energy exhibit a higher group velocity compared to the CBC case,
resulting in a lower phonon density of states (DOS) in the low-frequency range that controls low-field electron transport. However, the symmetry of the
modes yields a larger overlap integral, thus resulting in a lower electron mobility. On the contrary, when assuming CBCs, of the two branches seen at low
frequency, one represents a shear modes that couples weakly to the electrons via the small deformation potential $\Delta_{\rm sh}$ (see 
Table~\ref{phonon_parameters}). This results in a larger electron mobility. Indeed, we have calculated the low-field electron mobility in 
intrinsic 3$\times$3 SiNWs at zero field (so that the states $\ket{\nu}$ are Bloch waves) using the discretized form of the Kubo-Greenwood's expression 
in one dimension: 
\begin{equation}
\mu = \frac{e}{\pi n k_{\rm B}T} \int {\rm d} k \ \tau_{\rm p}(k) \ \upsilon_{\rm g}(k)^{2} f_{\rm FD}(k)[1-f_{\rm FD}(k)] \ ,
\label{eq:low_field_mobility}
\end{equation}
where $\tau_{\rm p}(k)$ is the momentum relaxation time, $n$ the carrier density, and $\upsilon_{\rm g}(k)=(1/\hbar) {\rm d}E/{\rm d}k$ is the group velocity. 
When assuming CBCs, we have obtained an electron mobility of 210~cm$^2$/Vs, and of 25~cm$^2$/Vs assuming FSBCs. Tienda-Luna {\it et al.}~\cite{tienda2013effect}
have also reached the qualitatively similar conclusion that FSBCs result in a lower electron mobility.

\begin{figure*}[tb]
\centerline{\includegraphics[width=17cm]{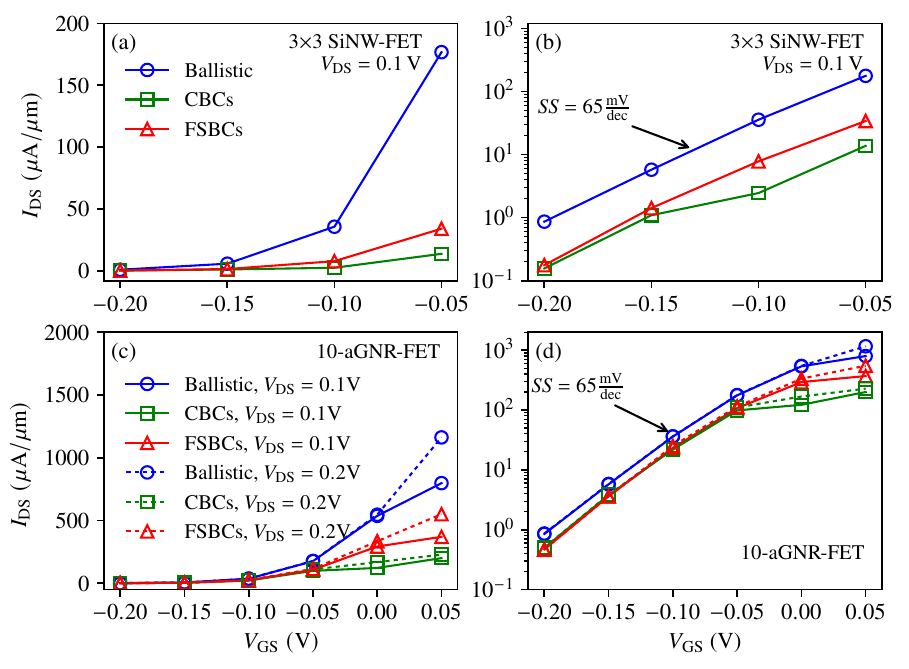}}
    \caption{Transfer characteristics of the 3$\times$3 SiNW-FET and the 10-aGNR-FET, focusing on the above-threshold (frames (a) and (c)) and 
             subthreshold regions (frames (b) and (d)). The current has been normalized to the width of the NW and of the NR. The blue lines with circles 
             refer to ballistic transport simulation, the green lines with squares to transport with electron/confined-phonon scattering assuming CBCs, the 
             red lines with triangles to transport that includes electron/confined-phonon scattering assuming FSBCs. The solid and dashed lines in (c) and 
             (d) show the current for a drain-source bias of 0.1~V and 0.2~V, respectively.} 
    \label{fig:transfer_characteristics_compare}
\end{figure*}
\begin{figure}[tb]
\centerline{\includegraphics[scale=1]{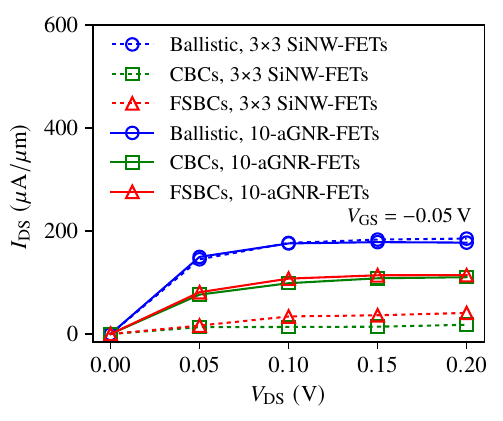}}
\caption{\label{fig:Id_Vd_compare} $I_{\rm DS}-V_{\rm DS}$ characteristics at $V_{\rm GS} = -0.05$~V of the 3$\times$3 SiNW-FET and the 10-aGNR-FET. 
             The dashed lines refer to the 3$\times$3 SiNW-FET, the solid lines to the 10-aGNR-FET}
\end{figure}
The diagonal elements of the density matrix in the case of CBCs and FSBCs are shown in Fig.~\ref{fig:density_matrix_occupation}; that is, the occupation 
$\rho_{\mu}$ of the electronic states plotted as a function of energy. Electron/confined-phonon scattering alters significantly the distribution of states 
injected from the source (black triangles). The energy-resolved density is shown in Fig.~\ref{fig:energy-resolved-density}. This is obtained by multiplying 
the local density of states (LDOS) by the occupation factors $\rho_{\mu}$, solutions of the PME. Comparison with the ballistic limit (dashed lines)
shows how electrons injected from the source lose energy via collisions with phonons. Moreover, a larger electron accumulation around
the potential barrier in the channel is seen when assuming FSBCs, compared to CBCs. This results in a larger current in the case of 
FSBCs (39.1~nA) compared to CBCs (11.8~nA), despite the larger low-field mobility obtained assuming CBCs. As mentioned above, this can be understood from
symmetry/parity considerations, especially of the dilatational waves that play a predominant role for both boundary conditions, thanks to their stronger coupling to electrons due to the larger deformation potential. In the case of FSBCs, their shape function (the divergence of the ionic displacement, Eq.~(\ref{eq:shape_function_free_standing}) is symmetric, reaching a maximum at the center of the NW and vanishing at the surface. When calculating the 
low-field mobility, electrons populate mainly lowest-energy subband and transport controlled mainly by intra-subband scattering. Since the  the initial and 
final electron wavefunctions in the ground-state subband exhibit the same symmetry, peaking at the center of the NW, the overlap integral (matrix element) 
between the electronic states and the shape function is large, thus boosting the scattering rate and yielding a low mobility. On the contrary, in the case of CBCs, the shape function is cosine-like, Eq.~(\ref{eq:shape_function_clamped_surface}), resulting in a smaller matrix element and, so, in a higher 
mobility.~\cite{bnote2} On the contrary, in the off-equilibrium conditions seen in Fig.~\ref{fig:energy-resolved-density}, at the higher energies that 
electrons reach in SiNW-FETs (some as high as around 0.1~eV), inter-subband scattering dominates, since the electrons that contribute mostly to the current
occupying higher-energy subbands, Now the symmetry of the initial and final electron wavefunctions is reversed, resulting in smaller matrix elements and 
weaker scattering rates when assuming FSBCs, and so in a higher current, than when assuming CBCs.\\

\noindent {\it C2. 10-aGNR-FETs}\\

\noindent To calculate the electron/confined-phonon scattering rates for 10-aGNR-FETs, we assume a graphene mass density of  
$\rho_{\rm 2D} = 7.63 \times 10^{-7}$~kg/m$^3$ and, for the deformation potentials, $\Delta_{\rm dil}$, $\Delta_{\rm sh}$, and $(DK)_{\rm op}$,
the values given in Table~\ref{phonon_parameters} that have been obtained from an analytic approximation of the scattering rates 
fitted to the results of rigid-pseudoion calculations~\cite{Maxbook}. Again, we consider the lattice to remain at equilibrium at room temperature.
For aGNRs, only one index, $n$, labels the confined vibrational modes. The band structure of 10-aGNRs obtained from the empirical pseudopotentials
of Ref.~\cite{Kurokawa_1996} is shown in Fig.~21 of Ref.~\cite{Fischetti_2011}.

We show in Fig.~\ref{fig:10aGNR_scattering_BCs} the total average scattering rates for 10-aGNR-FETs, calculated using Eq.~(\ref{eq:average_rate}), for different
BCs. Unlike what we saw for 3$\times$3 SiNW-FETs, the scattering rates are generally higher across the entire energy range when assuming CBCs compared to FSBCs.
In particular, in the low-frequency range that controls the low-field electron mobility, we see in Fig.~\ref{fig:10aGNR_phonon_BC_empi} that one acoustic shear
phonon (dashed line) is present when assuming both CBCs and FSBCs. However, the group velocity of this branch is higher in the case of FSBCs and the resulting lower phonon density of states causes a lower scattering rate than in the case of CBCs. Moreover, the acoustic dilatational branch present when assuming FSBCs
(solid line in Fig.~\ref{fig:10aGNR_phonon_BC_empi}(b)), does not couple with electrons in the bottom subband because of the symmetry of the associated shape
function, Eq.~(\ref{eq:shape_function_free_standing}). Overall, this results in a higher electron mobility in FSBCs. Indeed, at room
temperature, the calculated phonon-limited electron mobility in intrinsic 10-aGNRs when assuming CBCs is 420~cm$^2$/Vs, while it is 520~cm$^2$/(Vs) for FSBCs, 
as calculated using Eq.~(\ref{eq:low_field_mobility}). A previous {\it ab initio} theoretical study of the phonon-limited mobility in freestanding intrinsic 
10-aGNRs assuming mixed clamped/freestanding BCs has resulted in a value of 500~cm$^2$/(Vs)~\cite{betti2011strong}that falls between the values we have 
obtained for CBCs and FSBCs. 

Figure~\ref{fig:10aGNR_occupation} shows the final occupations of the electronic states in the 10-aGNR-FET, as in Fig.~\ref{fig:density_matrix_occupation}. 
As in the case of the SiNW-FET, a shift is observed in the final occupation of states injected from the source and/or drain compared to the ballistic occupation. 
Similarly, Figs.~\ref{fig:10aGNR_spectral_density_current} (a)-(c) illustrate the energy-resolved density of the 10-aGNR-FET. 
Whereas the effect of scattering is clearly visible in the larger population of low-energy states on the drain side of the device, these figures
do not show clearly how the current depends on the phonon boundary conditions. Therefore, in Fig.~\ref{fig:10aGNR_spectral_density_current} (d) we also
plot the energy-resolved current that shows clearly how much scattering reduces the total drain current and that assuming CBCs yields the worst performance.
\subsection{\label{subsec:transport}{Device characteristics}}
\noindent We now consider the effect of the confined-phonon boundary conditions on the drain-current {\it vs.} drain bias ($I_{\rm DS}-V_{\rm DS}$) 
and drain-current {\it vs.} gate bias ($I_{\rm DS}-V_{\rm GS}$) characteristics of the 3$\times$3 SiNW-FET and 10-aGNR-FET. 
Figure~\ref{fig:transfer_characteristics_compare} shows the transfer characteristics of these devices accounting for electron/confined-phonon scattering 
with different phonon boundary conditions. The effect of electron/confined-phonon scattering is clearly visible. This figure also shows that CBCs suffer a greater negative impact. The sub-threshold characteristics, as can be seen from Fig.~\ref{fig:transfer_characteristics_compare}(d) for the 10-aGNR-FET, remain relatively unaffected in both clamped-surface and free-standing boundary conditions. For the 3$\times$3 SiNW-FET the off-state behavior under CBCs degrades more significantly in the presence of electron/confined-phonon scattering, as shown in Fig.~\ref{fig:transfer_characteristics_compare}(b).

Finally, we show in Fig.~\ref{fig:Id_Vd_compare} the $I_{\rm DS}-V_{\rm DS}$ characteristics of the 3$\times$3 SiNW-FET and 10-aGNR-FET for  
$V_{\rm GS}=-0.05$~V. The figure clearly shows that 3$\times$3 SiNW-FETs are more significantly affected by electron/confined-phonon 
scattering than 10-aGNR-FETs, as it may be expected from the fact that in NW the phonons are affected by confinement in two dimensions, thus increasing the
overlap integrals with the electron wavefunctions (that is, the electron-phonon matrix elements). Additionally, in both cases, FSBCs affect 
electron/confined-phonon scattering by a smaller amount than CBCs.
\subsection{\label{subsec:10aGNR_self_consistency}{A self-consistent calculation}}
\noindent In principle, the final task involves performing self-consistent iteration by solving the Schr\"{o}dinger, PME, and Poisson equations with the
charge density calculated from the occupation $\rho_{\mu}$ obtained from the solution of the PME, Eq.~(\ref{eq:steady_equation}),
$n(\bf r) = \sum_{\mu} \rho_{\mu} \vert \braket{{\bf r}|\mu} \vert^{2}$. However, this process is computationally expensive. Obviously, it is necessary to assess
the magnitude of the differences between treating the PME as a post-processing step or, instead, incorporating it into the full self-consistent
Schr\"{o}dinger/Poisson/PME loop, even if only for a single voltage bias point. Therefore, we performed about ten self-consistent iterations for a 10-aGNR-FET 
with a gate bias $V_{\rm GS} = -0.1$~V and a drain bias $V_{\rm DS} = 0.1$~V, with an under-relaxation parameter $\lambda$ ranging from 0.5 to 0.9 (that is:
$\lambda V_{\rm new} + (1-\lambda)V_{\rm old} \rightarrow V_{\rm new}$  where $V_{\rm old}$ is the potential used in the previous iteration and $V_{\rm new}$
is the solution of the Poisson equation with the updated charge density). Although we were not able to reach convergence in a affordable small number of iterations,
after 10 iterations we found a root-mean-square error of the potential of a few meVs. The potential showed oscillations from one iteration to the next, changing
mainly in the drain-extension region and not in the all-important source/channel junction that controls the current. As a consequence, the drain current, 
about 36~$\mu$A/$\mu$m assuming ballistic transport, dropped to about 25~$\mu$A/$\mu$m when solving the PME for the first time and, as the iterative procedure
progressed, exhibited oscillations around this value of an amplitude of  about $\pm$5\% around this value, showing that the charge redistribution induced by
scattering has a small effect.  
\section{\label{sec:conclusion}{Conclusions}}
\noindent To deal with phonons confined in 1D structures we have employed a simple model that matches the elastic-continuum approximation at long wavelengths and 
relies on the folding of the bulk dispersion at short wavelengths. We have studied the dispersion and symmetries of these phonons assuming two extreme 
cases for the phonon boundary conditions: clamped-surface (CBCs) and free-standing-surface (FSBCs). Using a formulation of quantum transport based on 
the PME to treat the inelastic scattering of electrons with these confined phonons in one-dimensional systems, we have performed simulations of
3$\times$3 SiNW-FETs and 10-aGNR-FETs. Since acoustic bulk phonons exhibit a higher group velocity, we have not considered them, expecting
a weaker coupling to electrons, as indeed found by Donetti {\it et al.}~\cite{Donetti_2006} in thin Si films.

Our main result consists in finding that, even when describing confined phonons with our simple model, the main factor that controls electron transport is 
the enhanced scattering rate for interactions with such confined phonons. Ramayya {\it el al.}~\cite{Ramayya_2008} have attributed this enhancement 
to the higher phonon density of states, $\mathcal{D}_{\rm ph}(\hbar \omega)$, due either to a low group velocity, $\upsilon_{\rm g}$, or to the presence of 
more branches. This is evident from the presence of $\vert {\rm d}\omega/{\rm d}Q \vert^{-1} \sim \mathcal{D}_{\rm ph}(\hbar \omega)$ in the expression for the
scattering rates, Eqs.~(\ref{eq:acoustic_contact_normal}) and (\ref{eq:optic_contact_normal}). However, we have found that the phonon density of states is not 
the only factor that controls transport: As shown in Fig.~\ref{fig:10aGNR_phonon_BC_empi}, in 10-aGNRs, although the group velocity of the low-energy acoustic 
shear branch is higher when assuming FSBCs, the presence of an additional acoustic dilatational branch increases the phonon DOS. However, the smaller overlap
integrals (matrix elements) implied by FSBCs result in a lower electron mobility, 420~cm$^2$/(Vs), for CBCs compared to 520~cm$^2$/(Vs) for FSBCs. Similarly,
Fig.~\ref{fig:SiNW_phonon_BC_empi} shows that in SiNWs, CBCs yield a higher acoustic phonon density of states at low frequencies. Yet, we found that CBCs 
result in a higher room-temperature electron mobility, 210~cm$^2$/(Vs), than FSBCs, 25~cm$^2$/(Vs), as also found in the case of thin Si films~\cite{Donetti_2006} and Si NWs~\cite{tienda2013effect}. This is the result of a larger overlap integral (the matrix element) between the phonon
displacement field (the shape function, $\mathcal{G}_{n,m}(x,y)$), and the initial and final electron wavefunctions in the case of FSBCs. 
The symmetry/parity of the modes (the phonon `shape function') also affects strongly the inter-subband transitions that occur in off-equilibrium electron
transport at higher energy. This is an effect that may escape investigations based on the elastic-continuum approximation. Since in 3$\times$3 SiNWs 
the matrix elements associated with these transitions are larger assuming FSBCs, we found rather surprisingly that the performance of SiNW-FETs is actually
better when assuming FSBCs, despite the fact that they yield a lower low-field electron mobility: In GAA FETs with 7-nm gate length, FSBCs yield an 
on-current that is almost three times larger than what is obtained assuming CBCs. We also found that 10-aGNR-FETs are less affected by phonon confinement, 
while maintaining a similar subthreshold swing compared to 3$\times$3 SiNW-FETs.
\acknowledgments
\noindent We acknowledge the help of the Texas Advanced Computing Center (TACC) for having provided computing resources 
and Texas Instruments for the Endowment that has provided financial support. One of us (BC) also acknowledges Yaoqiao Hu's help with the DFT calculations 
and Shoaib Mansoori and Dallin O. Nielsen for sharing their expertise on Quantum ESPRESSO. 
\appendix
\section{\label{Appen.A}{The elastic continuum model}}
\noindent In this appendix we consider the elastic continuum model applied to SiNWs and aGNRs analyzing the symmetry of the vibrational modes. We show that 
the `shape functions' we employ to describe the ionic displacement, Eqs.~(\ref{eq:shape_function_clamped_surface}) and
(\ref{eq:shape_function_free_standing}), do indeed capture the basic symmetry and parity properties of this model in the simpler case of aGNRs. However,
a similar procedure can be used also in the case of SiNWs. We consider only acoustic phonons, following the treatment of Donetti {\it et al.}~\cite{Donetti_2006}, 
a treatment that, in turn, is based on the original work by Bannov {\it et al.}~\cite{Bannov_1994}. 
\vspace*{-0.55cm}
\subsection{\label{Appen.A1:SiNW continuum model}Acoustic phonons in rectangular-cross-section nanowires}
\vspace*{-0.35cm}
\noindent Considering for simplicity a transversally isotropic medium (that is, characterized by a single transverse sound velocity, $c_{\rm t}$),
the ionic displacement field ${\bf u}({\bf r})$ satisfies the equation:
\begin{equation}
\frac{\partial^{2} {\bf u}}{\partial t^{2}} = c^{2}_{\rm t} \nabla^{2} {\bf u} + (c^{2}_{\rm l}-c^{2}_{\rm t}) \nabla ( \nabla \cdot {\bf u}) \ ,
\label{eq:continuum_ac_1}
\end{equation}
where $c_{\rm l}$ and $c_{\rm t}$ are the longitudinal and transverse velocities, respectively. 
Reference~\cite{Bannov_1994} expresses the sound velocities as $c_{\rm l} = (\lambda+2\mu)/\rho$ and $c_{\rm t} = \mu/\rho$, where $\rho$ is the
bulk mass density, as usual, and $\lambda$ and $\mu$ are the Lam\'{e} constants. They are implicitly defined in terms of the stress and strain tensors,
$\boldsymbol{\sigma}$ and $\boldsymbol{\epsilon}$, by the relation, 
$\boldsymbol{\sigma} = 2 \mu \boldsymbol{\epsilon} + \lambda \mbox{Tr} (\boldsymbol{\epsilon}) {\bf I}$, where ${\bf I}$ is the identity (unit) tensor. 
We look for solutions, normalized to the area $A$, of the form:
\vspace*{-0.15cm}
\begin{equation}
{\bf u}({\bf R},z,t) = \sum_{Q,n,m} {\bf w}_{Q;n,m}({\bf R}) \ e^{iQz - i \omega_{n,m}t} \ ,
\label{eq:continuum_ac_2}
\end{equation}
representing phonons of frequency $\omega_{n,m}$ propagating freely along the axial direction $z$ and confined on the $(x,y)$ plane. The indices
$n$ and $m$ result from solving Eq.~(\ref{eq:continuum_ac_1}) subject to the appropriate free-standing or clamped-surface boundary conditions.
In Eq.~(\ref{eq:continuum_ac_2}), ${\bf R}$ represents the position on the $(x,y)$ cross-sectional plane.
Having found these solutions, the quantized phonon field can be expressed as:
\vspace{-0.25cm}
\begin{multline}
\widehat{{\bf u}}({\bf R},z,t) = \sum_{Q,n,m} \left ( \dfrac{\hbar}{2 \rho_{\rm b} \omega_{n,m}} \right )^{1/2} \\
                        \times \ \left ( \widehat{b}^{\dagger}_{Q;n,m} + \widehat{b}_{Q;n,m} \right )
                              {\bf w}_{Q;n,m}({\bf R}) \ e^{iQz - i \omega_{n,m}t} \ ,
\label{eq:continuum_ac_3}
\end{multline}
(where $\widehat{b}_{Q;n,m}$ and $\widehat{b}^{\dagger}_{Q;n,m}$ are the phonon annihilation and creation operators and $\rho_{\rm b}$ is rhe bulk)
mass density) and the electron-phonon perturbation potential takes the form:
\vspace{-0.25cm}
\begin{multline}
\widehat{{\bf H}}({\bf R},z,t) = \Delta_{\rm ac} \nabla \cdot {\widehat{\bf u}}({\bf r}) = \\
                   \sum_{Q,n,m,\nu,\mu} \Delta_{\rm ac} \left ( \dfrac{\hbar}{2 \rho_{\rm b} \omega_{n,m}} \right )^{1/2} 
                      \widehat{c}^{\dagger}_{\rm \nu} \left ( \widehat{b}^{\dagger}_{Q;n,m} + \widehat{b}_{Q;n,m} \right ) \widehat{c}_{\mu} \\
                           \times \ \left [ iQ w^{\parallel}_{Q;n,m}({\bf R}) + \nabla_{\rm 2D} \cdot {\bf w}^{\perp}_{Q;n,m}({\bf R}) \right ] \ 
                                 e^{iQz - i \omega_{n,m} t}, 
\label{eq:continuum_ac_4}
\end{multline}
\vspace{-0.15cm}
where $\widehat{c}_{\mu}$ and $\widehat{c}^{\dagger}_{\mu}$ are the electron annihilation and creation operators, and
$w^{\parallel}_{Q;n,m}$ and  ${\bf w}^{\perp}_{Q;n,m}$ are the components of ${\bf w}_{Q;n,m}$ along the axial and transverse directions, respectively. 
Insert Eq.~(\ref{eq:continuum_ac_2}) into Eq.~(\ref{eq:continuum_ac_1}) we obtain the following system of equations:

\vspace{-0.15cm}
\begin{multline}
- \omega^{2} w_{x} = \left [ c_{\rm l}^{2} \frac{\partial^{2}} {\partial x^{2}} 
                             + c_{\rm t}^{2} \left ( \frac{\partial^{2}}{\partial y^{2}} - Q^{2} \right ) \right ] w_{x} + \\
            (c_{\rm l}^{2} - c_{\rm t}^{2}) \left ( \frac{\partial^{2}w_{y}}{\partial x \partial y} + i Q  \frac{\partial w_{z}}{\partial x} \right )  
\label{eq:continuum_ac_5a}
\end{multline}
\vspace{-0.25cm}
\begin{multline}
- \omega^{2} w_{y} = \left [ c_{\rm l}^{2} \frac{\partial^{2}} {\partial y^{2}} 
                             + c_{\rm t}^{2} \left ( \frac{\partial^{2}}{\partial x^{2}} - Q^{2} \right ) \right ] w_{y} + \\
            (c_{\rm l}^{2} - c_{\rm t}^{2}) \left ( \frac{\partial^{2}w_{x}}{\partial x \partial y} + i Q  \frac{\partial w_{z}}{\partial y} \right )  
\label{eq:continuum_ac_5b}
\end{multline}
\vspace{-0.25cm}
\begin{multline}
- \omega^{2} w_{z} = \left [ c_{\rm t}^{2} \left ( \frac{\partial^{2}}{\partial x^{2}} + \frac{\partial^{2}}{\partial y^{2}} \right ) 
                             - c_{\rm l}^{2} Q^{2} \right  ] w_{z} + \\
                          i Q  (c_{\rm l}^{2} - c_{\rm t}^{2}) \left ( \frac{\partial w_{x}}{\partial x} + \frac{\partial w_{y}}{\partial y} \right ) \ .   
\label{eq:continuum_ac_5c}
\end{multline}

This problem has been solved by Nishiguchi~\cite{Nishiguchi_1997} and by Ayedh and Wacker~\cite{Ayedh_2011}, but only numerically. In Ref.~\cite{Ayedh_2011} the problem is solved numerically with the finite element method. In Ref.~\cite{Nishiguchi_1997}, the unknowns are expanded on the
basis of functions of the type $(x/L_{\rm x})^{n} (y/L_{\rm y})^{m}$ (where $L_{\rm x} \times L_{\rm y}$ is the area of the rectangular cross section of the 
nanowire and with the origin of the coordinates at the center of the nanowire). This results in a
rather large eigenvalue problem that cannot be solved analytically. Here, we consider
nanowires with rectangular cross sections and draw some conclusions from the analysis given in Ref.~\cite{Nishiguchi_1997} of the symmetry of the 
modes. 

Comparing the analysis of Ref.~\cite{Nishiguchi_1997} with the basis functions $\mathcal{G}_{n,m}$ used in Sec.~\ref{subsec:phonon dispersion}, we see that our simple model provides reasonably accurate results identifying various modes as follows 
(what follows applies to the case of clamped BCs; for FSBCs, 'even' and 'odd' must be swapped):
\vspace*{-0.25cm}
\begin{widetext}
\begin{equation}
\label{eq:continuum_ac_8}
\begin{array}{llll}
\mbox{{parity:}}=(++); & m \ {\mbox {even}}  \ \ n \ {\mbox {even:}}  & \mbox{dilatational modes}     & \omega(q) \sim \omega^{\rm (LA)}(q) \\
\mbox{{parity:}}=(+-); & m \ {\mbox {odd}}  \ \ n \ {\mbox {even:}}   & \mbox{flexural modes along }y & \omega(q) \sim q^{2} \\
\mbox{{parity:}}=(-+); & m \ {\mbox {even}} \ \ n \ {\mbox {odd:}}    & \mbox{flexural modes along }x & \omega(q) \sim q^{2} \\ 
\mbox{{parity:}}=(--); & m \ {\mbox {odd}} \ \ n \ {\mbox {odd:}}     & \mbox{shear/torsional modes}  & \omega(q) \sim \omega^{\rm (TA)}(q) \ .   
\end{array}
\end{equation} 
\end{widetext}

The anti-symmetric transverse components and symmetric longitudinal components of the ionic displacement ${\bf u}$ are of even parity, 
whereas symmetric transverse components and anti-symmetric longitudinal components have odd parity. The shear/torsional modes are 
linear combinations of the two flexural modes. Therefore, there are only three linearly independent sets of modes, and we can consider the 
dilatational mode, the shear/torsional mode, and one flexural mode as the basis set of modes. 

The two flexural modes correspond to the out-of-plane (ZA) modes in two-dimensional structures and exhibit the same parabolic behavior at small $Q$. 
Indeed, Nishiguchi {\it et al.}~\cite{Nishiguchi_1997}, in their fully anisotropic model, find that the dispersion of the lowest-energy flexural modes is given by 
$\omega(Q) = Q^{2} \left[YI_{\rm x,y}/(\rho A)\right]^{1/2}$, where $Y$ is the Young's modulus, and $I_{\rm x,y}=AL_{\rm {x/y}}/12$ are 
the moments of inertia of the cross-section. Similar to what occurs in two-dimensional (2D) layers, if the structure is symmetric under inversion along the
$x$ or $y$ axis, these modes do not couple to electrons and can be ignored. The torsional modes produce vanishing matrix elements for inter-subband transitions;
however, they can contribute to intra-subband processes involving wave functions of opposite parity. The dilatation modes dominate intra-subband processes 
and can also significant contribute to scattering processes involving electron wave functions of the same parity. 
These properties are correctly reflected in the simple model employed here.
\vspace*{-0.55cm}
\subsection{\label{Appen.A2:aGNR continuum model}Acoustic phonons in nanoribbons}
\vspace*{-0.25cm}
\noindent In the case of nanoribbons, following Donetti {\it et al.}~\cite{Donetti_2006}, we may find relatively easily an approximate analytic expression 
for the phonon displacement field and show that the shape function $\mathcal{G}_{m}(y)$ used here is indeed consistent with the result of the elastic
continuum approximation. 

Let's consider solutions, normalized to the width $W$, of the form:
\begin{equation}
{\bf u}(y,z,t) = \sum_{Q,n} {\bf w}_{Q;n}(y) \ e^{iQz - i \omega_{n}t} \ ,
\label{eq:continuum_ac_2NR}
\end{equation}
representing, as usual, phonons of frequency $\omega_{n}$ propagating freely along the axial direction $z$ and confined along the $y$ direction.
Then, the electron-phonon Hamiltonian takes the form:

\vspace{-0.25cm}
\begin{multline}
\widehat{{\bf H}} = 
    \int_{A} {\rm d}{\bf R} \ \widehat{\psi}^{\dagger} ({\bf R}) \ [ \Delta_{\rm ac} \nabla \cdot {\widehat{\bf u}}({\bf R}) ] \ \widehat{\psi}({\bf R}) = \\
                \sum_{Q,n,\nu,\mu} \Delta_{\rm ac} \left ( \dfrac{\hbar}{2 \rho_{\rm 2D} \omega_{n}} \right )^{1/2} 
                       \widehat{c}^{\dagger}_{\rm \nu}  \left ( \widehat{b}^{\dagger}_{Q;n} + \widehat{b}_{Q;n} \right ) \widehat{c}_{\mu} \\
                           \times \ \left [ iQ w_{z;Q;n}(y) + \frac{{\rm d}}{{\rm d}y} w_{y;Q;n}(y) \right ] \ ,
\label{eq:continuum_ac_4NR}
\end{multline}
where now the surface mass-density of the layer, $\rho_{\rm 2D}$, appears in the Hamiltonian, since we assume a ribbon of zero thickness.

The system of equations we must solve now takes the form:
\begin{equation}
- \omega^{2} w_{x} = c_{\rm t}^{2} \left ( \frac{{\rm d}^{2}}{{\rm d} y^{2}} - Q^{2} \right )  w_{x} 
\label{eq:continuum_ac_5aNR}
\end{equation}
\begin{equation}
- \omega^{2} w_{y} = \left [ c_{\rm l}^{2} \frac{{\rm d}^{2}} {{\rm d} y^{2}} - c_{\rm t}^{2} Q^{2} \right ] w_{y} +
            i Q (c_{\rm l}^{2} - c_{\rm t}^{2}) \frac{{\rm d} w_{z}}{{\rm d} y} 
\label{eq:continuum_ac_5bNR}
\end{equation}
\begin{equation}
- \omega^{2} w_{z} = \left [ c_{\rm t}^{2} \frac{{\rm d}^{2}}{{\rm d} y^{2}} - c_{\rm l}^{2} Q^{2} \right  ] w_{z} +
                          i Q  (c_{\rm l}^{2} - c_{\rm t}^{2})  \frac{{\rm d} w_{y}}{{\rm d} y}  \ .   
\label{eq:continuum_ac_5cNR}
\end{equation}
Note that the out-of-plane modes (ZA phonons), whose displacement is represented by $w_{x}$, are decoupled from both from the other vibrational modes
and from the electrons, since they do not appear in the electron-phonon Hamiltonian, Eq.~(\ref{eq:continuum_ac_4NR}). This is the result 
of having ignored any dependence on the out-of-plane coordinate, $x$, which is equivalent to assuming a $\boldsymbol{\sigma}_{\rm h}$-symmetric layer. 
Thus, we may ignore them. 
Incidentally, note that in this simple transversally-isotropic model, the frequency of these modes is 
$\omega^{\rm (ZA)}_{n}(Q) = c_{\rm t}[Q^{2}+(n\pi/W)^{2}]^{1/2}$. More accurate models for vertically clamped layers (such as supported 2D layers)
that account for the $x$ dependence of the shape functions~\cite{Amorim_2013} give, instead, a dispersion of the form 
$\omega^{\rm (ZA)}(Q) = [\alpha^{2}Q^{4}+\omega_{0}^{2}]^{1/2}$, where $\omega_{0} \approx \sqrt{g/\rho_{\rm 2D}}$ 
is the zero-point frequency determined by the coupling force $g$ between the 2D layer and the clamping materials, and 
$\alpha \approx \sqrt{\kappa/\rho_{\rm 2D}}$, where $\kappa$ is the bending rigidity of the 2D layer.

Equations~(\ref{eq:continuum_ac_5bNR}) and (\ref{eq:continuum_ac_5cNR}) can be solved semi-analytically obtaining shear and dilation waves:
For the dilatational waves, let's look for solutions that are anti-symmetric along the transverse $y$ direction, corresponding to opposite ionic displacements
at opposite edges (like breathing modes, solutions that are of even parity, symmetric under reflections around the ribbon axis). Therefore, we look for solutions of the form (once again following the procedure of Ref.~\cite{Donetti_2006}):

\begin{equation}
\left \{
\begin{array}{ll}
w_{y,n}(y) =  & - A_{n}(Q) k_{\rm l}  \sin \left [ k_{\rm l} \left ( y - \dfrac{W}{2}  \right ) \right ] \\
              & + B_{n}(Q) Q          \sin \left [ k_{\rm t} \left ( y - \dfrac{W}{2}  \right ) \right ] \\
              \\ 
w_{z,n}(y) =  & i A_{n}(Q) Q          \cos \left [ k_{\rm l} \left ( y - \dfrac{W}{2}  \right ) \right ] \\
              &+ i B_{n}(Q) k_{\rm t} \cos \left [ k_{\rm t} \left ( y - \dfrac{W}{2}  \right ) \right ]  \ ,
\end{array} \right.
\label{eq:continuum_ac_7NR}
\end{equation}
having omitted to add an eigenmode-index $n$ to the longitudinal and transverse wave vectors $k_{\rm l}$ and $k_{\rm t}$.
These parameters, $k_{\rm l}$ and $k_{\rm t}$, are determined by the BCs ${\bf w}_{n}(0) = {\bf w}_{n}(W)=0$. These two equations give:
\vspace{-0.15cm}
\begin{equation}
Q^{2} \tan \left ( \dfrac {k_{\rm t}W}{2} \right ) = - k_{\rm l} k_{\rm t} \tan \left ( \dfrac {k_{\rm l}W}{2} \right ) \ ,
\label{eq:continuum_ac_8NR}
\end{equation}
and
\vspace{-0.15cm}
\begin{equation}
A_{n}(Q) \cos \left ( \dfrac{k_{\rm l}W}{2} \right ) = B_{n}(Q) k_{\rm t} \cos \left ( \dfrac{k_{\rm t}W}{2} \right ) \ .
\label{eq:continuum_ac_8aNR}
\end{equation}
Thus, $A$ and $B$ acquire a dependence on $Q$ and on an index $n$ that labels the infinite discrete solutions of Eq.~(\ref{eq:continuum_ac_8NR});
therefore, we have denoted them as $A_{n}(Q)$ and $B_{n}(Q)$. Inserting now the functions Eq.~(\ref{eq:continuum_ac_7NR}) into
Eqs.~(\ref{eq:continuum_ac_5bNR}) and (\ref{eq:continuum_ac_5cNR}), the resulting secular equation is:
\begin{equation}
\omega_{n}(Q) = c_{\rm l} \sqrt{Q^{2} + k_{{\rm l},n}(Q)^{2}} = c_{\rm t} \sqrt{Q^{2} + k_{{\rm t},n}(Q)^{2}} \ ,
\label{eq:continuum_ac_9NR}
\end{equation} 
For each $Q$, Eqs.~(\ref{eq:continuum_ac_8NR}) and (\ref{eq:continuum_ac_9NR}) give a discrete set of values $k_{{\rm l},n}(Q)$ and $k_{{\rm t},n}(Q)$.
The normalization of the shape functions implies:
\vspace{-0.25cm}
\begin{multline}
\int_{0}^{W} {\rm d} y \ [ w_{y,n}(y)^{2} + w_{z,n}(y)^{2} ] = \\
  \frac{W}{2} \left \{ A_{n}(Q)^{2} [Q^{2} + k_{{\rm l},n}(Q)^{2}] +  B_{n}(Q)^{2} [Q^{2}+k_{{\rm t},n}(Q)^{2}] \right. \\
  \left.                   - 2 A_{n}(Q) B_{n}(Q) Q [k_{{\rm l},n}(Q) + k_{{\rm t},n}(Q) ] \right \} = 1 \ . 
\label{eq:continuum_ac_11NR}
\end{multline}
Using Eq.~(\ref{eq:continuum_ac_8aNR}) to eliminate $B_{n}(Q)$, this equation allows us to determine $A_{n}(Q)$.  
Finally, the interaction terms inside the square bracket in Eq.~(\ref{eq:continuum_ac_4NR}) now becomes:
\vspace{-0.25cm}
\begin{multline}
iQ w_{z;Q;n}(y) + \frac{{\rm d}}{{\rm d}y} w_{y;Q;n}(y) = \\
                            -A_{n}(Q) [ Q^{2} + k_{{\rm l},n}(Q)^{2} ] \ 
                                       \cos \left [ k_{{\rm l},n}(Q) \left ( y - \dfrac{W}{2}  \right ) \right ] 
\label{eq:continuum_ac_10NR}
\end{multline}
Although the result cannot be expressed in closed form (mainly because Eq.~(\ref{eq:continuum_ac_8NR}) is transcendental), it 
is helpful and interesting because it shows that we can use the simpler `fully empirical model' described in Sec.~\ref{subsec:phonon dispersion}, provided we: 
{\bf 1}. approximate the phonon dispersion, $\omega_{n}(Q)$, with the result of the secular equation, Eq.~(\ref{eq:continuum_ac_9NR});
{\bf 2}. normalize the phonon shape functions according to Eq.~(\ref{eq:continuum_ac_11NR}); and, finally,
{\bf 3}. replace the term $\overline{q}_{n} \mathcal{G}_{n}(y) \approx (2/W)^{1/2} \overline{q}_{n}\cos(k_{n}y)$  in Eqs.~(\ref{eq:acoustic_contact_normal})
with the result of Eq.~(\ref{eq:continuum_ac_10NR}), thus accounting for the fact that the phonons are not purely longitudinal. This a result of the fact 
that the Poisson ratio of the material requires a finite and nonzero ratio $w_{z}/w_{y}$.

Let's consider the case in which neither $k_{{\rm l},n}$ nor $k_{{\rm t},n}$ vanish in the limit of $Q \rightarrow 0$ (if one of them does, 
then Eq.~(\ref{eq:continuum_ac_9NR}) forces both of them to vanish). Then, the long-wavelength limit
is defined as $Q^{2} \ll k_{{\rm l},n} k_{{\rm t},n}$. In this limit, for each anti-symmetric mode $n$,
Eq.~(\ref{eq:continuum_ac_8NR}) requires $k_{{\rm l},n} \approx n \pi/W - \alpha Q^{2} \rightarrow n \pi/W$ with $n$ even (nonzero, or we would not have 
$Q^{2} \ll k_{{\rm t},n} k_{{\rm t},n}$). This ensures that the right-hand side of Eq.~(\ref{eq:continuum_ac_8NR}) approaches zero as $Q \rightarrow 0$.
Furthermore, Eq.~(\ref{eq:continuum_ac_9NR}) determines the value of the constant $\alpha = [2W/(sn^{2}\pi^{2})] \tan(sn\pi/2)$, 
where $s=c_{\rm l}/c_{\rm t}$. Thus, $k_{{\rm l},n}$ does not change significantly  from $n\pi/W$ as long as 
$Q < [s^{1/2} (n\pi)^{3/2}/(2^{1/2}W)]/\tan(sn\pi/2)^{1/2}$. 

The range of $Q$ covered by this condition, and thus the range of validity of the `long-wavelength' region, depends on the value of the parameter $\alpha$. 
This is  highly sensitive to the ratio $s$ and a small correction to the frequency of the dilatation waves results in a significant correction to 
the dispersion of the shear waves (or vice versa, depending on the value of $s$). Therefore, the validity of this approximate expression may vary from 
a small-$Q$ region of the 1D Brillouin zone to more than a half of it, even for the lowest-energy mode $n=2$. 
We have not evaluated directly the magnitude of higher-order terms, such as those of order $Q^{4}$ or beyond. However, it has been
verified that no term of order $Q^{3}$ exists. Notably, in the fully isotropic case $s=1$, $\alpha =0$, and the expected behavior of fully 
decoupled modes with $k_{{\rm l},n} = n \pi/W$ for all $Q$ is recovered. (The symmetric solutions corresponding to shear waves will be discussed in the 
following.)

Equation~(\ref{eq:continuum_ac_8aNR}) indicates that $B_{n}(Q) \rightarrow 0$. Therefore, from Eq.~(\ref{eq:continuum_ac_11NR}), 
$A_{n}(Q) \rightarrow (2/W)^{1/2}[ Q^{2} + n^{2}\pi^{2}/W^{2} ]^{-1/2}$ and, evaluating the interaction term, Eq.~(\ref{eq:continuum_ac_10NR}), we obtain:
\vspace{-0.35cm}
\begin{multline}
- A_{n}(Q) [ Q^{2} + k_{{\rm l},n}(Q)^{2} ] \ \cos \left [ k_{{\rm l},n}(Q) \left ( y - \dfrac{W}{2}  \right ) \right ] \\ \rightarrow
 \left ( \dfrac{2}{W} \right)^{1/2} [ Q^{2} + n^{2}\pi^{2}/W^{2} ]^{1/2} \ \cos \left (\dfrac{n \pi}{W} y  \right )  \ ,
\label{eq:continuum_ac_12NR}
\end{multline}
for even $n$. This is exactly the result of the empirical model described in Sec.~\ref{subsec:phonon dispersion}. Therefore, for narrow ribbons 
and even $n$, this fully empirical model approximates quite well the properties of the lowest-$n$ (anti-symmetric) dilatational waves predicted 
by the elastic continuum model. 

The situation for the shear waves is similar. We can look for solutions for which the ionic displacement is (mirror) anti-symmetric along the 
longitudinal $z$ direction (thus yielding a shear strain), thus for solutions of the form:
\vspace{-0.15cm}
\begin{equation}
\left \{
\begin{array}{ll}
w_{y,n}(y) =  &  A_{n}(Q) k_{\rm l} \cos \left [ k_{\rm l} \left ( y - \dfrac{W}{2}  \right ) \right ] \\
              &- B_{n}(Q) Q         \cos \left [ k_{\rm t} \left ( y - \dfrac{W}{2}  \right ) \right ] \\
              \\
w_{z,n}(y) =  &i A_{n}(Q) Q         \sin \left [ k_{\rm l} \left ( y - \dfrac{W}{2}  \right ) \right ] \\
              &+ i B_{n}(Q) k_{\rm t} \sin \left [ k_{\rm t} \left ( y - \dfrac{W}{2}  \right ) \right ]  \ .
\end{array} \right.
\label{eq:continuum_ac_7NRs}
\end{equation}
The boundary conditions now imply:
\begin{equation}
Q^{2} \tan \left ( \dfrac {k_{\rm l}W}{2} \right ) = - k_{\rm l} k_{\rm t} \tan \left ( \dfrac {k_{\rm t}W}{2} \right ) \ ,
\label{eq:continuum_ac_8NRs}
\end{equation}
and the secular equation yields a dispersion identical to Eq.~(\ref{eq:continuum_ac_9NR}) with a normalization condition 
identical to Eq.~(\ref{eq:continuum_ac_11NR}). The interaction terms now can be written as:
\begin{equation}
 A_{n}(Q) [ Q^{2} + k_{{\rm l},n}(Q)^{2} ] \ \sin \left [ k_{{\rm l},n}(Q) \left ( y - \dfrac{W}{2}  \right ) \right ] \ .
\label{eq:continuum_ac_10NRs}
\end{equation}
In the small-$Q$ limit, a solution of Eqs.~(\ref{eq:continuum_ac_9NR}) and (\ref{eq:continuum_ac_8NRs}) is 
$k_{{\rm l},n} \approx n\pi/W - [2W/(sn^{2}\pi^{2})]Q^{2}/\tan(sn\pi/2)$, where now $n$ is an odd integer. This can be seen following a procedure
completely analogous with the procedure just outlined in the case of dilatational waves. In this case, the interaction term behaves asymptotically as:
\begin{multline}
 A_{n}(Q) [ Q^{2} + k_{{\rm l},n}(Q)^{2} ] \ \sin \left [ k_{{\rm l},n}(Q) \left ( y - \dfrac{W}{2}  \right ) \right ] \rightarrow \\
  - \left ( \dfrac{2}{W} \right)^{1/2} [ Q^{2} + n^{2}\pi^{2}/W^{2} ]^{1/2} \ \cos \left (\dfrac{n \pi}{W} y  \right )  \ ,
\label{eq:continuum_ac_11NRs}
\end{multline} 
for $n$ odd. Once again, this is the same form obtained using the empirical model of Sec.~\ref{subsec:phonon dispersion} (except for an immaterial sign), 
but now describing `transverse' shear waves, symmetric along the transverse $y$ direction, anti-symmetric along the longitudinal direction $z$.
As long as $Q < [s^{1/2}(n\pi)^{3/2}/(2^{1/2}W)] \tan(sn\pi/2)^{1/2}$, the small-$Q$ limit is accurate.

Note also that in the limit of $Q \ll k_{\rm l} k_{\rm t}$ we have considered, both the dilatational and the shear modes have an `optical' nature near $Q=0$: 
Their frequency remains nonzero even in this limit, since the amplitude of the phonon displacement, $\sin(k_{n}y)$, of the $n=0$ mode vanishes and the 
lowest-energy phonon branch has a frequency $c_{\rm t} \pi/L$ at $Q=0$. This result has been obtained also by Donetti {\it et al.}~\cite{Donetti_2006}.
The cause of this behavior can be understood from Eq.~(\ref{eq:continuum_ac_8NR}): Assuming $k_{\rm l} \sim \alpha Q$ as $Q \rightarrow 0$, we find that Eq.~(\ref{eq:continuum_ac_8NR}) admits a solution only for imaginary $\alpha$. In other words, the purely acoustic-like mode is evanescent.
Such a purely acoustic-like mode (that we would label as the $(n,m)= (0,0)$ mode) exists when assuming, instead, freestanding-surface boundary conditions.

Finally, the discussion above has been limited to the case of Dirichlet boundary conditions (CBCs). The case of freestanding BCs is similar, with the 
notable exception of the presence of the $(n,m)=(0,0)$ acoustic-like mode, as previously remarked. The model presented in Sec.~\ref{subsec:phonon dispersion} 
can be shown to remain sufficiently accurate in the small-$Q$ limit. However, as mentioned earlier, the identification of the physical nature of the modes
requires interchanging `even' and `odd': Even-$n$ modes correspond to symmetric shear waves, while odd-$n$ modes correspond to anti-symmetric dilatational waves.


\bibliography{Confined_phonons_ref}

\providecommand{\noopsort}[1]{}\providecommand{\singleletter}[1]{#1}%
\begin{thebibliography}{99}%
\makeatletter
\providecommand \@ifxundefined [1]{%
 \@ifx{#1\undefined}
}%
\providecommand \@ifnum [1]{%
 \ifnum #1\expandafter \@firstoftwo
 \else \expandafter \@secondoftwo
 \fi
}%
\providecommand \@ifx [1]{%
 \ifx #1\expandafter \@firstoftwo
 \else \expandafter \@secondoftwo
 \fi
}%
\providecommand \natexlab [1]{#1}%
\providecommand \enquote  [1]{``#1''}%
\providecommand \bibnamefont  [1]{#1}%
\providecommand \bibfnamefont [1]{#1}%
\providecommand \citenamefont [1]{#1}%
\providecommand \href@noop [0]{\@secondoftwo}%
\providecommand \href [0]{\begingroup \@sanitize@url \@href}%
\providecommand \@href[1]{\@@startlink{#1}\@@href}%
\providecommand \@@href[1]{\endgroup#1\@@endlink}%
\providecommand \@sanitize@url [0]{\catcode `\\12\catcode `\$12\catcode
  `\&12\catcode `\#12\catcode `\^12\catcode `\_12\catcode `\%12\relax}%
\providecommand \@@startlink[1]{}%
\providecommand \@@endlink[0]{}%
\providecommand \url  [0]{\begingroup\@sanitize@url \@url }%
\providecommand \@url [1]{\endgroup\@href {#1}{\urlprefix }}%
\providecommand \urlprefix  [0]{URL }%
\providecommand \Eprint [0]{\href }%
\providecommand \doibase [0]{https://doi.org/}%
\providecommand \selectlanguage [0]{\@gobble}%
\providecommand \bibinfo  [0]{\@secondoftwo}%
\providecommand \bibfield  [0]{\@secondoftwo}%
\providecommand \translation [1]{[#1]}%
\providecommand \BibitemOpen [0]{}%
\providecommand \bibitemStop [0]{}%
\providecommand \bibitemNoStop [0]{.\EOS\space}%
\providecommand \EOS [0]{\spacefactor3000\relax}%
\providecommand \BibitemShut  [1]{\csname bibitem#1\endcsname}%
\let\auto@bib@innerbib\@empty
\bibitem [{\citenamefont {Cui}\ \emph {et~al.}(2000)\citenamefont {Cui},
  \citenamefont {Duan}, \citenamefont {Hu},\ and\ \citenamefont
  {Lieber}}]{ycui2000}%
  \BibitemOpen
  \bibfield  {author} {\bibinfo {author} {\bibfnamefont {Y.}~\bibnamefont
  {Cui}}, \bibinfo {author} {\bibfnamefont {X.}~\bibnamefont {Duan}}, \bibinfo
  {author} {\bibfnamefont {J.}~\bibnamefont {Hu}},\ and\ \bibinfo {author}
  {\bibfnamefont {C.~M.}\ \bibnamefont {Lieber}},\ }\bibfield  {title}
  {\bibinfo {title} {{\it Doping and Electrical Transport in Silicon
  Nanowires}},\ }\href {https://doi.org/https://doi.org/10.1021/jp0009305}
  {\bibfield  {journal} {\bibinfo  {journal} {J. Phys. Chem. B}\ }\textbf
  {\bibinfo {volume} {104}},\ \bibinfo {pages} {5213} (\bibinfo {year}
  {2000})}\BibitemShut {NoStop}%
\bibitem [{\citenamefont {Cui}\ and\ \citenamefont {Lieber}(2001)}]{Ycui2001}%
  \BibitemOpen
  \bibfield  {author} {\bibinfo {author} {\bibfnamefont {Y.}~\bibnamefont
  {Cui}}\ and\ \bibinfo {author} {\bibfnamefont {C.~M.}\ \bibnamefont
  {Lieber}},\ }\bibfield  {title} {\bibinfo {title} {{\it Functional Nanoscale
  Electronic Devices Assembled Using Silicon Nanowire Building Blocks}},\
  }\href {https://doi.org/https://doi.org/10.1126/science.291.5505.851}
  {\bibfield  {journal} {\bibinfo  {journal} {Science}\ }\textbf {\bibinfo
  {volume} {291}},\ \bibinfo {pages} {851} (\bibinfo {year}
  {2001})}\BibitemShut {NoStop}%
\bibitem [{\citenamefont {Cui}\ \emph {et~al.}(2003)\citenamefont {Cui},
  \citenamefont {Zhong}, \citenamefont {Wang}, \citenamefont {Wang},\ and\
  \citenamefont {Lieber}}]{Ycui2003}%
  \BibitemOpen
  \bibfield  {author} {\bibinfo {author} {\bibfnamefont {Y.}~\bibnamefont
  {Cui}}, \bibinfo {author} {\bibfnamefont {Z.}~\bibnamefont {Zhong}}, \bibinfo
  {author} {\bibfnamefont {D.}~\bibnamefont {Wang}}, \bibinfo {author}
  {\bibfnamefont {W.~U.}\ \bibnamefont {Wang}},\ and\ \bibinfo {author}
  {\bibfnamefont {C.~M.}\ \bibnamefont {Lieber}},\ }\bibfield  {title}
  {\bibinfo {title} {{\it High Performance Silicon Nanowire Field Effect
  Transistors}},\ }\href {https://doi.org/https://doi.org/10.1021/nl025875l}
  {\bibfield  {journal} {\bibinfo  {journal} {Nano Lett.}\ }\textbf {\bibinfo
  {volume} {3}},\ \bibinfo {pages} {149} (\bibinfo {year} {2003})}\BibitemShut
  {NoStop}%
\bibitem [{\citenamefont {Jie}\ \emph {et~al.}(2008)\citenamefont {Jie},
  \citenamefont {Zhang}, \citenamefont {Peng}, \citenamefont {Yuan},
  \citenamefont {Lee},\ and\ \citenamefont {Lee}}]{JJie2008}%
  \BibitemOpen
  \bibfield  {author} {\bibinfo {author} {\bibfnamefont {J.}~\bibnamefont
  {Jie}}, \bibinfo {author} {\bibfnamefont {W.}~\bibnamefont {Zhang}}, \bibinfo
  {author} {\bibfnamefont {K.}~\bibnamefont {Peng}}, \bibinfo {author}
  {\bibfnamefont {G.}~\bibnamefont {Yuan}}, \bibinfo {author} {\bibfnamefont
  {C.~S.}\ \bibnamefont {Lee}},\ and\ \bibinfo {author} {\bibfnamefont {S.-T.}\
  \bibnamefont {Lee}},\ }\bibfield  {title} {\bibinfo {title} {{\it
  Surface-Dominated Transport Properties of Silicon Nanowires}},\ }\href
  {https://doi.org/https://doi.org/10.1002/adfm.200800399} {\bibfield
  {journal} {\bibinfo  {journal} {Adv. Funct. Mater.}\ }\textbf {\bibinfo
  {volume} {18}},\ \bibinfo {pages} {3251} (\bibinfo {year}
  {2008})}\BibitemShut {NoStop}%
\bibitem [{\citenamefont {Fukata}(2009)}]{nfukata2009}%
  \BibitemOpen
  \bibfield  {author} {\bibinfo {author} {\bibfnamefont {N.}~\bibnamefont
  {Fukata}},\ }\bibfield  {title} {\bibinfo {title} {{\it Impurity Doping in
  Silicon Nanowires}},\ }\href
  {https://doi.org/https://doi.org/10.1002/adma.200900376} {\bibfield
  {journal} {\bibinfo  {journal} {Adv. Mater.}\ }\textbf {\bibinfo {volume}
  {21(27)}},\ \bibinfo {pages} {2829} (\bibinfo {year} {2009})}\BibitemShut
  {NoStop}%
\bibitem [{\citenamefont {Singh}\ \emph {et~al.}(2006)\citenamefont {Singh},
  \citenamefont {Lim}, \citenamefont {Fang}, \citenamefont {Rustagi},
  \citenamefont {Bera}, \citenamefont {Agarwal}, \citenamefont {Tung},
  \citenamefont {Hoe}, \citenamefont {Omampuliyur}, \citenamefont {Tripathi},
  \citenamefont {Adeyeye}, \citenamefont {Lo}, \citenamefont
  {Balasubramanian},\ and\ \citenamefont {Kwong}}]{nsingh2006}%
  \BibitemOpen
  \bibfield  {author} {\bibinfo {author} {\bibfnamefont {N.}~\bibnamefont
  {Singh}}, \bibinfo {author} {\bibfnamefont {F.~Y.}\ \bibnamefont {Lim}},
  \bibinfo {author} {\bibfnamefont {W.~W.}\ \bibnamefont {Fang}}, \bibinfo
  {author} {\bibfnamefont {S.~C.}\ \bibnamefont {Rustagi}}, \bibinfo {author}
  {\bibfnamefont {L.~K.}\ \bibnamefont {Bera}}, \bibinfo {author}
  {\bibfnamefont {A.}~\bibnamefont {Agarwal}}, \bibinfo {author} {\bibfnamefont
  {C.~H.}\ \bibnamefont {Tung}}, \bibinfo {author} {\bibfnamefont {K.~M.}\
  \bibnamefont {Hoe}}, \bibinfo {author} {\bibfnamefont {S.~R.}\ \bibnamefont
  {Omampuliyur}}, \bibinfo {author} {\bibfnamefont {D.}~\bibnamefont
  {Tripathi}}, \bibinfo {author} {\bibfnamefont {A.~O.}\ \bibnamefont
  {Adeyeye}}, \bibinfo {author} {\bibfnamefont {G.~Q.}\ \bibnamefont {Lo}},
  \bibinfo {author} {\bibfnamefont {N.}~\bibnamefont {Balasubramanian}},\ and\
  \bibinfo {author} {\bibfnamefont {D.~L.}\ \bibnamefont {Kwong}},\ }\bibfield
  {title} {\bibinfo {title} {{\it Ultra-narrow Silicon Nanowire Gate-All-Around
  CMOS Devices: Impact of Diameter, Channel-Orientation and Low Temperature on
  Device Performance}},\ }in\ \href
  {https://doi.org/https://doi.org/10.1109/IEDM.2006.346840} {\emph {\bibinfo
  {booktitle} {2006 IEEE International Electron Devices Meeting (IEDM)}}}\
  (\bibinfo {organization} {IEEE},\ \bibinfo {year} {2006})\ pp.\ \bibinfo
  {pages} {1--4}\BibitemShut {NoStop}%
\bibitem [{\citenamefont {Bangsaruntip}\ \emph {et~al.}(2009)\citenamefont
  {Bangsaruntip}, \citenamefont {Cohen}, \citenamefont {Majumdar},
  \citenamefont {Zhang}, \citenamefont {Engelmann}, \citenamefont {Fuller},
  \citenamefont {Gignac}, \citenamefont {Mittal}, \citenamefont {Newbury},
  \citenamefont {Guillorn}, \citenamefont {Barwicz}, \citenamefont {Sekaric},
  \citenamefont {Frank},\ and\ \citenamefont {Sleight}}]{sbang2009}%
  \BibitemOpen
  \bibfield  {author} {\bibinfo {author} {\bibfnamefont {S.}~\bibnamefont
  {Bangsaruntip}}, \bibinfo {author} {\bibfnamefont {G.~M.}\ \bibnamefont
  {Cohen}}, \bibinfo {author} {\bibfnamefont {A.}~\bibnamefont {Majumdar}},
  \bibinfo {author} {\bibfnamefont {Y.}~\bibnamefont {Zhang}}, \bibinfo
  {author} {\bibfnamefont {S.~U.}\ \bibnamefont {Engelmann}}, \bibinfo {author}
  {\bibfnamefont {N.~C.~M.}\ \bibnamefont {Fuller}}, \bibinfo {author}
  {\bibfnamefont {L.~M.}\ \bibnamefont {Gignac}}, \bibinfo {author}
  {\bibfnamefont {S.}~\bibnamefont {Mittal}}, \bibinfo {author} {\bibfnamefont
  {J.~S.}\ \bibnamefont {Newbury}}, \bibinfo {author} {\bibfnamefont
  {M.}~\bibnamefont {Guillorn}}, \bibinfo {author} {\bibfnamefont
  {T.}~\bibnamefont {Barwicz}}, \bibinfo {author} {\bibfnamefont
  {L.}~\bibnamefont {Sekaric}}, \bibinfo {author} {\bibfnamefont {M.~M.}\
  \bibnamefont {Frank}},\ and\ \bibinfo {author} {\bibfnamefont {J.~W.}\
  \bibnamefont {Sleight}},\ }\bibfield  {title} {\bibinfo {title} {{\it High
  performance and highly uniform gate-all-around silicon nanowire MOSFETs with
  wire size dependent scaling}},\ }in\ \href
  {https://doi.org/https://doi.org/10.1109/IEDM.2009.5424364} {\emph {\bibinfo
  {booktitle} {2009 IEEE International Electron Devices Meeting (IEDM)}}}\
  (\bibinfo {organization} {IEEE},\ \bibinfo {year} {2009})\ pp.\ \bibinfo
  {pages} {1--4}\BibitemShut {NoStop}%
\bibitem [{\citenamefont {Li}\ \emph {et~al.}(2009)\citenamefont {Li},
  \citenamefont {Yeo}, \citenamefont {Suk}, \citenamefont {Yeoh}, \citenamefont
  {Kim}, \citenamefont {Chung}, \citenamefont {Oh},\ and\ \citenamefont
  {Lee}}]{mlee2009}%
  \BibitemOpen
  \bibfield  {author} {\bibinfo {author} {\bibfnamefont {M.}~\bibnamefont
  {Li}}, \bibinfo {author} {\bibfnamefont {K.~H.}\ \bibnamefont {Yeo}},
  \bibinfo {author} {\bibfnamefont {S.~D.}\ \bibnamefont {Suk}}, \bibinfo
  {author} {\bibfnamefont {Y.~Y.}\ \bibnamefont {Yeoh}}, \bibinfo {author}
  {\bibfnamefont {D.-W.}\ \bibnamefont {Kim}}, \bibinfo {author} {\bibfnamefont
  {T.~Y.}\ \bibnamefont {Chung}}, \bibinfo {author} {\bibfnamefont {K.~S.}\
  \bibnamefont {Oh}},\ and\ \bibinfo {author} {\bibfnamefont {W.-S.}\
  \bibnamefont {Lee}},\ }\bibfield  {title} {\bibinfo {title} {{\it Sub-10 nm
  gate-all-around CMOS nanowire transistors on bulk Si substrate}},\ }in\ \href
  {https://ieeexplore.ieee.org/document/5200646} {\emph {\bibinfo {booktitle}
  {2009 Symposium on VLSI Technology}}}\ (\bibinfo {organization} {IEEE},\
  \bibinfo {year} {2009})\ pp.\ \bibinfo {pages} {94--95}\BibitemShut {NoStop}%
\bibitem [{\citenamefont {Bolotin}\ \emph {et~al.}(2008)\citenamefont
  {Bolotin}, \citenamefont {Sikes}, \citenamefont {Jiang}, \citenamefont
  {Klima}, \citenamefont {Fudenberg}, \citenamefont {Hone}, \citenamefont
  {Kim},\ and\ \citenamefont {Stormer}}]{Bolotin_2008}%
  \BibitemOpen
  \bibfield  {author} {\bibinfo {author} {\bibfnamefont {K.~I.}\ \bibnamefont
  {Bolotin}}, \bibinfo {author} {\bibfnamefont {K.~J.}\ \bibnamefont {Sikes}},
  \bibinfo {author} {\bibfnamefont {Z.}~\bibnamefont {Jiang}}, \bibinfo
  {author} {\bibfnamefont {M.}~\bibnamefont {Klima}}, \bibinfo {author}
  {\bibfnamefont {G.}~\bibnamefont {Fudenberg}}, \bibinfo {author}
  {\bibfnamefont {J.}~\bibnamefont {Hone}}, \bibinfo {author} {\bibfnamefont
  {P.}~\bibnamefont {Kim}},\ and\ \bibinfo {author} {\bibfnamefont {H.~L.}\
  \bibnamefont {Stormer}},\ }\bibfield  {title} {\bibinfo {title} {{\it
  Ultrahigh electron mobility in suspended graphene}},\ }\href
  {https://doi.org/https://doi.org/10.1016/j.ssc.2008.02.024} {\bibfield
  {journal} {\bibinfo  {journal} {Solid State Commun.}\ }\textbf {\bibinfo
  {volume} {146}},\ \bibinfo {pages} {351} (\bibinfo {year}
  {2008})}\BibitemShut {NoStop}%
\bibitem [{\citenamefont {Son}\ \emph {et~al.}(2006)\citenamefont {Son},
  \citenamefont {Cohen},\ and\ \citenamefont {Louie}}]{Son_2006}%
  \BibitemOpen
  \bibfield  {author} {\bibinfo {author} {\bibfnamefont {Y.-W.}\ \bibnamefont
  {Son}}, \bibinfo {author} {\bibfnamefont {M.~L.}\ \bibnamefont {Cohen}},\
  and\ \bibinfo {author} {\bibfnamefont {S.~G.}\ \bibnamefont {Louie}},\
  }\bibfield  {title} {\bibinfo {title} {{\it Energy Gaps in Graphene
  Nanoribbons}},\ }\href
  {https://doi.org/https://doi.org/10.1103/PhysRevLett.97.216803} {\bibfield
  {journal} {\bibinfo  {journal} {Phys. Rev. Lett.}\ }\textbf {\bibinfo
  {volume} {97}},\ \bibinfo {pages} {216803} (\bibinfo {year}
  {2006})}\BibitemShut {NoStop}%
\bibitem [{\citenamefont {Ruffieux}\ \emph {et~al.}(2012)\citenamefont
  {Ruffieux}, \citenamefont {Cai}, \citenamefont {Plumb}, \citenamefont
  {Patthey}, \citenamefont {Prezzi}, \citenamefont {Ferretti}, \citenamefont
  {Molinari}, \citenamefont {Feng}, \citenamefont {Müllen}, \citenamefont
  {Pignedoli} \emph {et~al.}}]{ruffieux2012electronic}%
  \BibitemOpen
  \bibfield  {author} {\bibinfo {author} {\bibfnamefont {P.}~\bibnamefont
  {Ruffieux}}, \bibinfo {author} {\bibfnamefont {J.}~\bibnamefont {Cai}},
  \bibinfo {author} {\bibfnamefont {N.~C.}\ \bibnamefont {Plumb}}, \bibinfo
  {author} {\bibfnamefont {L.}~\bibnamefont {Patthey}}, \bibinfo {author}
  {\bibfnamefont {D.}~\bibnamefont {Prezzi}}, \bibinfo {author} {\bibfnamefont
  {A.}~\bibnamefont {Ferretti}}, \bibinfo {author} {\bibfnamefont
  {E.}~\bibnamefont {Molinari}}, \bibinfo {author} {\bibfnamefont
  {X.}~\bibnamefont {Feng}}, \bibinfo {author} {\bibfnamefont {K.}~\bibnamefont
  {Müllen}}, \bibinfo {author} {\bibfnamefont {C.~A.}\ \bibnamefont
  {Pignedoli}}, \emph {et~al.},\ }\bibfield  {title} {\bibinfo {title} {{\it
  Electronic structure of atomically precise graphene nanoribbons}},\ }\href
  {https://doi.org/https://doi.org/10.1021/nn3021376} {\bibfield  {journal}
  {\bibinfo  {journal} {ACS Nano}\ }\textbf {\bibinfo {volume} {6}},\ \bibinfo
  {pages} {6930} (\bibinfo {year} {2012})}\BibitemShut {NoStop}%
\bibitem [{\citenamefont {Huang}\ \emph {et~al.}(2012)\citenamefont {Huang},
  \citenamefont {Wei}, \citenamefont {Sun}, \citenamefont {Wong}, \citenamefont
  {Feng}, \citenamefont {Neto},\ and\ \citenamefont
  {Wee}}]{huang2012spatially}%
  \BibitemOpen
  \bibfield  {author} {\bibinfo {author} {\bibfnamefont {H.}~\bibnamefont
  {Huang}}, \bibinfo {author} {\bibfnamefont {D.}~\bibnamefont {Wei}}, \bibinfo
  {author} {\bibfnamefont {J.}~\bibnamefont {Sun}}, \bibinfo {author}
  {\bibfnamefont {S.~L.}\ \bibnamefont {Wong}}, \bibinfo {author}
  {\bibfnamefont {Y.~P.}\ \bibnamefont {Feng}}, \bibinfo {author}
  {\bibfnamefont {A.~C.}\ \bibnamefont {Neto}},\ and\ \bibinfo {author}
  {\bibfnamefont {A.~T.~S.}\ \bibnamefont {Wee}},\ }\bibfield  {title}
  {\bibinfo {title} {{\it Spatially resolved electronic structures of
  atomically precise armchair graphene nanoribbons}},\ }\href
  {https://doi.org/https://doi.org/10.1038/srep00983} {\bibfield  {journal}
  {\bibinfo  {journal} {Sci. Rep.}\ }\textbf {\bibinfo {volume} {2}},\ \bibinfo
  {pages} {983} (\bibinfo {year} {2012})}\BibitemShut {NoStop}%
\bibitem [{\citenamefont {Dinh}\ \emph {et~al.}(2024)\citenamefont {Dinh},
  \citenamefont {Yusufoglu}, \citenamefont {Yumigeta}, \citenamefont {Kinikar},
  \citenamefont {Sweepe}, \citenamefont {Zeszut}, \citenamefont {Chang},
  \citenamefont {Copic}, \citenamefont {Janssen}, \citenamefont {Holloway},
  \citenamefont {Battaglia}, \citenamefont {Kuntubek}, \citenamefont {Zahin},
  \citenamefont {Lin}, \citenamefont {Vandenberghe}, \citenamefont {LeRoy},
  \citenamefont {M{\"{u}}llen}, \citenamefont {Fasel}, \citenamefont
  {Borin~Barin},\ and\ \citenamefont {Mutlu}}]{Dinh_2024}%
  \BibitemOpen
  \bibfield  {author} {\bibinfo {author} {\bibfnamefont {C.}~\bibnamefont
  {Dinh}}, \bibinfo {author} {\bibfnamefont {M.}~\bibnamefont {Yusufoglu}},
  \bibinfo {author} {\bibfnamefont {K.}~\bibnamefont {Yumigeta}}, \bibinfo
  {author} {\bibfnamefont {A.}~\bibnamefont {Kinikar}}, \bibinfo {author}
  {\bibfnamefont {T.}~\bibnamefont {Sweepe}}, \bibinfo {author} {\bibfnamefont
  {Z.}~\bibnamefont {Zeszut}}, \bibinfo {author} {\bibfnamefont {Y.-J.}\
  \bibnamefont {Chang}}, \bibinfo {author} {\bibfnamefont {C.}~\bibnamefont
  {Copic}}, \bibinfo {author} {\bibfnamefont {S.}~\bibnamefont {Janssen}},
  \bibinfo {author} {\bibfnamefont {R.}~\bibnamefont {Holloway}}, \bibinfo
  {author} {\bibfnamefont {J.}~\bibnamefont {Battaglia}}, \bibinfo {author}
  {\bibfnamefont {A.}~\bibnamefont {Kuntubek}}, \bibinfo {author}
  {\bibfnamefont {F.}~\bibnamefont {Zahin}}, \bibinfo {author} {\bibfnamefont
  {Y.~C.}\ \bibnamefont {Lin}}, \bibinfo {author} {\bibfnamefont {W.~G.}\
  \bibnamefont {Vandenberghe}}, \bibinfo {author} {\bibfnamefont {B.~J.}\
  \bibnamefont {LeRoy}}, \bibinfo {author} {\bibfnamefont {K.}~\bibnamefont
  {M{\"{u}}llen}}, \bibinfo {author} {\bibfnamefont {R.}~\bibnamefont {Fasel}},
  \bibinfo {author} {\bibfnamefont {G.}~\bibnamefont {Borin~Barin}},\ and\
  \bibinfo {author} {\bibfnamefont {Z.}~\bibnamefont {Mutlu}},\ }\bibfield
  {title} {\bibinfo {title} {{\it Atomically Precise Graphene Nanoribbon
  Transistors with Long-Term Stability and Reliability}},\ }\href
  {https://doi.org/https://doi.org/10.1021/acsnano.4c04097} {\bibfield
  {journal} {\bibinfo  {journal} {ACS Nano}\ }\textbf {\bibinfo {volume}
  {18}},\ \bibinfo {pages} {22949} (\bibinfo {year} {2024})}\BibitemShut
  {NoStop}%
\bibitem [{\citenamefont {Tian}\ \emph {et~al.}(2023)\citenamefont {Tian},
  \citenamefont {Miao}, \citenamefont {Zhao},\ and\ \citenamefont
  {Wang}}]{Tian_2023}%
  \BibitemOpen
  \bibfield  {author} {\bibinfo {author} {\bibfnamefont {C.}~\bibnamefont
  {Tian}}, \bibinfo {author} {\bibfnamefont {W.}~\bibnamefont {Miao}}, \bibinfo
  {author} {\bibfnamefont {L.}~\bibnamefont {Zhao}},\ and\ \bibinfo {author}
  {\bibfnamefont {J.}~\bibnamefont {Wang}},\ }\bibfield  {title} {\bibinfo
  {title} {{\it Graphene nanoribbons: Current status and challenges as
  quasi-one-dimensional nanomaterials}},\ }\href
  {https://doi.org/https://doi.org/10.1016/j.revip.2023.100082} {\bibfield
  {journal} {\bibinfo  {journal} {Rev. Phys.}\ }\textbf {\bibinfo {volume}
  {10}},\ \bibinfo {pages} {100082} (\bibinfo {year} {2023})}\BibitemShut
  {NoStop}%
\bibitem [{\citenamefont {Lou}\ \emph {et~al.}(2024)\citenamefont {Lou},
  \citenamefont {Lyu}, \citenamefont {Zhou}, \citenamefont {Shen},
  \citenamefont {Chen},\ and\ \citenamefont {Shi}}]{Lou_2024}%
  \BibitemOpen
  \bibfield  {author} {\bibinfo {author} {\bibfnamefont {S.}~\bibnamefont
  {Lou}}, \bibinfo {author} {\bibfnamefont {B.}~\bibnamefont {Lyu}}, \bibinfo
  {author} {\bibfnamefont {X.}~\bibnamefont {Zhou}}, \bibinfo {author}
  {\bibfnamefont {P.}~\bibnamefont {Shen}}, \bibinfo {author} {\bibfnamefont
  {J.}~\bibnamefont {Chen}},\ and\ \bibinfo {author} {\bibfnamefont
  {Z.}~\bibnamefont {Shi}},\ }\bibfield  {title} {\bibinfo {title} {{\it
  Graphene nanoribbons: current status, challenges and opportunities}},\ }\href
  {https://doi.org/https://doi.org/0.1007/s44214-024-00050-8} {\bibfield
  {journal} {\bibinfo  {journal} {Quantum Front.}\ }\textbf {\bibinfo {volume}
  {3}},\ \bibinfo {pages} {3} (\bibinfo {year} {2024})}\BibitemShut {NoStop}%
\bibitem [{\citenamefont {Gunst}\ \emph {et~al.}(2016)\citenamefont {Gunst},
  \citenamefont {Markussen}, \citenamefont {Stokbro},\ and\ \citenamefont
  {Brandbyge}}]{Gunst_2016}%
  \BibitemOpen
  \bibfield  {author} {\bibinfo {author} {\bibfnamefont {T.}~\bibnamefont
  {Gunst}}, \bibinfo {author} {\bibfnamefont {T.}~\bibnamefont {Markussen}},
  \bibinfo {author} {\bibfnamefont {K.}~\bibnamefont {Stokbro}},\ and\ \bibinfo
  {author} {\bibfnamefont {M.}~\bibnamefont {Brandbyge}},\ }\bibfield  {title}
  {\bibinfo {title} {{\it First-principles method for electron-phonon coupling
  and electron mobility: Applications to two-dimensional materials}},\ }\href
  {https://doi.org/https://doi.org/10.1103/PhysRevB.93.035414} {\bibfield
  {journal} {\bibinfo  {journal} {Phys. Rev. B}\ }\textbf {\bibinfo {volume}
  {93}},\ \bibinfo {pages} {035414} (\bibinfo {year} {2016})}\BibitemShut
  {NoStop}%
\bibitem [{\citenamefont {Ponc{\'{e}}}\ \emph {et~al.}(2020)\citenamefont
  {Ponc{\'{e}}}, \citenamefont {Li}, \citenamefont {Reichardt},\ and\
  \citenamefont {Giustino}}]{Ponce_2020}%
  \BibitemOpen
  \bibfield  {author} {\bibinfo {author} {\bibfnamefont {S.}~\bibnamefont
  {Ponc{\'{e}}}}, \bibinfo {author} {\bibfnamefont {W.}~\bibnamefont {Li}},
  \bibinfo {author} {\bibfnamefont {S.}~\bibnamefont {Reichardt}},\ and\
  \bibinfo {author} {\bibfnamefont {F.}~\bibnamefont {Giustino}},\ }\bibfield
  {title} {\bibinfo {title} {{\it First-principles calculations of charge
  carrier mobility and conductivity in bulk semiconductors and two-dimensional
  materials}},\ }\href
  {https://doi.org/https://doi.org/10.1088/1361-6633/ab6a43} {\bibfield
  {journal} {\bibinfo  {journal} {Rep. Prog. Phys.}\ }\textbf {\bibinfo
  {volume} {83}},\ \bibinfo {pages} {036501} (\bibinfo {year}
  {2020})}\BibitemShut {NoStop}%
\bibitem [{\citenamefont {Afzalian}\ \emph {et~al.}(2021)\citenamefont
  {Afzalian}, \citenamefont {Akhoundi}, \citenamefont {Gaddemane},
  \citenamefont {Duflou},\ and\ \citenamefont {Houssa}}]{Afzalian_2021}%
  \BibitemOpen
  \bibfield  {author} {\bibinfo {author} {\bibfnamefont {A.}~\bibnamefont
  {Afzalian}}, \bibinfo {author} {\bibfnamefont {E.}~\bibnamefont {Akhoundi}},
  \bibinfo {author} {\bibfnamefont {G.}~\bibnamefont {Gaddemane}}, \bibinfo
  {author} {\bibfnamefont {R.}~\bibnamefont {Duflou}},\ and\ \bibinfo {author}
  {\bibfnamefont {M.}~\bibnamefont {Houssa}},\ }\bibfield  {title} {\bibinfo
  {title} {{\it Advanced DFT-NEGF Transport Techniques for Novel 2-D Material
  and Device Exploration Including HfS{$_2$}/WSe{$_2$} van der Waals
  Heterojunction TFET and WTe{$_2$}/WS{$_2$} Metal/Semiconductor Contact}},\
  }\href {https://doi.org/https://doi.org/10.1109/TED.2021.3078412} {\bibfield
  {journal} {\bibinfo  {journal} {IEEE Trans. Electron Devices}\ }\textbf
  {\bibinfo {volume} {68}},\ \bibinfo {pages} {5372} (\bibinfo {year}
  {2021})}\BibitemShut {NoStop}%
\bibitem [{\citenamefont {Campi}\ \emph {et~al.}(2023)\citenamefont {Campi},
  \citenamefont {Mounet}, \citenamefont {Gibertini}, \citenamefont {Pizzi},\
  and\ \citenamefont {Marzari}}]{Campi_2023}%
  \BibitemOpen
  \bibfield  {author} {\bibinfo {author} {\bibfnamefont {D.}~\bibnamefont
  {Campi}}, \bibinfo {author} {\bibfnamefont {N.}~\bibnamefont {Mounet}},
  \bibinfo {author} {\bibfnamefont {M.}~\bibnamefont {Gibertini}}, \bibinfo
  {author} {\bibfnamefont {G.}~\bibnamefont {Pizzi}},\ and\ \bibinfo {author}
  {\bibfnamefont {N.}~\bibnamefont {Marzari}},\ }\bibfield  {title} {\bibinfo
  {title} {{\it Expansion of the Materials Cloud 2D Database}},\ }\href
  {https://doi.org/https://10.1021/acsnano.2c11510} {\bibfield  {journal}
  {\bibinfo  {journal} {ACS Nano}\ }\textbf {\bibinfo {volume} {17}},\ \bibinfo
  {pages} {11268} (\bibinfo {year} {2023})}\BibitemShut {NoStop}%
\bibitem [{\citenamefont {Luisier}\ \emph {et~al.}(2023)\citenamefont
  {Luisier}, \citenamefont {Klinkert}, \citenamefont {Fiore}, \citenamefont
  {Backman}, \citenamefont {Lee}, \citenamefont {Stieger},\ and\ \citenamefont
  {Szab{\'{o}}}}]{Cresti_2023}%
  \BibitemOpen
  \bibfield  {author} {\bibinfo {author} {\bibfnamefont {M.}~\bibnamefont
  {Luisier}}, \bibinfo {author} {\bibfnamefont {C.}~\bibnamefont {Klinkert}},
  \bibinfo {author} {\bibfnamefont {S.}~\bibnamefont {Fiore}}, \bibinfo
  {author} {\bibfnamefont {J.}~\bibnamefont {Backman}}, \bibinfo {author}
  {\bibfnamefont {Y.}~\bibnamefont {Lee}}, \bibinfo {author} {\bibfnamefont
  {C.}~\bibnamefont {Stieger}},\ and\ \bibinfo {author} {\bibfnamefont
  {{\'{A}}.}~\bibnamefont {Szab{\'{o}}}},\ }\bibfield  {title} {\bibinfo
  {title} {{\it Field-Effect Transistors Based on 2D Materials: A Modeling
  Perspective}},\ }in\ \href
  {https://doi.org/https://doi.org/10.1002/9781394228713} {\emph {\bibinfo
  {booktitle} {\it Beyond CMOS -- State of the Art and Trends}}},\ \bibinfo
  {editor} {edited by\ \bibinfo {editor} {\bibfnamefont {A.}~\bibnamefont
  {Cresti}}}\ (\bibinfo  {publisher} {{ISTE - John Wiley {\&} Sons}},\ \bibinfo
  {address} {London - Hoboken, New Jersey},\ \bibinfo {year} {2023})\
  Chap.~\bibinfo {chapter} {2}, pp.\ \bibinfo {pages} {33--78}\BibitemShut
  {NoStop}%
\bibitem [{\citenamefont {Wang}\ \emph {et~al.}(2004)\citenamefont {Wang},
  \citenamefont {Polizzi},\ and\ \citenamefont {Lundstrom}}]{Wang_2004}%
  \BibitemOpen
  \bibfield  {author} {\bibinfo {author} {\bibfnamefont {J.}~\bibnamefont
  {Wang}}, \bibinfo {author} {\bibfnamefont {E.}~\bibnamefont {Polizzi}},\ and\
  \bibinfo {author} {\bibfnamefont {M.}~\bibnamefont {Lundstrom}},\ }\bibfield
  {title} {\bibinfo {title} {{\it A three-dimensional quantum simulation of
  silicon nanowire transistors with the effective-mass approximation}},\ }\href
  {https://doi.org/https://doi.org/10.1063/1.1769089} {\bibfield  {journal}
  {\bibinfo  {journal} {J. Appl. Phys.}\ }\textbf {\bibinfo {volume} {96}},\
  \bibinfo {pages} {2192–2203} (\bibinfo {year} {2004})}\BibitemShut
  {NoStop}%
\bibitem [{\citenamefont {Ng}\ \emph {et~al.}(2007)\citenamefont {Ng},
  \citenamefont {Zhou}, \citenamefont {Yang}, \citenamefont {Sim},
  \citenamefont {Tan},\ and\ \citenamefont {Wu}}]{Ng_2007}%
  \BibitemOpen
  \bibfield  {author} {\bibinfo {author} {\bibfnamefont {M.-F.}\ \bibnamefont
  {Ng}}, \bibinfo {author} {\bibfnamefont {L.}~\bibnamefont {Zhou}}, \bibinfo
  {author} {\bibfnamefont {S.-W.}\ \bibnamefont {Yang}}, \bibinfo {author}
  {\bibfnamefont {L.~Y.}\ \bibnamefont {Sim}}, \bibinfo {author} {\bibfnamefont
  {V.~B.~C.}\ \bibnamefont {Tan}},\ and\ \bibinfo {author} {\bibfnamefont
  {P.}~\bibnamefont {Wu}},\ }\bibfield  {title} {\bibinfo {title} {{\it
  Theoretical investigation of silicon nanowires: Methodology, geometry,
  surface modification, and electrical conductivity using a multiscale
  approach}},\ }\href
  {https://doi.org/https://doi.org/10.1103/PhysRevB.76.155435} {\bibfield
  {journal} {\bibinfo  {journal} {Phys. Rev. B}\ }\textbf {\bibinfo {volume}
  {76}},\ \bibinfo {pages} {155435} (\bibinfo {year} {2007})}\BibitemShut
  {NoStop}%
\bibitem [{\citenamefont {Jin}\ \emph {et~al.}(2007)\citenamefont {Jin},
  \citenamefont {Fischetti},\ and\ \citenamefont {Tang}}]{sJin2007}%
  \BibitemOpen
  \bibfield  {author} {\bibinfo {author} {\bibfnamefont {S.}~\bibnamefont
  {Jin}}, \bibinfo {author} {\bibfnamefont {M.~V.}\ \bibnamefont {Fischetti}},\
  and\ \bibinfo {author} {\bibfnamefont {T.-w.}\ \bibnamefont {Tang}},\
  }\bibfield  {title} {\bibinfo {title} {{\it Modeling of electron mobility in
  gated silicon nanowires at room temperature: Surface roughness scattering,
  dielectric screening, and band nonparabolicity}},\ }\href
  {https://doi.org/https://doi.org/10.1063/1.2802586} {\bibfield  {journal}
  {\bibinfo  {journal} {J. Appl. Phys.}\ }\textbf {\bibinfo {volume} {102}},\
  \bibinfo {pages} {083715} (\bibinfo {year} {2007})}\BibitemShut {NoStop}%
\bibitem [{\citenamefont {Schenk}\ and\ \citenamefont
  {Luisier}(2008)}]{Schenk_2008}%
  \BibitemOpen
  \bibfield  {author} {\bibinfo {author} {\bibfnamefont {A.}~\bibnamefont
  {Schenk}}\ and\ \bibinfo {author} {\bibfnamefont {M.}~\bibnamefont
  {Luisier}},\ }\bibfield  {title} {\bibinfo {title} {{\it Three-dimensional
  quantum simulation of silicon nanowires}},\ }in\ \href
  {https://doi.org/https://doi.org/10.1109/SNW.2008.5418465} {\emph {\bibinfo
  {booktitle} {2008 IEEE Silicon Nanoelectronics Workshop}}}\ (\bibinfo {year}
  {2008})\ pp.\ \bibinfo {pages} {1--2}\BibitemShut {NoStop}%
\bibitem [{\citenamefont {Rurali}\ \emph {et~al.}(2008)\citenamefont {Rurali},
  \citenamefont {Markussen}, \citenamefont {Su{\~n}{\'e}}, \citenamefont
  {Brandbyge},\ and\ \citenamefont {Jauho}}]{Rurali_2008}%
  \BibitemOpen
  \bibfield  {author} {\bibinfo {author} {\bibfnamefont {R.}~\bibnamefont
  {Rurali}}, \bibinfo {author} {\bibfnamefont {T.}~\bibnamefont {Markussen}},
  \bibinfo {author} {\bibfnamefont {J.}~\bibnamefont {Su{\~n}{\'e}}}, \bibinfo
  {author} {\bibfnamefont {M.}~\bibnamefont {Brandbyge}},\ and\ \bibinfo
  {author} {\bibfnamefont {A.-P.}\ \bibnamefont {Jauho}},\ }\bibfield  {title}
  {\bibinfo {title} {{\it Modeling Transport in Ultrathin Si Nanowires: Charged
  versus Neutral Impurities}},\ }\href
  {https://doi.org/https://doi.org/110.1021/nl801409m} {\bibfield  {journal}
  {\bibinfo  {journal} {Nano Lett.}\ }\textbf {\bibinfo {volume} {8}},\
  \bibinfo {pages} {2825} (\bibinfo {year} {2008})}\BibitemShut {NoStop}%
\bibitem [{\citenamefont {Luisier}\ and\ \citenamefont
  {Klimeck}(2009)}]{Luisier_2009}%
  \BibitemOpen
  \bibfield  {author} {\bibinfo {author} {\bibfnamefont {M.}~\bibnamefont
  {Luisier}}\ and\ \bibinfo {author} {\bibfnamefont {G.}~\bibnamefont
  {Klimeck}},\ }\bibfield  {title} {\bibinfo {title} {{\it A three-dimensional
  quantum simulation of silicon nanowire transistors with the effective-mass
  approximation}},\ }\href
  {https://doi.org/https://doi.org/10.1103/PhysRevB.80.155430} {\bibfield
  {journal} {\bibinfo  {journal} {Phys. Rev. B}\ }\textbf {\bibinfo {volume}
  {80}},\ \bibinfo {pages} {155430} (\bibinfo {year} {2009})}\BibitemShut
  {NoStop}%
\bibitem [{\citenamefont {Dong}\ \emph {et~al.}(2011)\citenamefont {Dong},
  \citenamefont {Li}, \citenamefont {Sun}, \citenamefont {Zhang},\ and\
  \citenamefont {Li}}]{Dong_2011}%
  \BibitemOpen
  \bibfield  {author} {\bibinfo {author} {\bibfnamefont {J.~C.}\ \bibnamefont
  {Dong}}, \bibinfo {author} {\bibfnamefont {H.}~\bibnamefont {Li}}, \bibinfo
  {author} {\bibfnamefont {F.~W.}\ \bibnamefont {Sun}}, \bibinfo {author}
  {\bibfnamefont {K.}~\bibnamefont {Zhang}},\ and\ \bibinfo {author}
  {\bibfnamefont {Y.~F.}\ \bibnamefont {Li}},\ }\bibfield  {title} {\bibinfo
  {title} {{\it Theoretical Study of the Properties of Si Nanowire Electronic
  Devices}},\ }\href {https://doi.org/https://doi.org/10.1021/jp2007045}
  {\bibfield  {journal} {\bibinfo  {journal} {J. Phys. Chem. C}\ }\textbf
  {\bibinfo {volume} {115}},\ \bibinfo {pages} {13901} (\bibinfo {year}
  {2011})}\BibitemShut {NoStop}%
\bibitem [{\citenamefont {Fang}\ \emph {et~al.}(2016)\citenamefont {Fang},
  \citenamefont {Vandenberghe}, \citenamefont {Fu},\ and\ \citenamefont
  {Fischetti}}]{fang2016pseudopotential}%
  \BibitemOpen
  \bibfield  {author} {\bibinfo {author} {\bibfnamefont {J.}~\bibnamefont
  {Fang}}, \bibinfo {author} {\bibfnamefont {W.~G.}\ \bibnamefont
  {Vandenberghe}}, \bibinfo {author} {\bibfnamefont {B.}~\bibnamefont {Fu}},\
  and\ \bibinfo {author} {\bibfnamefont {M.~V.}\ \bibnamefont {Fischetti}},\
  }\bibfield  {title} {\bibinfo {title} {{\it Pseudopotential-based electron
  quantum transport: Theoretical formulation and application to nanometer-scale
  silicon nanowire transistors}},\ }\href
  {https://doi.org/https://doi.org/10.1063/1.4939963} {\bibfield  {journal}
  {\bibinfo  {journal} {J. Appl. Phys.}\ }\textbf {\bibinfo {volume} {119}},\
  \bibinfo {pages} {035701} (\bibinfo {year} {2016})}\BibitemShut {NoStop}%
\bibitem [{\citenamefont {Georgiev}\ \emph {et~al.}(2017)\citenamefont
  {Georgiev}, \citenamefont {Mirza}, \citenamefont {Dochioiu}, \citenamefont
  {Adamu-Lema}, \citenamefont {Amoroso}, \citenamefont {Towie}, \citenamefont
  {Riddet}, \citenamefont {MacLaren}, \citenamefont {Asenov},\ and\
  \citenamefont {Paul}}]{Georgiev_2017}%
  \BibitemOpen
  \bibfield  {author} {\bibinfo {author} {\bibfnamefont {V.~P.}\ \bibnamefont
  {Georgiev}}, \bibinfo {author} {\bibfnamefont {M.~M.}\ \bibnamefont {Mirza}},
  \bibinfo {author} {\bibfnamefont {A.-I.}\ \bibnamefont {Dochioiu}}, \bibinfo
  {author} {\bibfnamefont {F.}~\bibnamefont {Adamu-Lema}}, \bibinfo {author}
  {\bibfnamefont {S.~M.}\ \bibnamefont {Amoroso}}, \bibinfo {author}
  {\bibfnamefont {E.}~\bibnamefont {Towie}}, \bibinfo {author} {\bibfnamefont
  {C.}~\bibnamefont {Riddet}}, \bibinfo {author} {\bibfnamefont {D.~A.}\
  \bibnamefont {MacLaren}}, \bibinfo {author} {\bibfnamefont {A.}~\bibnamefont
  {Asenov}},\ and\ \bibinfo {author} {\bibfnamefont {D.~J.}\ \bibnamefont
  {Paul}},\ }\bibfield  {title} {\bibinfo {title} {{\it Experimental and
  Simulation Study of Silicon Nanowire Transistors Using Heavily Doped
  Channels}},\ }\href {https://doi.org/10.1109/TNANO.2017.2665691} {\bibfield
  {journal} {\bibinfo  {journal} {IEEE Trans. Nanotechnol.}\ }\textbf {\bibinfo
  {volume} {16}},\ \bibinfo {pages} {727} (\bibinfo {year} {2017})}\BibitemShut
  {NoStop}%
\bibitem [{\citenamefont {Liang}\ \emph {et~al.}(2007)\citenamefont {Liang},
  \citenamefont {Neophytou}, \citenamefont {Lundstrom},\ and\ \citenamefont
  {Nikonov}}]{liang2007ballistic}%
  \BibitemOpen
  \bibfield  {author} {\bibinfo {author} {\bibfnamefont {G.}~\bibnamefont
  {Liang}}, \bibinfo {author} {\bibfnamefont {N.}~\bibnamefont {Neophytou}},
  \bibinfo {author} {\bibfnamefont {M.~S.}\ \bibnamefont {Lundstrom}},\ and\
  \bibinfo {author} {\bibfnamefont {D.~E.}\ \bibnamefont {Nikonov}},\
  }\bibfield  {title} {\bibinfo {title} {{\it Ballistic graphene nanoribbon
  metal-oxide-semiconductor field-effect transistors: A full real-space quantum
  transport simulation}},\ }\href
  {https://doi.org/https://doi.org/10.1063/1.2775917} {\bibfield  {journal}
  {\bibinfo  {journal} {J. Appl. Phys.}\ }\textbf {\bibinfo {volume} {102}},\
  \bibinfo {pages} {054307} (\bibinfo {year} {2007})}\BibitemShut {NoStop}%
\bibitem [{\citenamefont {Ouyang}\ \emph {et~al.}(2007)\citenamefont {Ouyang},
  \citenamefont {Yoon},\ and\ \citenamefont {Guo}}]{ouyang2007scaling}%
  \BibitemOpen
  \bibfield  {author} {\bibinfo {author} {\bibfnamefont {Y.}~\bibnamefont
  {Ouyang}}, \bibinfo {author} {\bibfnamefont {Y.}~\bibnamefont {Yoon}},\ and\
  \bibinfo {author} {\bibfnamefont {J.}~\bibnamefont {Guo}},\ }\bibfield
  {title} {\bibinfo {title} {{\it Scaling behaviors of graphene nanoribbon
  FETs: A three-dimensional quantum simulation study}},\ }\href
  {https://doi.org/10.1109/TED.2007.902692} {\bibfield  {journal} {\bibinfo
  {journal} {IEEE Trans. Electron Devices}\ }\textbf {\bibinfo {volume} {54}},\
  \bibinfo {pages} {2223} (\bibinfo {year} {2007})}\BibitemShut {NoStop}%
\bibitem [{\citenamefont {Fiori}\ and\ \citenamefont
  {Iannaccone}(2007)}]{Fiori_2007}%
  \BibitemOpen
  \bibfield  {author} {\bibinfo {author} {\bibfnamefont {G.}~\bibnamefont
  {Fiori}}\ and\ \bibinfo {author} {\bibfnamefont {G.}~\bibnamefont
  {Iannaccone}},\ }\bibfield  {title} {\bibinfo {title} {{\it Simulation of
  Graphene Nanoribbon Field-Effect Transistors}},\ }\href
  {https://doi.org/https://doi.org/10.1109/LED.2007.901680} {\bibfield
  {journal} {\bibinfo  {journal} {IEEE Electron Device Lett.}\ }\textbf
  {\bibinfo {volume} {28}},\ \bibinfo {pages} {760} (\bibinfo {year}
  {2007})}\BibitemShut {NoStop}%
\bibitem [{\citenamefont {Ouyang}\ \emph {et~al.}(2008)\citenamefont {Ouyang},
  \citenamefont {Wang}, \citenamefont {Dai},\ and\ \citenamefont
  {Guo}}]{ouyang2008carrier}%
  \BibitemOpen
  \bibfield  {author} {\bibinfo {author} {\bibfnamefont {Y.}~\bibnamefont
  {Ouyang}}, \bibinfo {author} {\bibfnamefont {X.}~\bibnamefont {Wang}},
  \bibinfo {author} {\bibfnamefont {H.}~\bibnamefont {Dai}},\ and\ \bibinfo
  {author} {\bibfnamefont {J.}~\bibnamefont {Guo}},\ }\bibfield  {title}
  {\bibinfo {title} {{\it Carrier scattering in graphene nanoribbon
  field-effect transistors}},\ }\href
  {https://doi.org/https://doi.org/10.1063/1.2949749} {\bibfield  {journal}
  {\bibinfo  {journal} {Appl. Phys. Lett.}\ }\textbf {\bibinfo {volume} {92}},\
  \bibinfo {pages} {243124} (\bibinfo {year} {2008})}\BibitemShut {NoStop}%
\bibitem [{\citenamefont {Yoon}\ \emph {et~al.}(2008)\citenamefont {Yoon},
  \citenamefont {Fiori}, \citenamefont {Hong}, \citenamefont {Iannaccone},\
  and\ \citenamefont {Guo}}]{yoon2008performance}%
  \BibitemOpen
  \bibfield  {author} {\bibinfo {author} {\bibfnamefont {Y.}~\bibnamefont
  {Yoon}}, \bibinfo {author} {\bibfnamefont {G.}~\bibnamefont {Fiori}},
  \bibinfo {author} {\bibfnamefont {S.}~\bibnamefont {Hong}}, \bibinfo {author}
  {\bibfnamefont {G.}~\bibnamefont {Iannaccone}},\ and\ \bibinfo {author}
  {\bibfnamefont {J.}~\bibnamefont {Guo}},\ }\bibfield  {title} {\bibinfo
  {title} {{\it Performance comparison of graphene nanoribbon FETs with
  Schottky contacts and doped reservoirs}},\ }\href
  {https://doi.org/https://doi.org/10.1109/TED.2008.928021} {\bibfield
  {journal} {\bibinfo  {journal} {IEEE Trans. Electron Devices}\ }\textbf
  {\bibinfo {volume} {55}},\ \bibinfo {pages} {2314} (\bibinfo {year}
  {2008})}\BibitemShut {NoStop}%
\bibitem [{\citenamefont {Lu}\ and\ \citenamefont {Guo}(2010)}]{lu2010local}%
  \BibitemOpen
  \bibfield  {author} {\bibinfo {author} {\bibfnamefont {Y.}~\bibnamefont
  {Lu}}\ and\ \bibinfo {author} {\bibfnamefont {J.}~\bibnamefont {Guo}},\
  }\bibfield  {title} {\bibinfo {title} {{\it Local strain in tunneling
  transistors based on graphene nanoribbons}},\ }\href
  {https://doi.org/https://doi.org/10.1063/1.3479915} {\bibfield  {journal}
  {\bibinfo  {journal} {Appl. Phys. Lett.}\ }\textbf {\bibinfo {volume} {97}},\
  \bibinfo {pages} {073105} (\bibinfo {year} {2010})}\BibitemShut {NoStop}%
\bibitem [{\citenamefont {Yoon}\ \emph {et~al.}(2011)\citenamefont {Yoon},
  \citenamefont {Nikonov},\ and\ \citenamefont {Salahuddin}}]{yoon2011role}%
  \BibitemOpen
  \bibfield  {author} {\bibinfo {author} {\bibfnamefont {Y.}~\bibnamefont
  {Yoon}}, \bibinfo {author} {\bibfnamefont {D.~E.}\ \bibnamefont {Nikonov}},\
  and\ \bibinfo {author} {\bibfnamefont {S.}~\bibnamefont {Salahuddin}},\
  }\bibfield  {title} {\bibinfo {title} {{\it Role of phonon scattering in
  graphene nanoribbon transistors: Nonequilibrium Green’s function method
  with real space approach}},\ }\href
  {https://doi.org/https://doi.org/10.1063/1.3589365} {\bibfield  {journal}
  {\bibinfo  {journal} {Appl. Phys. Lett.}\ }\textbf {\bibinfo {volume} {98}},\
  \bibinfo {pages} {203503} (\bibinfo {year} {2011})}\BibitemShut {NoStop}%
\bibitem [{\citenamefont {Yoon}\ and\ \citenamefont
  {Salahuddin}(2012)}]{yoon2012dissipative}%
  \BibitemOpen
  \bibfield  {author} {\bibinfo {author} {\bibfnamefont {Y.}~\bibnamefont
  {Yoon}}\ and\ \bibinfo {author} {\bibfnamefont {S.}~\bibnamefont
  {Salahuddin}},\ }\bibfield  {title} {\bibinfo {title} {{\it Dissipative
  transport in rough edge graphene nanoribbon tunnel transistors}},\ }\href
  {https://doi.org/https://doi.org/10.1063/1.4772532} {\bibfield  {journal}
  {\bibinfo  {journal} {Appl. Phys. Lett.}\ }\textbf {\bibinfo {volume}
  {101}},\ \bibinfo {pages} {263501} (\bibinfo {year} {2012})}\BibitemShut
  {NoStop}%
\bibitem [{\citenamefont {Zhao}\ \emph {et~al.}(2009)\citenamefont {Zhao},
  \citenamefont {Chauhan},\ and\ \citenamefont {Guo}}]{zhao2009computational}%
  \BibitemOpen
  \bibfield  {author} {\bibinfo {author} {\bibfnamefont {P.}~\bibnamefont
  {Zhao}}, \bibinfo {author} {\bibfnamefont {J.}~\bibnamefont {Chauhan}},\ and\
  \bibinfo {author} {\bibfnamefont {J.}~\bibnamefont {Guo}},\ }\bibfield
  {title} {\bibinfo {title} {{\it Computational study of tunneling transistor
  based on graphene nanoribbon}},\ }\href
  {https://doi.org/https://doi.org/10.1021/nl803176x} {\bibfield  {journal}
  {\bibinfo  {journal} {Nano Lett.}\ }\textbf {\bibinfo {volume} {9}},\
  \bibinfo {pages} {684} (\bibinfo {year} {2009})}\BibitemShut {NoStop}%
\bibitem [{\citenamefont {Betti}\ \emph {et~al.}(2011)\citenamefont {Betti},
  \citenamefont {Fiori},\ and\ \citenamefont {Iannaccone}}]{betti2011strong}%
  \BibitemOpen
  \bibfield  {author} {\bibinfo {author} {\bibfnamefont {A.}~\bibnamefont
  {Betti}}, \bibinfo {author} {\bibfnamefont {G.}~\bibnamefont {Fiori}},\ and\
  \bibinfo {author} {\bibfnamefont {G.}~\bibnamefont {Iannaccone}},\ }\bibfield
   {title} {\bibinfo {title} {{\it Strong mobility degradation in ideal
  graphene nanoribbons due to phonon scattering}},\ }\href
  {https://doi.org/https://doi.org/10.1063/1.3587627} {\bibfield  {journal}
  {\bibinfo  {journal} {Appl. Phys. Lett.}\ }\textbf {\bibinfo {volume} {98}},\
  \bibinfo {pages} {212111} (\bibinfo {year} {2011})}\BibitemShut {NoStop}%
\bibitem [{\citenamefont {Grassi}\ \emph {et~al.}(2013)\citenamefont {Grassi},
  \citenamefont {Gnudi}, \citenamefont {Imperiale}, \citenamefont {Gnani},
  \citenamefont {Reggiani},\ and\ \citenamefont {Baccarani}}]{grassi2013mode}%
  \BibitemOpen
  \bibfield  {author} {\bibinfo {author} {\bibfnamefont {R.}~\bibnamefont
  {Grassi}}, \bibinfo {author} {\bibfnamefont {A.}~\bibnamefont {Gnudi}},
  \bibinfo {author} {\bibfnamefont {I.}~\bibnamefont {Imperiale}}, \bibinfo
  {author} {\bibfnamefont {E.}~\bibnamefont {Gnani}}, \bibinfo {author}
  {\bibfnamefont {S.}~\bibnamefont {Reggiani}},\ and\ \bibinfo {author}
  {\bibfnamefont {G.}~\bibnamefont {Baccarani}},\ }\bibfield  {title} {\bibinfo
  {title} {{\it Mode space approach for tight-binding transport simulations in
  graphene nanoribbon field-effect transistors including phonon scattering}},\
  }\href {https://doi.org/https://doi.org/10.1063/1.4800900} {\bibfield
  {journal} {\bibinfo  {journal} {J. Appl. Phys.}\ }\textbf {\bibinfo {volume}
  {113}},\ \bibinfo {pages} {144506} (\bibinfo {year} {2013})}\BibitemShut
  {NoStop}%
\bibitem [{\citenamefont {Fischetti}\ \emph {et~al.}(2013)\citenamefont
  {Fischetti}, \citenamefont {Kim}, \citenamefont {Narayanan}, \citenamefont
  {Ong}, \citenamefont {Sachs}, \citenamefont {Ferry},\ and\ \citenamefont
  {Aboud}}]{fischetti2013pseudopotential}%
  \BibitemOpen
  \bibfield  {author} {\bibinfo {author} {\bibfnamefont {M.~V.}\ \bibnamefont
  {Fischetti}}, \bibinfo {author} {\bibfnamefont {J.}~\bibnamefont {Kim}},
  \bibinfo {author} {\bibfnamefont {S.}~\bibnamefont {Narayanan}}, \bibinfo
  {author} {\bibfnamefont {Z.-Y.}\ \bibnamefont {Ong}}, \bibinfo {author}
  {\bibfnamefont {C.}~\bibnamefont {Sachs}}, \bibinfo {author} {\bibfnamefont
  {D.~K.}\ \bibnamefont {Ferry}},\ and\ \bibinfo {author} {\bibfnamefont
  {S.~J.}\ \bibnamefont {Aboud}},\ }\bibfield  {title} {\bibinfo {title} {{\it
  Pseudopotential-based studies of electron transport in graphene and graphene
  nanoribbons}},\ }\href
  {https://doi.org/https://doi.org/10.1088/0953-8984/25/47/473202} {\bibfield
  {journal} {\bibinfo  {journal} {J. Phy: Condens. Matter}\ }\textbf {\bibinfo
  {volume} {25}},\ \bibinfo {pages} {473202} (\bibinfo {year}
  {2013})}\BibitemShut {NoStop}%
\bibitem [{\citenamefont {Yousefvand}\ \emph {et~al.}(2017)\citenamefont
  {Yousefvand}, \citenamefont {Ahmadi},\ and\ \citenamefont
  {Meshginqalam}}]{yousefvand2017analytical}%
  \BibitemOpen
  \bibfield  {author} {\bibinfo {author} {\bibfnamefont {A.}~\bibnamefont
  {Yousefvand}}, \bibinfo {author} {\bibfnamefont {M.~T.}\ \bibnamefont
  {Ahmadi}},\ and\ \bibinfo {author} {\bibfnamefont {B.}~\bibnamefont
  {Meshginqalam}},\ }\bibfield  {title} {\bibinfo {title} {{\it Analytical
  Modeling of Acoustic Phonon-Limited Mobility in Strained Graphene
  Nanoribbons}},\ }\href
  {https://doi.org/https://doi.org/10.1007/s11664-017-5698-z} {\bibfield
  {journal} {\bibinfo  {journal} {J. Electron. Mater.}\ }\textbf {\bibinfo
  {volume} {46}},\ \bibinfo {pages} {6553} (\bibinfo {year}
  {2017})}\BibitemShut {NoStop}%
\bibitem [{\citenamefont {Rudan}\ \emph {et~al.}(2022)\citenamefont {Rudan},
  \citenamefont {Brunetti},\ and\ \citenamefont {Reggiani}}]{Springer_2022}%
  \BibitemOpen
  \bibfield  {author} {\bibinfo {author} {\bibfnamefont {M.}~\bibnamefont
  {Rudan}}, \bibinfo {author} {\bibfnamefont {R.}~\bibnamefont {Brunetti}},\
  and\ \bibinfo {author} {\bibfnamefont {E.}~\bibnamefont {Reggiani},
  \bibfnamefont {Susanna}},\ }\href
  {https://doi.org/https://doi.org/10.1007/978-3-030-79827-7} {\emph {\bibinfo
  {title} {{\it Springer Handbook of Semiconductor Devices}}}}\ (\bibinfo
  {publisher} {Springer Cham},\ \bibinfo {address} {Switzerland},\ \bibinfo
  {year} {2022})\BibitemShut {NoStop}%
\bibitem [{\citenamefont {Kadanoff}\ and\ \citenamefont
  {Baym}(1962)}]{Kadanoff_1962}%
  \BibitemOpen
  \bibfield  {author} {\bibinfo {author} {\bibfnamefont {L.~P.}\ \bibnamefont
  {Kadanoff}}\ and\ \bibinfo {author} {\bibfnamefont {G.}~\bibnamefont
  {Baym}},\ }\href {https://doi.org/https://doi.org/10.1201/9780429493218}
  {\emph {\bibinfo {title} {{\it Quantum Statistical Mechanics: Green's
  Function Methods in Equilibrium and Nonequilibrium Problems}}}}\ (\bibinfo
  {publisher} {Benjamin},\ \bibinfo {address} {New York},\ \bibinfo {year}
  {1962})\BibitemShut {NoStop}%
\bibitem [{\citenamefont {Keldysh}(1965)}]{Keldysh_1965}%
  \BibitemOpen
  \bibfield  {author} {\bibinfo {author} {\bibfnamefont {L.~V.}\ \bibnamefont
  {Keldysh}},\ }\bibfield  {title} {\bibinfo {title} {{\it Diagram Technique
  for Nonequilibrium Processes}},\ }\href@noop {} {\bibfield  {journal}
  {\bibinfo  {journal} {Sov. Phys. JETP}\ }\textbf {\bibinfo {volume} {20}},\
  \bibinfo {pages} {1018} (\bibinfo {year} {1965})},\ \bibinfo {note} {[Zh.
  Eksp. Theor. Fiz.\ {\bf 47}, 1515 (1964)]}\BibitemShut {NoStop}%
\bibitem [{\citenamefont {Lake}\ \emph {et~al.}(1997)\citenamefont {Lake},
  \citenamefont {Klimeck}, \citenamefont {Bowen},\ and\ \citenamefont
  {Jovanovic}}]{Lake_1997}%
  \BibitemOpen
  \bibfield  {author} {\bibinfo {author} {\bibfnamefont {R.}~\bibnamefont
  {Lake}}, \bibinfo {author} {\bibfnamefont {G.}~\bibnamefont {Klimeck}},
  \bibinfo {author} {\bibfnamefont {R.~C.}\ \bibnamefont {Bowen}},\ and\
  \bibinfo {author} {\bibfnamefont {D.}~\bibnamefont {Jovanovic}},\ }\bibfield
  {title} {\bibinfo {title} {{\it Single and multiband modeling of quantum
  electron transport through layered semiconductor devices}},\ }\href
  {https://doi.org/https://doi.org/10.1063/1.365394} {\bibfield  {journal}
  {\bibinfo  {journal} {J. Appl. Phys.}\ }\textbf {\bibinfo {volume} {81}},\
  \bibinfo {pages} {7845} (\bibinfo {year} {1997})}\BibitemShut {NoStop}%
\bibitem [{\citenamefont {Backman}\ \emph {et~al.}(2023)\citenamefont
  {Backman}, \citenamefont {Lee},\ and\ \citenamefont
  {Luisier}}]{Luisier_2023}%
  \BibitemOpen
  \bibfield  {author} {\bibinfo {author} {\bibfnamefont {J.}~\bibnamefont
  {Backman}}, \bibinfo {author} {\bibfnamefont {Y.}~\bibnamefont {Lee}},\ and\
  \bibinfo {author} {\bibfnamefont {M.}~\bibnamefont {Luisier}},\ }\bibfield
  {title} {\bibinfo {title} {{\it Phonon-Limited Transport in 2D Materials: A
  Unified Approach for ab initio Mobility and Current Calculations}},\ }\href
  {https://doi.org/https://doi.org/10.48550/arXiv.2312.00577} {\bibfield
  {journal} {\bibinfo  {journal} {arXiv:2312.00577 [cond-mat.mes-hall]}\ }
  (\bibinfo {year} {2023})}\BibitemShut {NoStop}%
\bibitem [{\citenamefont {Grassi}\ \emph {et~al.}(2009)\citenamefont {Grassi},
  \citenamefont {Gnudi}, \citenamefont {Gnani}, \citenamefont {Reggiani},\ and\
  \citenamefont {Baccarani}}]{grassi2009investigation}%
  \BibitemOpen
  \bibfield  {author} {\bibinfo {author} {\bibfnamefont {R.}~\bibnamefont
  {Grassi}}, \bibinfo {author} {\bibfnamefont {A.}~\bibnamefont {Gnudi}},
  \bibinfo {author} {\bibfnamefont {E.}~\bibnamefont {Gnani}}, \bibinfo
  {author} {\bibfnamefont {S.}~\bibnamefont {Reggiani}},\ and\ \bibinfo
  {author} {\bibfnamefont {G.}~\bibnamefont {Baccarani}},\ }\bibfield  {title}
  {\bibinfo {title} {{\it An investigation of performance limits of
  conventional and tunneling graphene-based transistors}},\ }\href
  {https://doi.org/https://doi.org/10.1007/s10825-009-0282-2} {\bibfield
  {journal} {\bibinfo  {journal} {J. comput. Electron.}\ }\textbf {\bibinfo
  {volume} {8}},\ \bibinfo {pages} {441} (\bibinfo {year} {2009})}\BibitemShut
  {NoStop}%
\bibitem [{\citenamefont {Mante}\ \emph {et~al.}(2018)\citenamefont {Mante},
  \citenamefont {Belliard},\ and\ \citenamefont {Perrin}}]{Mante_2018}%
  \BibitemOpen
  \bibfield  {author} {\bibinfo {author} {\bibfnamefont {P.-A.}\ \bibnamefont
  {Mante}}, \bibinfo {author} {\bibfnamefont {L.}~\bibnamefont {Belliard}},\
  and\ \bibinfo {author} {\bibfnamefont {B.}~\bibnamefont {Perrin}},\
  }\bibfield  {title} {\bibinfo {title} {{\it Acoustic phonons in nanowires
  probed by ultrafast pump-probe spectroscopy}},\ }\href
  {https://doi.org/https://doi.org/10.1515/nanoph-2018-0069} {\bibfield
  {journal} {\bibinfo  {journal} {Nanophotonics}\ }\textbf {\bibinfo {volume}
  {7}},\ \bibinfo {pages} {1759} (\bibinfo {year} {2018})}\BibitemShut
  {NoStop}%
\bibitem [{\citenamefont {Sotomayor~Torres}\ \emph {et~al.}(2004)\citenamefont
  {Sotomayor~Torres}, \citenamefont {Zwick}, \citenamefont {Poinsotte},
  \citenamefont {Groenen}, \citenamefont {Prunnila}, \citenamefont {Ahopelto},
  \citenamefont {Mlayah},\ and\ \citenamefont
  {Paillard}}]{torres2004observations}%
  \BibitemOpen
  \bibfield  {author} {\bibinfo {author} {\bibfnamefont {C.~M.}\ \bibnamefont
  {Sotomayor~Torres}}, \bibinfo {author} {\bibfnamefont {A.}~\bibnamefont
  {Zwick}}, \bibinfo {author} {\bibfnamefont {F.}~\bibnamefont {Poinsotte}},
  \bibinfo {author} {\bibfnamefont {J.}~\bibnamefont {Groenen}}, \bibinfo
  {author} {\bibfnamefont {M.}~\bibnamefont {Prunnila}}, \bibinfo {author}
  {\bibfnamefont {J.}~\bibnamefont {Ahopelto}}, \bibinfo {author}
  {\bibfnamefont {A.}~\bibnamefont {Mlayah}},\ and\ \bibinfo {author}
  {\bibfnamefont {V.}~\bibnamefont {Paillard}},\ }\bibfield  {title} {\bibinfo
  {title} {{\it Observations of confined acoustic phonons in silicon
  membranes}},\ }\href {https://doi.org/https://doi.org/10.1002/pssc.200405313}
  {\bibfield  {journal} {\bibinfo  {journal} {Phys. Status Solidi (c)}\
  }\textbf {\bibinfo {volume} {1}},\ \bibinfo {pages} {2609} (\bibinfo {year}
  {2004})}\BibitemShut {NoStop}%
\bibitem [{\citenamefont {Nishiguchi}\ \emph {et~al.}(1997)\citenamefont
  {Nishiguchi}, \citenamefont {Ando},\ and\ \citenamefont
  {Wybourne}}]{Nishiguchi_1997}%
  \BibitemOpen
  \bibfield  {author} {\bibinfo {author} {\bibfnamefont {N.}~\bibnamefont
  {Nishiguchi}}, \bibinfo {author} {\bibfnamefont {Y.}~\bibnamefont {Ando}},\
  and\ \bibinfo {author} {\bibfnamefont {N.~M.}\ \bibnamefont {Wybourne}},\
  }\bibfield  {title} {\bibinfo {title} {{\it Acoustic phonon modes of
  rectangular quantum wires}},\ }\href
  {https://doi.org/https://doi.org/10.1088/0953-8984/9/27/007} {\bibfield
  {journal} {\bibinfo  {journal} {J. Phys.: Condens. Matter}\ }\textbf
  {\bibinfo {volume} {9}},\ \bibinfo {pages} {5751} (\bibinfo {year}
  {1997})}\BibitemShut {NoStop}%
\bibitem [{\citenamefont {Donetti}\ \emph {et~al.}(2006)\citenamefont
  {Donetti}, \citenamefont {G{\'{a}}miz}, \citenamefont {Rold{\'{a}}n},\ and\
  \citenamefont {Godoy}}]{Donetti_2006}%
  \BibitemOpen
  \bibfield  {author} {\bibinfo {author} {\bibfnamefont {L.}~\bibnamefont
  {Donetti}}, \bibinfo {author} {\bibfnamefont {F.}~\bibnamefont
  {G{\'{a}}miz}}, \bibinfo {author} {\bibfnamefont {J.~B.}\ \bibnamefont
  {Rold{\'{a}}n}},\ and\ \bibinfo {author} {\bibfnamefont {A.}~\bibnamefont
  {Godoy}},\ }\bibfield  {title} {\bibinfo {title} {{\it Acoustic phonon
  confinement in silicon nanolayers: Effect on electron mobility}},\ }\href
  {https://doi.org/https://doi.org/10.1063/1.2208849} {\bibfield  {journal}
  {\bibinfo  {journal} {J. Appl. Phys.}\ }\textbf {\bibinfo {volume} {100}},\
  \bibinfo {pages} {013701} (\bibinfo {year} {2006})}\BibitemShut {NoStop}%
\bibitem [{\citenamefont {Ramayya}\ \emph {et~al.}(2008)\citenamefont
  {Ramayya}, \citenamefont {Vasileska}, \citenamefont {Goodnick},\ and\
  \citenamefont {Knezevic}}]{Ramayya_2008}%
  \BibitemOpen
  \bibfield  {author} {\bibinfo {author} {\bibfnamefont {E.~B.}\ \bibnamefont
  {Ramayya}}, \bibinfo {author} {\bibfnamefont {D.}~\bibnamefont {Vasileska}},
  \bibinfo {author} {\bibfnamefont {S.~M.}\ \bibnamefont {Goodnick}},\ and\
  \bibinfo {author} {\bibfnamefont {I.}~\bibnamefont {Knezevic}},\ }\bibfield
  {title} {\bibinfo {title} {{\it Electron transport in silicon nanowires: The
  role of acoustic phonon confinement and surface roughness scattering}},\
  }\href {https://doi.org/https://doi.org/10.1063/1.2977758} {\bibfield
  {journal} {\bibinfo  {journal} {J. Appl. Phys.}\ }\textbf {\bibinfo {volume}
  {104}},\ \bibinfo {pages} {063711} (\bibinfo {year} {2008})}\BibitemShut
  {NoStop}%
\bibitem [{\citenamefont {Tienda-Luna}\ \emph {et~al.}(2013)\citenamefont
  {Tienda-Luna}, \citenamefont {Ruiz}, \citenamefont {Godoy}, \citenamefont
  {Donetti}, \citenamefont {Martinez-Blanque},\ and\ \citenamefont
  {G\'{a}miz}}]{tienda2013effect}%
  \BibitemOpen
  \bibfield  {author} {\bibinfo {author} {\bibfnamefont {I.}~\bibnamefont
  {Tienda-Luna}}, \bibinfo {author} {\bibfnamefont {F.}~\bibnamefont {Ruiz}},
  \bibinfo {author} {\bibfnamefont {A.}~\bibnamefont {Godoy}}, \bibinfo
  {author} {\bibfnamefont {L.}~\bibnamefont {Donetti}}, \bibinfo {author}
  {\bibfnamefont {C.}~\bibnamefont {Martinez-Blanque}},\ and\ \bibinfo {author}
  {\bibfnamefont {F.}~\bibnamefont {G\'{a}miz}},\ }\bibfield  {title} {\bibinfo
  {title} {{\it Effect of confined acoustic phonons on the electron mobility of
  rectangular nanowires}},\ }\href
  {https://doi.org/https://doi.org/10.1063/1.4825210} {\bibfield  {journal}
  {\bibinfo  {journal} {Appl. Phys. Lett.}\ }\textbf {\bibinfo {volume}
  {103}},\ \bibinfo {pages} {163107} (\bibinfo {year} {2013})}\BibitemShut
  {NoStop}%
\bibitem [{\citenamefont {Mickevi\v{c}ius}\ and\ \citenamefont
  {Mitin}(1993)}]{mickevivcius1993acoustic}%
  \BibitemOpen
  \bibfield  {author} {\bibinfo {author} {\bibfnamefont {R.}~\bibnamefont
  {Mickevi\v{c}ius}}\ and\ \bibinfo {author} {\bibfnamefont {V.}~\bibnamefont
  {Mitin}},\ }\bibfield  {title} {\bibinfo {title} {{\it Acoustic-phonon
  scattering in a rectangular quantum wire}},\ }\href
  {https://doi.org/https://doi.org/10.1103/PhysRevB.48.17194} {\bibfield
  {journal} {\bibinfo  {journal} {Phy. Rev. B}\ }\textbf {\bibinfo {volume}
  {48}},\ \bibinfo {pages} {17194} (\bibinfo {year} {1993})}\BibitemShut
  {NoStop}%
\bibitem [{\citenamefont {Bannov}\ \emph {et~al.}(1994)\citenamefont {Bannov},
  \citenamefont {Mitin},\ and\ \citenamefont {Stroscio}}]{Bannov_1994}%
  \BibitemOpen
  \bibfield  {author} {\bibinfo {author} {\bibfnamefont {N.}~\bibnamefont
  {Bannov}}, \bibinfo {author} {\bibfnamefont {V.}~\bibnamefont {Mitin}},\ and\
  \bibinfo {author} {\bibfnamefont {M.}~\bibnamefont {Stroscio}},\ }\bibfield
  {title} {\bibinfo {title} {{\it Confined acoustic phonons in a free-standing
  quantum well and their interaction with electrons}},\ }\href
  {https://doi.org/https://doi.org/10.1063/1.2433149} {\bibfield  {journal}
  {\bibinfo  {journal} {Phys. Stat. Sol. (b)}\ }\textbf {\bibinfo {volume}
  {183}},\ \bibinfo {pages} {131} (\bibinfo {year} {1994})}\BibitemShut
  {NoStop}%
\bibitem [{\citenamefont {Ridley}\ \emph {et~al.}(2000)\citenamefont {Ridley},
  \citenamefont {Zakhleniuk},\ and\ \citenamefont {Babiker}}]{Ridley_2000}%
  \BibitemOpen
  \bibfield  {author} {\bibinfo {author} {\bibfnamefont {B.~K.}\ \bibnamefont
  {Ridley}}, \bibinfo {author} {\bibfnamefont {N.}~\bibnamefont {Zakhleniuk}},\
  and\ \bibinfo {author} {\bibfnamefont {M.}~\bibnamefont {Babiker}},\
  }\bibfield  {title} {\bibinfo {title} {{\it Continuum model for acoustic and
  optical phonons in heterostructure}},\ }\href
  {https://doi.org/https://doi.org/10.1016/S0038-1098(00)00335-5} {\bibfield
  {journal} {\bibinfo  {journal} {Solid State Commun.}\ }\textbf {\bibinfo
  {volume} {116}},\ \bibinfo {pages} {385} (\bibinfo {year}
  {2000})}\BibitemShut {NoStop}%
\bibitem [{\citenamefont {P\'{e}rez-Alvarez}\ and\ \citenamefont
  {Trallero-Giner}(1997)}]{Perez_1997}%
  \BibitemOpen
  \bibfield  {author} {\bibinfo {author} {\bibfnamefont {R.}~\bibnamefont
  {P\'{e}rez-Alvarez}}\ and\ \bibinfo {author} {\bibfnamefont {C.}~\bibnamefont
  {Trallero-Giner}},\ }\bibfield  {title} {\bibinfo {title} {{\it Planar
  vibrational modes in semiconductors: A simple model}},\ }\href
  {https://doi.org/https://doi.org/10.1088/0031-8949/56/4/013} {\bibfield
  {journal} {\bibinfo  {journal} {Phys. Scr.}\ }\textbf {\bibinfo {volume}
  {56}},\ \bibinfo {pages} {407} (\bibinfo {year} {1997})}\BibitemShut
  {NoStop}%
\bibitem [{\citenamefont {Karamitaheri}\ \emph {et~al.}(2013)\citenamefont
  {Karamitaheri}, \citenamefont {Neophytou}, \citenamefont {Karami~Taheri},
  \citenamefont {Faez},\ and\ \citenamefont {Kosina}}]{Karamitaheri_2013}%
  \BibitemOpen
  \bibfield  {author} {\bibinfo {author} {\bibfnamefont {H.}~\bibnamefont
  {Karamitaheri}}, \bibinfo {author} {\bibfnamefont {N.}~\bibnamefont
  {Neophytou}}, \bibinfo {author} {\bibfnamefont {M.}~\bibnamefont
  {Karami~Taheri}}, \bibinfo {author} {\bibfnamefont {R.}~\bibnamefont
  {Faez}},\ and\ \bibinfo {author} {\bibfnamefont {H.}~\bibnamefont {Kosina}},\
  }\bibfield  {title} {\bibinfo {title} {{\it Calculation of Confined Phonon
  Spectrum in Narrow Silicon Nanowires Using the Valence Force Field
  Methode}},\ }\href
  {https://doi.org/https://doi.org/10.1007/s11664-013-2533-z} {\bibfield
  {journal} {\bibinfo  {journal} {J. Electron. Mater.}\ }\textbf {\bibinfo
  {volume} {42}},\ \bibinfo {pages} {2091} (\bibinfo {year}
  {2013})}\BibitemShut {NoStop}%
\bibitem [{\citenamefont {Kane}(1985)}]{Kane_1985}%
  \BibitemOpen
  \bibfield  {author} {\bibinfo {author} {\bibfnamefont {E.~O.}\ \bibnamefont
  {Kane}},\ }\bibfield  {title} {\bibinfo {title} {{\it Phonon spectra of
  diamond and zinc-blende semiconductors}},\ }\href
  {https://doi.org/https://doi.org/10.1103/PhysRevB.31.7865} {\bibfield
  {journal} {\bibinfo  {journal} {Phys. Rev. B}\ }\textbf {\bibinfo {volume}
  {31}},\ \bibinfo {pages} {7865} (\bibinfo {year} {1985})}\BibitemShut
  {NoStop}%
\bibitem [{\citenamefont {Paul}\ \emph {et~al.}(2010)\citenamefont {Paul},
  \citenamefont {Luisier},\ and\ \citenamefont {Klimeck}}]{Paul_2010}%
  \BibitemOpen
  \bibfield  {author} {\bibinfo {author} {\bibfnamefont {A.}~\bibnamefont
  {Paul}}, \bibinfo {author} {\bibfnamefont {M.}~\bibnamefont {Luisier}},\ and\
  \bibinfo {author} {\bibfnamefont {G.}~\bibnamefont {Klimeck}},\ }\bibfield
  {title} {\bibinfo {title} {{\it Modified valence force field approach for
  phonon dispersion: from zinc-blende bulk to nanowires - Methodology and
  computational details}},\ }\href
  {https://doi.org/https://doi.org/10.1007/s10825-010-0332-9} {\bibfield
  {journal} {\bibinfo  {journal} {J. Comput. Electron.}\ }\textbf {\bibinfo
  {volume} {9}},\ \bibinfo {pages} {160} (\bibinfo {year} {2010})}\BibitemShut
  {NoStop}%
\bibitem [{\citenamefont {Ye}\ \emph {et~al.}(2015)\citenamefont {Ye},
  \citenamefont {Cao}, \citenamefont {Yao}, \citenamefont {Feng},\ and\
  \citenamefont {Ruan}}]{Ye_2015a}%
  \BibitemOpen
  \bibfield  {author} {\bibinfo {author} {\bibfnamefont {Z.-Q.}\ \bibnamefont
  {Ye}}, \bibinfo {author} {\bibfnamefont {B.-Y.}\ \bibnamefont {Cao}},
  \bibinfo {author} {\bibfnamefont {W.-J.}\ \bibnamefont {Yao}}, \bibinfo
  {author} {\bibfnamefont {T.}~\bibnamefont {Feng}},\ and\ \bibinfo {author}
  {\bibfnamefont {X.~R.}\ \bibnamefont {Ruan}},\ }\bibfield  {title} {\bibinfo
  {title} {{\it Spectral phonon thermal properties in graphene nanoribbons}},\
  }\href {https://doi.org/https://doi.org/10.1016/j.carbon.2015.06.008}
  {\bibfield  {journal} {\bibinfo  {journal} {Carbon}\ }\textbf {\bibinfo
  {volume} {93}},\ \bibinfo {pages} {915} (\bibinfo {year} {2015})}\BibitemShut
  {NoStop}%
\bibitem [{\citenamefont {Jannatul~Islam}\ \emph {et~al.}(2017)\citenamefont
  {Jannatul~Islam}, \citenamefont {Islam}, \citenamefont {Islam},\ and\
  \citenamefont {Bhuiyan}}]{Islam_2017}%
  \BibitemOpen
  \bibfield  {author} {\bibinfo {author} {\bibfnamefont {A.~S.~M.}\
  \bibnamefont {Jannatul~Islam}}, \bibinfo {author} {\bibfnamefont {M.~R.}\
  \bibnamefont {Islam}}, \bibinfo {author} {\bibfnamefont {M.~S.}\ \bibnamefont
  {Islam}},\ and\ \bibinfo {author} {\bibfnamefont {A.~G.}\ \bibnamefont
  {Bhuiyan}},\ }\bibfield  {title} {\bibinfo {title} {{\it Numerical simulation
  of vibrational properties of AGNR with vacancy and stone wales defects}},\
  }in\ \href {https://doi.org/https://doi.org/10.1109/EICT.2017.8275235} {\emph
  {\bibinfo {booktitle} {2017 3rd International Conference on Electrical
  Information and Communication Technology (EICT)}}}\ (\bibinfo {year} {2017})\
  pp.\ \bibinfo {pages} {1--4}\BibitemShut {NoStop}%
\bibitem [{\citenamefont {Gillen}\ \emph {et~al.}(2009)\citenamefont {Gillen},
  \citenamefont {Mohr}, \citenamefont {Maultzsch},\ and\ \citenamefont
  {Thomsen}}]{Gillen_2009}%
  \BibitemOpen
  \bibfield  {author} {\bibinfo {author} {\bibfnamefont {R.}~\bibnamefont
  {Gillen}}, \bibinfo {author} {\bibfnamefont {M.}~\bibnamefont {Mohr}},
  \bibinfo {author} {\bibfnamefont {J.}~\bibnamefont {Maultzsch}},\ and\
  \bibinfo {author} {\bibfnamefont {C.}~\bibnamefont {Thomsen}},\ }\bibfield
  {title} {\bibinfo {title} {{\it Lattice vibrations in graphene nanoribbons
  from density functional theory}},\ }\href
  {https://doi.org/https://doi.org/10.1002/pssb.200982343} {\bibfield
  {journal} {\bibinfo  {journal} {Phys. Status Solidi (b)}\ }\textbf {\bibinfo
  {volume} {246}},\ \bibinfo {pages} {2577} (\bibinfo {year}
  {2009})}\BibitemShut {NoStop}%
\bibitem [{\citenamefont {Zhang}\ \emph {et~al.}(2014)\citenamefont {Zhang},
  \citenamefont {Heid}, \citenamefont {Bohnen}, \citenamefont {Sheng},\ and\
  \citenamefont {Chan}}]{Zhang_2014}%
  \BibitemOpen
  \bibfield  {author} {\bibinfo {author} {\bibfnamefont {T.}~\bibnamefont
  {Zhang}}, \bibinfo {author} {\bibfnamefont {R.}~\bibnamefont {Heid}},
  \bibinfo {author} {\bibfnamefont {K.-P.}\ \bibnamefont {Bohnen}}, \bibinfo
  {author} {\bibfnamefont {P.}~\bibnamefont {Sheng}},\ and\ \bibinfo {author}
  {\bibfnamefont {C.~T.}\ \bibnamefont {Chan}},\ }\bibfield  {title} {\bibinfo
  {title} {{\it Phonon spectrum and electron-phonon coupling in zigzag graphene
  nanoribbons}},\ }\href
  {https://doi.org/https://doi.org/10.1103/PhysRevB.89.205404} {\bibfield
  {journal} {\bibinfo  {journal} {Phys. Rev. B}\ }\textbf {\bibinfo {volume}
  {89}},\ \bibinfo {pages} {205404} (\bibinfo {year} {2014})}\BibitemShut
  {NoStop}%
\bibitem [{\citenamefont {Chamberlain}\ \emph {et~al.}(1993)\citenamefont
  {Chamberlain}, \citenamefont {Cardona},\ and\ \citenamefont
  {Ridley}}]{Chamberlain_1993}%
  \BibitemOpen
  \bibfield  {author} {\bibinfo {author} {\bibfnamefont {M.}~\bibnamefont
  {Chamberlain}}, \bibinfo {author} {\bibfnamefont {M.}~\bibnamefont
  {Cardona}},\ and\ \bibinfo {author} {\bibfnamefont {B.}~\bibnamefont
  {Ridley}},\ }\bibfield  {title} {\bibinfo {title} {Optical modes in gaas/alas
  superlattices},\ }\href
  {https://doi.org/https://doi.org/10.1103/PhysRevB.48.14356} {\bibfield
  {journal} {\bibinfo  {journal} {Phys. Rev. B}\ }\textbf {\bibinfo {volume}
  {48}},\ \bibinfo {pages} {14356} (\bibinfo {year} {1993})}\BibitemShut
  {NoStop}%
\bibitem [{\citenamefont {Gao}\ \emph {et~al.}(2008)\citenamefont {Gao},
  \citenamefont {Botez},\ and\ \citenamefont {Knezevic}}]{Xao_2008}%
  \BibitemOpen
  \bibfield  {author} {\bibinfo {author} {\bibfnamefont {X.}~\bibnamefont
  {Gao}}, \bibinfo {author} {\bibfnamefont {D.}~\bibnamefont {Botez}},\ and\
  \bibinfo {author} {\bibfnamefont {I.}~\bibnamefont {Knezevic}},\ }\bibfield
  {title} {\bibinfo {title} {{\it Phonon confinement and electron transport in
  GaAs-based quantum cascade structures}},\ }\href
  {https://doi.org/https://doi.org/10.1063/1.2899963} {\bibfield  {journal}
  {\bibinfo  {journal} {J. Appl. Phys.}\ }\textbf {\bibinfo {volume} {103}},\
  \bibinfo {pages} {073101} (\bibinfo {year} {2008})}\BibitemShut {NoStop}%
\bibitem [{\citenamefont {de~Le\'{o}n-P\'{e}rez}\ and\ \citenamefont
  {P\'{e}rez-Alvarez}(2000)}]{Perez_2000}%
  \BibitemOpen
  \bibfield  {author} {\bibinfo {author} {\bibfnamefont {F.}~\bibnamefont
  {de~Le\'{o}n-P\'{e}rez}}\ and\ \bibinfo {author} {\bibfnamefont
  {R.}~\bibnamefont {P\'{e}rez-Alvarez}},\ }\bibfield  {title} {\bibinfo
  {title} {{\it Long-wavelength nonpolar optical modes in semiconductor
  heterostructures: Continuum phenomenological model}},\ }\href
  {https://doi.org/https://doi.org/10.1103/PhysRevB.61.4820} {\bibfield
  {journal} {\bibinfo  {journal} {Phys. Rev. B}\ }\textbf {\bibinfo {volume}
  {61}},\ \bibinfo {pages} {4829} (\bibinfo {year} {2000})}\BibitemShut
  {NoStop}%
\bibitem [{\citenamefont {Comas}\ \emph {et~al.}(2007)\citenamefont {Comas},
  \citenamefont {Camps}, \citenamefont {Marques},\ and\ \citenamefont
  {Studart}}]{Comas_2007}%
  \BibitemOpen
  \bibfield  {author} {\bibinfo {author} {\bibfnamefont {F.}~\bibnamefont
  {Comas}}, \bibinfo {author} {\bibfnamefont {I.}~\bibnamefont {Camps}},
  \bibinfo {author} {\bibfnamefont {G.~E.}\ \bibnamefont {Marques}},\ and\
  \bibinfo {author} {\bibfnamefont {N.}~\bibnamefont {Studart}},\ }\bibfield
  {title} {\bibinfo {title} {{\it Dispersion of confined optical phonons in
  semiconductor nanowires in the framework of a continuum approach}},\ }\href
  {https://doi.org/https://doi.org/10.1063/1.2433149} {\bibfield  {journal}
  {\bibinfo  {journal} {J. Appl. Phys.}\ }\textbf {\bibinfo {volume} {101}},\
  \bibinfo {pages} {033525} (\bibinfo {year} {2007})}\BibitemShut {NoStop}%
\bibitem [{\citenamefont {Fischetti}(1998)}]{Fischetti98jap}%
  \BibitemOpen
  \bibfield  {author} {\bibinfo {author} {\bibfnamefont {M.~V.}\ \bibnamefont
  {Fischetti}},\ }\bibfield  {title} {\bibinfo {title} {{\it Theory of electron
  transport in small semiconductor devices using the Pauli master equation}},\
  }\href {https://doi.org/https://doi.org/10.1063/1.367149} {\bibfield
  {journal} {\bibinfo  {journal} {J. Appl. Phys.}\ }\textbf {\bibinfo {volume}
  {83}},\ \bibinfo {pages} {270} (\bibinfo {year} {1998})}\BibitemShut
  {NoStop}%
\bibitem [{\citenamefont {Fischetti}(1999)}]{Fischetti99prb}%
  \BibitemOpen
  \bibfield  {author} {\bibinfo {author} {\bibfnamefont {M.~V.}\ \bibnamefont
  {Fischetti}},\ }\bibfield  {title} {\bibinfo {title} {{\it Master-equation
  approach to the study of electronic transport in small semiconductor
  devices}},\ }\href {https://doi.org/https://doi.org/10.1103/PhysRevB.59.4901}
  {\bibfield  {journal} {\bibinfo  {journal} {Phys. Rev. B}\ }\textbf {\bibinfo
  {volume} {59}},\ \bibinfo {pages} {4901} (\bibinfo {year}
  {1999})}\BibitemShut {NoStop}%
\bibitem [{\citenamefont {Van~de Put}\ \emph {et~al.}(2019)\citenamefont
  {Van~de Put}, \citenamefont {Fischetti},\ and\ \citenamefont
  {Vandenberghe}}]{Vandeput19computer}%
  \BibitemOpen
  \bibfield  {author} {\bibinfo {author} {\bibfnamefont {M.~L.}\ \bibnamefont
  {Van~de Put}}, \bibinfo {author} {\bibfnamefont {M.~V.}\ \bibnamefont
  {Fischetti}},\ and\ \bibinfo {author} {\bibfnamefont {W.~G.}\ \bibnamefont
  {Vandenberghe}},\ }\bibfield  {title} {\bibinfo {title} {{\it Scalable
  atomistic simulations of quantum electron transport using empirical
  pseudopotentials}},\ }\href
  {https://doi.org/https://doi.org/10.1016/j.cpc.2019.06.009} {\bibfield
  {journal} {\bibinfo  {journal} {Comput. Phys. Commun.}\ }\textbf {\bibinfo
  {volume} {244}},\ \bibinfo {pages} {156} (\bibinfo {year}
  {2019})}\BibitemShut {NoStop}%
\bibitem [{\citenamefont {van Hove}(1954)}]{vanHove_1954}%
  \BibitemOpen
  \bibfield  {author} {\bibinfo {author} {\bibfnamefont {L.}~\bibnamefont {van
  Hove}},\ }\bibfield  {title} {\bibinfo {title} {{\it Quantum-mechanical
  perturbations giving rise to a statistical transport equation}},\ }\href
  {https://doi.org/https://doi.org/10.1016/S0031-8914(54)92646-4} {\bibfield
  {journal} {\bibinfo  {journal} {Physica}\ }\textbf {\bibinfo {volume} {21}},\
  \bibinfo {pages} {517} (\bibinfo {year} {1954})}\BibitemShut {NoStop}%
\bibitem [{\citenamefont {Zwanwig}(1964)}]{Zwanzig_1964}%
  \BibitemOpen
  \bibfield  {author} {\bibinfo {author} {\bibfnamefont {R.}~\bibnamefont
  {Zwanwig}},\ }\bibfield  {title} {\bibinfo {title} {{\it On the identity of
  three generalized Master equations}},\ }\href
  {https://doi.org/https://doi.org/10.1016/0031-8914(64)90102-8} {\bibfield
  {journal} {\bibinfo  {journal} {Physica}\ }\textbf {\bibinfo {volume} {30}},\
  \bibinfo {pages} {1109} (\bibinfo {year} {1964})}\BibitemShut {NoStop}%
\bibitem [{\citenamefont {von Neumann}(1927)}]{vonNeumann_1927}%
  \BibitemOpen
  \bibfield  {author} {\bibinfo {author} {\bibfnamefont {J.}~\bibnamefont {von
  Neumann}},\ }\bibfield  {title} {\bibinfo {title} {{\it Nachrichten von der
  Gesellschaft der Wissenschaften zu G\"{o}ttingen, Mathematisch-Physikalische
  Klasse}},\ }\href
  {http://resolver.sub.uni-goettingen.de/purl?PPN252457811_1927} {\bibfield
  {journal} {\bibinfo  {journal} {G\"{o}ttinger Nachrichten}\ }\textbf
  {\bibinfo {volume} {1927}},\ \bibinfo {pages} {245} (\bibinfo {year}
  {1927})}\BibitemShut {NoStop}%
\bibitem [{\citenamefont {Landau}(1965)}]{Landau_1965}%
  \BibitemOpen
  \bibfield  {author} {\bibinfo {author} {\bibfnamefont {L.~D.}\ \bibnamefont
  {Landau}},\ }\bibfield  {title} {\bibinfo {title} {{\it The Damping Problem
  in Wave Mechanics}},\ }in\ \href
  {https://doi.org/https://doi.org/10.1016/B978-0-08-010586-4.50007-9} {\emph
  {\bibinfo {booktitle} {Collected Papers of L.D. Landau}}},\ \bibinfo {editor}
  {edited by\ \bibinfo {editor} {\bibfnamefont {D.}~\bibnamefont {Ter~Haar}}}\
  (\bibinfo  {publisher} {Pergamon},\ \bibinfo {year} {1965})\ pp.\ \bibinfo
  {pages} {8--18}\BibitemShut {NoStop}%
\bibitem [{\citenamefont {Fischetti}\ \emph {et~al.}(2011)\citenamefont
  {Fischetti}, \citenamefont {Fu}, \citenamefont {Narayanan},\ and\
  \citenamefont {Kim}}]{Fischetti_2011}%
  \BibitemOpen
  \bibfield  {author} {\bibinfo {author} {\bibfnamefont {M.~V.}\ \bibnamefont
  {Fischetti}}, \bibinfo {author} {\bibfnamefont {B.}~\bibnamefont {Fu}},
  \bibinfo {author} {\bibfnamefont {S.}~\bibnamefont {Narayanan}},\ and\
  \bibinfo {author} {\bibfnamefont {J.}~\bibnamefont {Kim}},\ }\bibinfo {title}
  {{\it Semiclassical and Quantum Electronic Transport in Nanometer-Scale
  Structures: Empirical Pseudopotential Band Structure, Monte Carlo Simulations
  and Pauli Master Equation}},\ in\ \href
  {https://doi.org/https://doi.org/10.1016/10.1007/978-1-4419-8840-9_3} {\emph
  {\bibinfo {booktitle} {Nano-Electronic Devices: Semiclassical and Quantum
  Transport Modeling}}},\ \bibinfo {editor} {edited by\ \bibinfo {editor}
  {\bibfnamefont {D.}~\bibnamefont {Vasileska}}\ and\ \bibinfo {editor}
  {\bibfnamefont {S.~M.}\ \bibnamefont {Goodnick}}}\ (\bibinfo  {publisher}
  {Springer New York},\ \bibinfo {address} {New York, NY},\ \bibinfo {year}
  {2011})\ pp.\ \bibinfo {pages} {183--247}\BibitemShut {NoStop}%
\bibitem [{\citenamefont {Van~de Put}\ \emph {et~al.}(2018)\citenamefont
  {Van~de Put}, \citenamefont {Fischetti},\ and\ \citenamefont
  {Vandenberghe}}]{Maarten18APSmeeting}%
  \BibitemOpen
  \bibfield  {author} {\bibinfo {author} {\bibfnamefont {M.}~\bibnamefont
  {Van~de Put}}, \bibinfo {author} {\bibfnamefont {M.}~\bibnamefont
  {Fischetti}},\ and\ \bibinfo {author} {\bibfnamefont {W.}~\bibnamefont
  {Vandenberghe}},\ }\bibfield  {title} {\bibinfo {title} {{\it Accelerated
  modeling of electron transport using Bloch waves}},\ }in\ \href
  {http://meetings.aps.org/link/BAPS.2018.MAR.P12.6} {\emph {\bibinfo
  {booktitle} {APS March Meeting Abstracts}}},\ Vol.\ \bibinfo {volume} {2018}\
  (\bibinfo {year} {2018})\ p.\ \bibinfo {pages} {P12.0006}\BibitemShut
  {NoStop}%
\bibitem [{\citenamefont {Lent}\ and\ \citenamefont
  {Kirkner}(1990)}]{Lent90JAP}%
  \BibitemOpen
  \bibfield  {author} {\bibinfo {author} {\bibfnamefont {C.~S.}\ \bibnamefont
  {Lent}}\ and\ \bibinfo {author} {\bibfnamefont {D.~J.}\ \bibnamefont
  {Kirkner}},\ }\bibfield  {title} {\bibinfo {title} {{\it The quantum
  transmitting boundary method}},\ }\href
  {https://doi.org/https://doi.org/10.1063/1.345156} {\bibfield  {journal}
  {\bibinfo  {journal} {J. Appl. Phys.}\ }\textbf {\bibinfo {volume} {67}},\
  \bibinfo {pages} {6353} (\bibinfo {year} {1990})}\BibitemShut {NoStop}%
\bibitem [{\citenamefont {Weber}(1977)}]{Weber_1977}%
  \BibitemOpen
  \bibfield  {author} {\bibinfo {author} {\bibfnamefont {W.}~\bibnamefont
  {Weber}},\ }\bibfield  {title} {\bibinfo {title} {{\it Adiabatic bond charge
  model for the phonons in diamond, Si, Ge, and $\alpha$-Sn}},\ }\href
  {https://doi.org/https://doi.org/10.1103/PhysRevB.15.4789} {\bibfield
  {journal} {\bibinfo  {journal} {Phys. Rev. B}\ }\textbf {\bibinfo {volume}
  {15}},\ \bibinfo {pages} {4789} (\bibinfo {year} {1977})}\BibitemShut
  {NoStop}%
\bibitem [{\citenamefont {Kunc}\ and\ \citenamefont
  {Nielsen}(1979)}]{Kunc_1979}%
  \BibitemOpen
  \bibfield  {author} {\bibinfo {author} {\bibfnamefont {K.}~\bibnamefont
  {Kunc}}\ and\ \bibinfo {author} {\bibfnamefont {O.~H.}\ \bibnamefont
  {Nielsen}},\ }\bibfield  {title} {\bibinfo {title} {{\it Lattice dynamics of
  zincblende structure compounds II. Shell model}},\ }\href
  {https://doi.org/https://doi.org/10.1016/0010-4655(79)90104-8} {\bibfield
  {journal} {\bibinfo  {journal} {Comp. Phys. Commun.}\ }\textbf {\bibinfo
  {volume} {17}},\ \bibinfo {pages} {413} (\bibinfo {year} {1979})}\BibitemShut
  {NoStop}%
\bibitem [{\citenamefont {Noffsinger}\ \emph {et~al.}(2010)\citenamefont
  {Noffsinger}, \citenamefont {Giustino}, \citenamefont {Malone}, \citenamefont
  {Park}, \citenamefont {Louie},\ and\ \citenamefont {Cohen}}]{EPW_2010}%
  \BibitemOpen
  \bibfield  {author} {\bibinfo {author} {\bibfnamefont {J.}~\bibnamefont
  {Noffsinger}}, \bibinfo {author} {\bibfnamefont {F.}~\bibnamefont
  {Giustino}}, \bibinfo {author} {\bibfnamefont {B.~D.}\ \bibnamefont
  {Malone}}, \bibinfo {author} {\bibfnamefont {C.-H.}\ \bibnamefont {Park}},
  \bibinfo {author} {\bibfnamefont {S.~G.}\ \bibnamefont {Louie}},\ and\
  \bibinfo {author} {\bibfnamefont {M.~L.}\ \bibnamefont {Cohen}},\ }\bibfield
  {title} {\bibinfo {title} {{\it EPW: A program for calculating the
  electron–phonon coupling using maximally localized Wannier functions}},\
  }\href {https://doi.org/https://doi.org/10.1016/j.cpc.2010.08.027} {\bibfield
   {journal} {\bibinfo  {journal} {Comp. Phys. Commun.}\ }\textbf {\bibinfo
  {volume} {181}},\ \bibinfo {pages} {2140} (\bibinfo {year}
  {2010})}\BibitemShut {NoStop}%
\bibitem [{\citenamefont {Landau}\ \emph {et~al.}(1986)\citenamefont {Landau},
  \citenamefont {Kosevich}, \citenamefont {Pitaevskii},\ and\ \citenamefont
  {Lifshitz}}]{Landau_1970}%
  \BibitemOpen
  \bibfield  {author} {\bibinfo {author} {\bibfnamefont {L.~D.}\ \bibnamefont
  {Landau}}, \bibinfo {author} {\bibfnamefont {A.~M.}\ \bibnamefont
  {Kosevich}}, \bibinfo {author} {\bibfnamefont {L.~P.}\ \bibnamefont
  {Pitaevskii}},\ and\ \bibinfo {author} {\bibfnamefont {E.~M.}\ \bibnamefont
  {Lifshitz}},\ }\href
  {https://doi.org/https://doi.org/10.1016/C2009-0-25521-8} {\emph {\bibinfo
  {title} {{\it Theory of Elasticity (Course of Theoretical Physics), Third
  edition}}}},\ Vol.~\bibinfo {volume} {7}\ (\bibinfo  {publisher}
  {Butterworth-Heinemann - Elsevier},\ \bibinfo {address} {Oxford, United
  Kingdom},\ \bibinfo {year} {1986})\BibitemShut {NoStop}%
\bibitem [{\citenamefont {Ridley}(1993)}]{Ridley_1993}%
  \BibitemOpen
  \bibfield  {author} {\bibinfo {author} {\bibfnamefont {B.}~\bibnamefont
  {Ridley}},\ }\bibfield  {title} {\bibinfo {title} {Electron-hybridon
  interaction in a quantum well},\ }\href
  {https://doi.org/https://doi.org/10.1103/PhysRevB.47.4592} {\bibfield
  {journal} {\bibinfo  {journal} {Phys. Rev. B}\ }\textbf {\bibinfo {volume}
  {47}},\ \bibinfo {pages} {4592} (\bibinfo {year} {1993})}\BibitemShut
  {NoStop}%
\bibitem [{\citenamefont {Fischetti}\ and\ \citenamefont
  {Vandenberghe}(2016{\natexlab{a}})}]{Fischetti_2016}%
  \BibitemOpen
  \bibfield  {author} {\bibinfo {author} {\bibfnamefont {M.~V.}\ \bibnamefont
  {Fischetti}}\ and\ \bibinfo {author} {\bibfnamefont {W.~G.}\ \bibnamefont
  {Vandenberghe}},\ }\bibfield  {title} {\bibinfo {title} {{\it Mermin-Wagner
  theorem, flexural modes, and degraded carrier mobility in two-dimensional
  crystals with broken horizontal mirror symmetry}},\ }\href
  {https://doi.org/https://doi.org/10.1103/PhysRevB.93.155413} {\bibfield
  {journal} {\bibinfo  {journal} {Phys. Rev. B}\ }\textbf {\bibinfo {volume}
  {93}},\ \bibinfo {pages} {155413} (\bibinfo {year}
  {2016}{\natexlab{a}})}\BibitemShut {NoStop}%
\bibitem [{cno()}]{cnote3}%
  \BibitemOpen
  \href@noop {} {}\bibinfo {note} {Strictly speaking, perhaps being excessively
  pedantic, a cell containing $N_{\rm a}$ atoms sustains 3 acoustic phonons and
  $3(N_{\rm a}-1)$ optical phonons. However, here we adopt the more physically
  meaningful terminology that is usually employed in the literature, accounting
  for the folding of the bands into the 1D Brillouin zone. Therefore, we label
  `acoustic/optical' the phonons that describe acoustic/optical modes in a unit
  cell, as in the bulk. We also ignore the higher-frequency phonons associated
  with the vibrations of the H atoms that terminate the surfaces/edges and the
  possible relaxation of the Si or C atoms at the surfaces or edges of the
  structure.}\BibitemShut {Stop}%
\bibitem [{\citenamefont {Fanciulli}\ \emph {et~al.}(2016)\citenamefont
  {Fanciulli}, \citenamefont {Belli}, \citenamefont {Paleari}, \citenamefont
  {Lamperti}, \citenamefont {M.},\ and\ \citenamefont
  {Pizio}}]{Fanciulli_2016}%
  \BibitemOpen
  \bibfield  {author} {\bibinfo {author} {\bibfnamefont {M.}~\bibnamefont
  {Fanciulli}}, \bibinfo {author} {\bibfnamefont {M.}~\bibnamefont {Belli}},
  \bibinfo {author} {\bibfnamefont {S.}~\bibnamefont {Paleari}}, \bibinfo
  {author} {\bibfnamefont {A.}~\bibnamefont {Lamperti}}, \bibinfo {author}
  {\bibfnamefont {S.}~\bibnamefont {M.}},\ and\ \bibinfo {author}
  {\bibfnamefont {A.}~\bibnamefont {Pizio}},\ }\bibfield  {title} {\bibinfo
  {title} {{\it Defects and Dopants in Silicon Nanowires Produced by
  Metal-Assisted Chemical Etching}},\ }\href
  {https://doi.org/https://doi.org/10.1149/2.0171604jss} {\bibfield  {journal}
  {\bibinfo  {journal} {ECS J. Solid State Sci. Technol.}\ }\textbf {\bibinfo
  {volume} {5}},\ \bibinfo {pages} {P3138} (\bibinfo {year}
  {2016})}\BibitemShut {NoStop}%
\bibitem [{\citenamefont {Veerbeek}\ \emph {et~al.}(2017)\citenamefont
  {Veerbeek}, \citenamefont {Ye}, \citenamefont {Vijselaar}, \citenamefont
  {Kudernac}, \citenamefont {van~der Wielb},\ and\ \citenamefont
  {Huskens}}]{Verbeeck_2017}%
  \BibitemOpen
  \bibfield  {author} {\bibinfo {author} {\bibfnamefont {J.}~\bibnamefont
  {Veerbeek}}, \bibinfo {author} {\bibfnamefont {L.}~\bibnamefont {Ye}},
  \bibinfo {author} {\bibfnamefont {W.}~\bibnamefont {Vijselaar}}, \bibinfo
  {author} {\bibfnamefont {T.}~\bibnamefont {Kudernac}}, \bibinfo {author}
  {\bibfnamefont {W.~G.}\ \bibnamefont {van~der Wielb}},\ and\ \bibinfo
  {author} {\bibfnamefont {J.}~\bibnamefont {Huskens}},\ }\bibfield  {title}
  {\bibinfo {title} {{\it Highly doped silicon nanowires by monolayer
  doping}},\ }\href {https://doi.org/https://doi.org/10.1039/c6nr07623h}
  {\bibfield  {journal} {\bibinfo  {journal} {Nanoscale}\ }\textbf {\bibinfo
  {volume} {9}},\ \bibinfo {pages} {2836} (\bibinfo {year} {2017})}\BibitemShut
  {NoStop}%
\bibitem [{\citenamefont {Choi}\ \emph {et~al.}(2021)\citenamefont {Choi},
  \citenamefont {Nipane}, \citenamefont {Kim}, \citenamefont {Ziffer},
  \citenamefont {Datta}, \citenamefont {Borah}, \citenamefont {Jung},
  \citenamefont {Kim}, \citenamefont {Rhodes}, \citenamefont {Jundal},
  \citenamefont {Lamport}, \citenamefont {Lee}, \citenamefont {Zangiabadi},
  \citenamefont {Nair}, \citenamefont {Taniguchi}, \citenamefont {Watanmabe},
  \citenamefont {Kymissis}, \citenamefont {Pasupathy}, \citenamefont {Lipson},
  \citenamefont {Zhu}, \citenamefont {Yoo}, \citenamefont {Hone},\ and\
  \citenamefont {Teherani}}]{Choi_2021}%
  \BibitemOpen
  \bibfield  {author} {\bibinfo {author} {\bibfnamefont {M.~S.}\ \bibnamefont
  {Choi}}, \bibinfo {author} {\bibfnamefont {A.}~\bibnamefont {Nipane}},
  \bibinfo {author} {\bibfnamefont {B.~S.~Y.}\ \bibnamefont {Kim}}, \bibinfo
  {author} {\bibfnamefont {M.~E.}\ \bibnamefont {Ziffer}}, \bibinfo {author}
  {\bibfnamefont {I.}~\bibnamefont {Datta}}, \bibinfo {author} {\bibfnamefont
  {A.}~\bibnamefont {Borah}}, \bibinfo {author} {\bibfnamefont
  {Y.}~\bibnamefont {Jung}}, \bibinfo {author} {\bibfnamefont {B.}~\bibnamefont
  {Kim}}, \bibinfo {author} {\bibfnamefont {D.}~\bibnamefont {Rhodes}},
  \bibinfo {author} {\bibfnamefont {A.}~\bibnamefont {Jundal}}, \bibinfo
  {author} {\bibfnamefont {Z.~A.}\ \bibnamefont {Lamport}}, \bibinfo {author}
  {\bibfnamefont {M.}~\bibnamefont {Lee}}, \bibinfo {author} {\bibfnamefont
  {A.}~\bibnamefont {Zangiabadi}}, \bibinfo {author} {\bibfnamefont {M.~N.}\
  \bibnamefont {Nair}}, \bibinfo {author} {\bibfnamefont {T.}~\bibnamefont
  {Taniguchi}}, \bibinfo {author} {\bibfnamefont {K.}~\bibnamefont
  {Watanmabe}}, \bibinfo {author} {\bibfnamefont {I.}~\bibnamefont {Kymissis}},
  \bibinfo {author} {\bibfnamefont {A.~N.}\ \bibnamefont {Pasupathy}}, \bibinfo
  {author} {\bibfnamefont {M.}~\bibnamefont {Lipson}}, \bibinfo {author}
  {\bibfnamefont {X.}~\bibnamefont {Zhu}}, \bibinfo {author} {\bibfnamefont
  {W.~J.}\ \bibnamefont {Yoo}}, \bibinfo {author} {\bibfnamefont
  {J.}~\bibnamefont {Hone}},\ and\ \bibinfo {author} {\bibfnamefont {J.~T.}\
  \bibnamefont {Teherani}},\ }\bibfield  {title} {\bibinfo {title} {{\it High
  carrier mobility in graphene doped using a monolayer of tungsten
  oxyselenide}},\ }\href
  {https://doi.org/https://doi.org/10.1038/s41928-021-00657-y} {\bibfield
  {journal} {\bibinfo  {journal} {Nat. Electron.}\ }\textbf {\bibinfo {volume}
  {8}},\ \bibinfo {pages} {731} (\bibinfo {year} {2021})}\BibitemShut {NoStop}%
\bibitem [{\citenamefont {Giannozzi}\ \emph {et~al.}(2009)\citenamefont
  {Giannozzi}, \citenamefont {Baroni}, \citenamefont {Bonini}, \citenamefont
  {Calandra}, \citenamefont {Car}, \citenamefont {Cavazzoni}, \citenamefont
  {Ceresoli}, \citenamefont {Chiarotti}, \citenamefont {Cococcioni},
  \citenamefont {Dabo}, \citenamefont {Dal~Corso}, \citenamefont
  {de~Gironcoli}, \citenamefont {Fabris}, \citenamefont {Fratesi},
  \citenamefont {Gebauer}, \citenamefont {Gerstmann}, \citenamefont
  {Gougoussis}, \citenamefont {Kokalj}, \citenamefont {Lazzeri}, \citenamefont
  {Martin-Samos}, \citenamefont {Marzari}, \citenamefont {Mauri}, \citenamefont
  {Mazzarello}, \citenamefont {Paolini}, \citenamefont {Pasquarello},
  \citenamefont {Paulatto}, \citenamefont {Sbraccia}, \citenamefont {Scandolo},
  \citenamefont {Sclauzero}, \citenamefont {Seitsonen}, \citenamefont
  {Smogunov}, \citenamefont {Paolo}, ,\ and\ \citenamefont
  {M~Wentzcovitch}}]{giannozzi2009quantum}%
  \BibitemOpen
  \bibfield  {author} {\bibinfo {author} {\bibfnamefont {P.}~\bibnamefont
  {Giannozzi}}, \bibinfo {author} {\bibfnamefont {S.}~\bibnamefont {Baroni}},
  \bibinfo {author} {\bibfnamefont {N.}~\bibnamefont {Bonini}}, \bibinfo
  {author} {\bibfnamefont {M.}~\bibnamefont {Calandra}}, \bibinfo {author}
  {\bibfnamefont {R.}~\bibnamefont {Car}}, \bibinfo {author} {\bibfnamefont
  {C.}~\bibnamefont {Cavazzoni}}, \bibinfo {author} {\bibfnamefont
  {D.}~\bibnamefont {Ceresoli}}, \bibinfo {author} {\bibfnamefont {G.~L.}\
  \bibnamefont {Chiarotti}}, \bibinfo {author} {\bibfnamefont {M.}~\bibnamefont
  {Cococcioni}}, \bibinfo {author} {\bibfnamefont {I.}~\bibnamefont {Dabo}},
  \bibinfo {author} {\bibfnamefont {A.}~\bibnamefont {Dal~Corso}}, \bibinfo
  {author} {\bibfnamefont {S.}~\bibnamefont {de~Gironcoli}}, \bibinfo {author}
  {\bibfnamefont {S.}~\bibnamefont {Fabris}}, \bibinfo {author} {\bibfnamefont
  {G.}~\bibnamefont {Fratesi}}, \bibinfo {author} {\bibfnamefont
  {R.}~\bibnamefont {Gebauer}}, \bibinfo {author} {\bibfnamefont
  {U.}~\bibnamefont {Gerstmann}}, \bibinfo {author} {\bibfnamefont
  {C.}~\bibnamefont {Gougoussis}}, \bibinfo {author} {\bibfnamefont
  {A.}~\bibnamefont {Kokalj}}, \bibinfo {author} {\bibfnamefont
  {M.}~\bibnamefont {Lazzeri}}, \bibinfo {author} {\bibfnamefont
  {L.}~\bibnamefont {Martin-Samos}}, \bibinfo {author} {\bibfnamefont
  {N.}~\bibnamefont {Marzari}}, \bibinfo {author} {\bibfnamefont
  {F.}~\bibnamefont {Mauri}}, \bibinfo {author} {\bibfnamefont
  {R.}~\bibnamefont {Mazzarello}}, \bibinfo {author} {\bibfnamefont
  {S.}~\bibnamefont {Paolini}}, \bibinfo {author} {\bibfnamefont
  {A.}~\bibnamefont {Pasquarello}}, \bibinfo {author} {\bibfnamefont
  {L.}~\bibnamefont {Paulatto}}, \bibinfo {author} {\bibfnamefont
  {C.}~\bibnamefont {Sbraccia}}, \bibinfo {author} {\bibfnamefont
  {S.}~\bibnamefont {Scandolo}}, \bibinfo {author} {\bibfnamefont
  {G.}~\bibnamefont {Sclauzero}}, \bibinfo {author} {\bibfnamefont {A.~P.}\
  \bibnamefont {Seitsonen}}, \bibinfo {author} {\bibfnamefont {A.}~\bibnamefont
  {Smogunov}}, \bibinfo {author} {\bibfnamefont {U.}~\bibnamefont {Paolo}}, ,\
  and\ \bibinfo {author} {\bibfnamefont {R.}~\bibnamefont {M~Wentzcovitch}},\
  }\bibfield  {title} {\bibinfo {title} {{\it QUANTUM ESPRESSO: a modular and
  open-source software project for quantum simulations of materials}},\ }\href
  {https://doi.org/https://doi.org/10.1088/0953-8984/21/39/395502} {\bibfield
  {journal} {\bibinfo  {journal} {J. Phys.: Condens. Matter}\ }\textbf
  {\bibinfo {volume} {21}},\ \bibinfo {pages} {395502} (\bibinfo {year}
  {2009})}\BibitemShut {NoStop}%
\bibitem [{\citenamefont {Perdew}\ \emph {et~al.}(1996)\citenamefont {Perdew},
  \citenamefont {Burke},\ and\ \citenamefont
  {Ernzerhof}}]{perdew1996generalized}%
  \BibitemOpen
  \bibfield  {author} {\bibinfo {author} {\bibfnamefont {J.~P.}\ \bibnamefont
  {Perdew}}, \bibinfo {author} {\bibfnamefont {K.}~\bibnamefont {Burke}},\ and\
  \bibinfo {author} {\bibfnamefont {M.}~\bibnamefont {Ernzerhof}},\ }\bibfield
  {title} {\bibinfo {title} {{\it Generalized gradient approximation made
  simple}},\ }\href
  {https://doi.org/https://doi.org/10.1103/PhysRevLett.77.3865} {\bibfield
  {journal} {\bibinfo  {journal} {Phys. Rev. Lett.}\ }\textbf {\bibinfo
  {volume} {77}},\ \bibinfo {pages} {3865} (\bibinfo {year}
  {1996})}\BibitemShut {NoStop}%
\bibitem [{\citenamefont {Fischetti}\ and\ \citenamefont
  {Laux}(1996)}]{fischetti_1996}%
  \BibitemOpen
  \bibfield  {author} {\bibinfo {author} {\bibfnamefont {M.~V.}\ \bibnamefont
  {Fischetti}}\ and\ \bibinfo {author} {\bibfnamefont {S.~E.}\ \bibnamefont
  {Laux}},\ }\bibfield  {title} {\bibinfo {title} {{\it Band structure,
  deformation potentials, and carrier mobility in strained Si, Ge, and SiGe
  alloys}},\ }\href {https://doi.org/https://doi.org/10.1063/1.363052}
  {\bibfield  {journal} {\bibinfo  {journal} {J. Appl. Phys.}\ }\textbf
  {\bibinfo {volume} {80}},\ \bibinfo {pages} {2234} (\bibinfo {year}
  {1996})}\BibitemShut {NoStop}%
\bibitem [{\citenamefont {Fischetti}\ and\ \citenamefont
  {Laux}(1988)}]{fischetti1988monte}%
  \BibitemOpen
  \bibfield  {author} {\bibinfo {author} {\bibfnamefont {M.~V.}\ \bibnamefont
  {Fischetti}}\ and\ \bibinfo {author} {\bibfnamefont {S.~E.}\ \bibnamefont
  {Laux}},\ }\bibfield  {title} {\bibinfo {title} {{\it Monte Carlo analysis of
  electron transport in small semiconductor devices including band-structure
  and space-charge effects}},\ }\href
  {https://doi.org/https://doi.org/10.1103/PhysRevB.38.9721} {\bibfield
  {journal} {\bibinfo  {journal} {Phys. Rev. B}\ }\textbf {\bibinfo {volume}
  {38}},\ \bibinfo {pages} {9721} (\bibinfo {year} {1988})}\BibitemShut
  {NoStop}%
\bibitem [{\citenamefont {Zhang}\ \emph {et~al.}(1993)\citenamefont {Zhang},
  \citenamefont {Yeh},\ and\ \citenamefont {Zunger}}]{Zunger_1993}%
  \BibitemOpen
  \bibfield  {author} {\bibinfo {author} {\bibfnamefont {S.~B.}\ \bibnamefont
  {Zhang}}, \bibinfo {author} {\bibfnamefont {C.-Y.}\ \bibnamefont {Yeh}},\
  and\ \bibinfo {author} {\bibfnamefont {A.}~\bibnamefont {Zunger}},\
  }\bibfield  {title} {\bibinfo {title} {{\it Electronic structure of
  semiconductor quantum films}},\ }\href
  {https://doi.org/https://doi.org/10.1103/PhysRevB.48.11204} {\bibfield
  {journal} {\bibinfo  {journal} {Phys. Rev. B}\ }\textbf {\bibinfo {volume}
  {48}},\ \bibinfo {pages} {11204} (\bibinfo {year} {1993})}\BibitemShut
  {NoStop}%
\bibitem [{bno()}]{bnote2}%
  \BibitemOpen
  \href@noop {} {}\bibinfo {note} {Although the scattering rates shown in
  Fig.~\ref{fig:SiNW_scattering_BCs} have been calculated assuming the presence
  of a drain bias, the zero-bias momentum relaxation rates used in
  Eq.~(\ref{eq:low_field_mobility}) exhibit qualitatively the same behavior,
  FSBCs resulting in weaker scattering than CBCs at electrone energies smaller
  than about 0.1~eV, but higher at larger energies.}\BibitemShut {Stop}%
\bibitem [{\citenamefont {Fischetti}\ and\ \citenamefont
  {Vandenberghe}(2016{\natexlab{b}})}]{Maxbook}%
  \BibitemOpen
  \bibfield  {author} {\bibinfo {author} {\bibfnamefont {M.~V.}\ \bibnamefont
  {Fischetti}}\ and\ \bibinfo {author} {\bibfnamefont {W.~G.}\ \bibnamefont
  {Vandenberghe}},\ }\href
  {https://doi.org/https://doi.org/10.1007/978-3-319-01101-1} {\emph {\bibinfo
  {title} {{\it Advanced Physics of Electron Transport in Semiconductors and
  Nanostructures}}}}\ (\bibinfo  {publisher} {Springer},\ \bibinfo {year}
  {2016})\BibitemShut {NoStop}%
\bibitem [{\citenamefont {Kurokawa}\ \emph {et~al.}(1996)\citenamefont
  {Kurokawa}, \citenamefont {Nomura}, \citenamefont {Takemori},\ and\
  \citenamefont {Aoyagi}}]{Kurokawa_1996}%
  \BibitemOpen
  \bibfield  {author} {\bibinfo {author} {\bibfnamefont {Y.}~\bibnamefont
  {Kurokawa}}, \bibinfo {author} {\bibfnamefont {S.}~\bibnamefont {Nomura}},
  \bibinfo {author} {\bibfnamefont {T.}~\bibnamefont {Takemori}},\ and\
  \bibinfo {author} {\bibfnamefont {Y.}~\bibnamefont {Aoyagi}},\ }\bibfield
  {title} {\bibinfo {title} {{\it Large-scale calculation of optical dielectric
  functions of diamond nanocrystallites}},\ }\href
  {https://doi.org/https://doi.org/10.1103/PhysRevB.61.12616} {\bibfield
  {journal} {\bibinfo  {journal} {Phys. Rev. B}\ }\textbf {\bibinfo {volume}
  {61}},\ \bibinfo {pages} {12616} (\bibinfo {year} {1996})}\BibitemShut
  {NoStop}%
\bibitem [{\citenamefont {Ayedh}\ and\ \citenamefont
  {Wacker}(2011)}]{Ayedh_2011}%
  \BibitemOpen
  \bibfield  {author} {\bibinfo {author} {\bibfnamefont {H.~M.}\ \bibnamefont
  {Ayedh}}\ and\ \bibinfo {author} {\bibfnamefont {A.}~\bibnamefont {Wacker}},\
  }\bibfield  {title} {\bibinfo {title} {{\it Acoustic phonons in nanowires
  with embedded heterostructures}},\ }\href
  {https://doi.org/https://doi.org/10.1155/2011/743846} {\bibfield  {journal}
  {\bibinfo  {journal} {J. Nanomater.}\ }\textbf {\bibinfo {volume} {2011}},\
  \bibinfo {pages} {743846} (\bibinfo {year} {2011})}\BibitemShut {NoStop}%
\bibitem [{\citenamefont {Amorim}\ and\ \citenamefont
  {Guinea}(2013)}]{Amorim_2013}%
  \BibitemOpen
  \bibfield  {author} {\bibinfo {author} {\bibfnamefont {B.}~\bibnamefont
  {Amorim}}\ and\ \bibinfo {author} {\bibfnamefont {F.}~\bibnamefont
  {Guinea}},\ }\bibfield  {title} {\bibinfo {title} {{\it Flexural mode of
  graphene on a substrate}},\ }\href
  {https://doi.org/https://doi.org/10.1103/PhysRevB.88.115418} {\bibfield
  {journal} {\bibinfo  {journal} {Phys. Rev. B}\ }\textbf {\bibinfo {volume}
  {88}},\ \bibinfo {pages} {115418} (\bibinfo {year} {2013})}\BibitemShut
  {NoStop}%
\end{thebibliography}%

\end{document}